\documentclass[11pt]{article}
\pdfoutput=1
\usepackage{amsmath,amssymb,epsfig,amsfonts}
\usepackage{graphicx}
\usepackage{subfig}
\usepackage[usenames, dvipsnames]{color}
\usepackage{hyperref}
\usepackage{cite}
\usepackage{multirow}
\usepackage{slashed}
\usepackage{verbatim}
\usepackage{float}
\usepackage{enumitem}
\usepackage{rotating}
\usepackage{xcolor}
\usepackage{multirow}
\usepackage{bm}
\usepackage[normalem]{ulem}
\usepackage{cleveref}
\usepackage{booktabs} 
\usepackage{url}

\numberwithin{equation}{section}

\definecolor{dark green}{rgb}{0.0, 0.65, 0.31}
\definecolor{k.gr}{rgb}{0.3, 0.73, 0.09}

\definecolor{bluebox}{RGB}{0,
102, 204}
\definecolor{blue2}{RGB}{0,
133, 204}
\definecolor{blue3}{RGB}{19,
153, 222}

\definecolor{blue4}{rgb}{0.12156863, 0.46666667, 0.70588235}

\def\a{\alpha}

\def\g{\gamma}
\def\G{\Gamma}
\def\d{\delta}

\def\D{\Delta}

\def\L{\Lambda}

\def\f{\phi}

\def\m{\mu}
\def\n{\nu}

\def\O{\Omega}

\def\r{\rho}

\def\S{\Sigma}

\def\nl{\newline}
\def\te{\text}

\graphicspath{ {./figures/} }

\newtheorem{definition}{Definition}

\newcommand{\be}{\begin{equation}}
\newcommand{\ee}{\end{equation}}
\newcommand{\ba}{\begin{aligned}}
\newcommand{\ea}{\end{aligned}}

\def\ii{{\text{i}}}
\def\exp{{\text{exp}}}
\def\d{{\text{d}}}

\def\CO{{\mathcal{O}}}

\usepackage[most]{tcolorbox}

\newcounter{conj}
\renewcommand{\theconj}{\arabic{conj}}
\newenvironment{conj}[2][]{%
\refstepcounter{conj}%
\begin{tcolorbox}[enhanced,attach boxed title to top left={yshift=-3mm,yshifttext=-1mm},
  colback=white,colframe=blue4!80!white,colbacktitle=blue4!30!white,coltitle=black,
  title=Conjecture~\theconj:~#1,fonttitle=\bfseries,
  boxed title style={size=small,colframe=blue4!80!white} ]
 \label{#2}
 }{ 
\end{tcolorbox}
}

\newenvironment{otherbox}[2][]{%
\begin{tcolorbox}[enhanced,attach boxed title to top left={yshift=-3mm,yshifttext=-1mm},
  colback=white,colframe=blue4!80!white,colbacktitle=white,coltitle=black,
  title=#1,fonttitle=\bfseries,
  boxed title style={size=small,colframe=blue4!80!white} ]
  \label{#2}
 }{
\end{tcolorbox}
}

\begin{document}

\thispagestyle{empty}
\vspace*{-2.5cm} 
\begin{flushright}
{\tt IFT-UAM/CSIC-21-10}\\
\end{flushright}

\vspace{1.2cm}
\begin{center}
\huge{\bf Lectures on the Swampland Program  \\ in String Compactifications }
\\[12mm] 

{ \Large QFT \& Geometry Summer School} \\
\medskip
{\large July 2020
}

 \vspace*{1.5cm}

\Large{Marieke van Beest$^1$, Jos\'e Calder\'on-Infante$^2$, Delaram Mirfendereski$^3$, Irene Valenzuela$^4$\\[8mm]}
\footnotesize{$^1$Mathematical Institute, University of Oxford,\\ Andrew-Wiles Building, Woodstock Road, Oxford, OX2 6GG, UK }\\
\footnotesize{$^2$Instituto de F\'{\i}sica Te\'orica IFT-UAM/CSIC,
C/ Nicol\'as Cabrera 13-15, 
28049 Madrid, Spain}\\ 
\footnotesize{$^3$ Physics Department, Boğaziçi University, 34342 Bebek / Istanbul, TURKEY}\\
\footnotesize{$^4$Jefferson Physical Laboratory, Harvard University, Cambridge, MA 02138, USA}\\

\vspace*{2cm}

\small{\bf Abstract} \\
\end{center}
\begin{center}
\begin{minipage}[h]{\textwidth}
The Swampland program aims to determine the constraints that an effective field theory must satisfy to be consistent with a UV embedding in a quantum gravity theory. Different proposals have been formulated in the form of Swampland conjectures. In these lecture notes, we provide a pedagogical introduction to the most important Swampland conjectures, their connections and their realization in string theory compactifications. The notes are based on the series of lectures given by Irene Valenzuela at the online \emph{QFT and Geometry summer school} in July 2020.
\end{minipage}
\end{center}

\newpage

\tableofcontents


\section{Introduction}

\emph{Not everything} is possible in quantum gravity (QG), in the sense that not every model constructed following the rules of quantum field theory will be consistent once gravity is considered at the quantum level.
This reflects the fact that obtaining a consistent quantum gravity theory is not so simple, and can still hide many surprises for the physics at lower energies. 


The Swampland program aims to determine the constraints that an effective field theory must satisfy to be consistent with a UV embedding in a quantum gravity theory. They are dubbed \emph{swampland constraints}, and the different proposals are formulated in the form of Swampland conjectures. The goal is to identify these constraints, gather evidence to prove (or disprove) them in a quantum gravity framework, provide a rationale to explain them in a model-independent way and understand their phenomenological implications for low energy physics. Although the notion of the swampland is in principle not restricted to string theory, the swampland conjectures are often motivated by or checked in string theory setups. Indeed, string theory provides a perfect framework to quantitatively and rigorously test the conjectures, improving our understanding of the possible string theory compactifications on the way. Interestingly, we have recently revealed that many of these conjectures are actually related, suggesting that they might simply be different faces of some more fundamental  quantum gravity principles yet to be uncovered.

The swampland constraints may also have important implications for Particle Physics and Cosmology. They can provide new guiding principles to constructing Beyond Standard Model theories and progress in High Energy Physics. They may also induce UV/IR mixing that breaks the expectation of scale separation and potentially provides new insights on the naturalness issues observed in our universe. Hence, the existence of a swampland is great news for Phenomenology!\\

In these lectures, we provide a pedagogical introduction to the most important Swampland conjectures, their connections and their realization in string theory compactifications. The notes are based on the series of lectures given by Dr. Irene Valenzuela at the online \emph{QFT and Geometry summer school} \cite{school} in July 2020.

 These lecture notes are not intended to be a review, so many references belonging to this research field will be missing. We will sometimes provide concrete references to particular papers within each topic, but we will mainly be using the more pedagogical references, which are most well-suited for students. For a more complete list of references of the Swampland program, we refer the reader to \cite{Palti:2019pca}, where the Swampland program is also reviewed and introduced. Other key references for lecture notes about the Swampland program are \cite{Brennan:2017rbf,Palti:2020mwc}.




\section{The Swampland Program}

\subsection{What is the Swampland?}

Effective field theories (EFTs) have proven to be very useful in the history of physics. They provide a description of the physical phenomena which is valid up to a certain energy scale denoted as the cut-off scale $\Lambda$. Above this scale, the EFT breaks down and needs to be modified, but below $\Lambda$ it should provide a valid description of the physics, both in agreement with the full theory and with the experiments. 

\begin{figure}[ht]
\centering
\includegraphics[width=\textwidth]{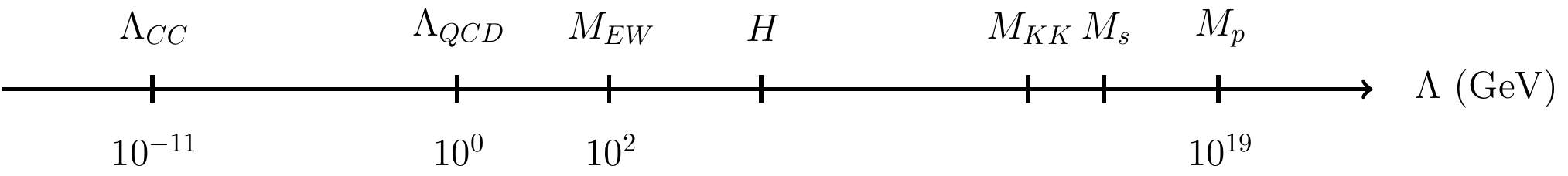}
\caption{Each EFT has an energy cut-off up to which the description is valid. Typical reference scales in high energy physics are indicated here, including the cosmological constant, QCD, electroweak symmetry breaking, and the Planck scale $M_p$. Also indicated is a possible estimation of the Hubble scale $H$ for inflation, the compactification scale $M_{KK}$ and the string scale $M_s$, at which stringy effects become important.}
\end{figure}
It is well-known that, given some UV theory, it is always possible to integrate out the UV degrees of freedom, and thereby obtain the low energy EFT. 
In this way, the effective Lagrangian of a $d$-dimensional theory can be divided into the renormalizable part and a tower of non-renormalizable operators (over the Planck mass $M_p$)
\be 
\mathcal{L}_{\text{eff}}=\mathcal{L}_{\text{ren}}+\sum_{n=d}^{\infty} \frac{\CO_n}{M_p^{n-d}}\,.
\ee

The question we are trying to answer with the Swampland program is whether this process can always be run in reverse; whether, \emph{starting from some quantum EFT weakly coupled to Einstein gravity, it can always be UV completed to a consistent theory of QG.}
The answer to this question turns out to be: no, we cannot UV complete any EFT in a way that is consistent with QG. This answer has its origin in the fact that not everything is possible in string theory. Even if the string landscape is huge, as we will discuss, it does not cover everything. The next question is, thus
: if this cannot always be done, what are the conditions that an EFT has to satisfy in order to make this possible?


In a sense, the fact that we can have a consistent EFT which becomes inconsistent when coupled to a gauge field or gravity is not a new thing. For instance, we know that there may appear new anomalies that need to be cancelled to guarantee consistency of the theory. To give an example, consider an EFT with an odd number of fermions in the fundamental representation of an $SU(2)$ flavour global symmetry. This EFT is perfectly sensible on its own, but when it is coupled to a dynamical gauge field that gauges the $SU(2)$ symmetry, the theory becomes inconsistent due to the presence of a Witten anomaly \cite{Witten:1982fp}.
We therefore have to require that an EFT coupled to a gauge field must be anomaly free. When the theory is coupled to gravity new anomalies will in general appear, but, interestingly, the absence of gravitational anomalies is not enough to guarantee quantum consistency of the theory. There are other constraints that one needs to satisfy, that simply seem to express the fact that arriving at a unitary theory of quantum gravity is not so simple. These additional constraints are called quantum gravity or \emph{Swampland constraints}. This brings us to our first definition of the lectures:
\\ 

\begin{otherbox}[The Swampland]{box:swamp}
Those apparently consistent (anomaly free) quantum EFTs that cannot be embedded in a UV consistent theory of quantum gravity \cite{Vafa:2005ui}.
\end{otherbox} 

Hence, we say that an EFT belongs to the Swampland when it does not satisfy the Swampland constraints, meaning that the theory cannot be UV completed in QG. The name was chosen in contrast to the \emph{Landscape}, which include those EFTs that are consistent with quantum gravity considerations.
The goal of the Swampland program is to determine where the boundary lies that separates the Swampland from the landscape.
From the point of view of string theory, the program attempts to classify the possible string geometries, i.e. string compactifications, to understand what can and cannot arise from string theory.

\begin{figure}[ht]
\centering
\includegraphics[width=0.7\textwidth]{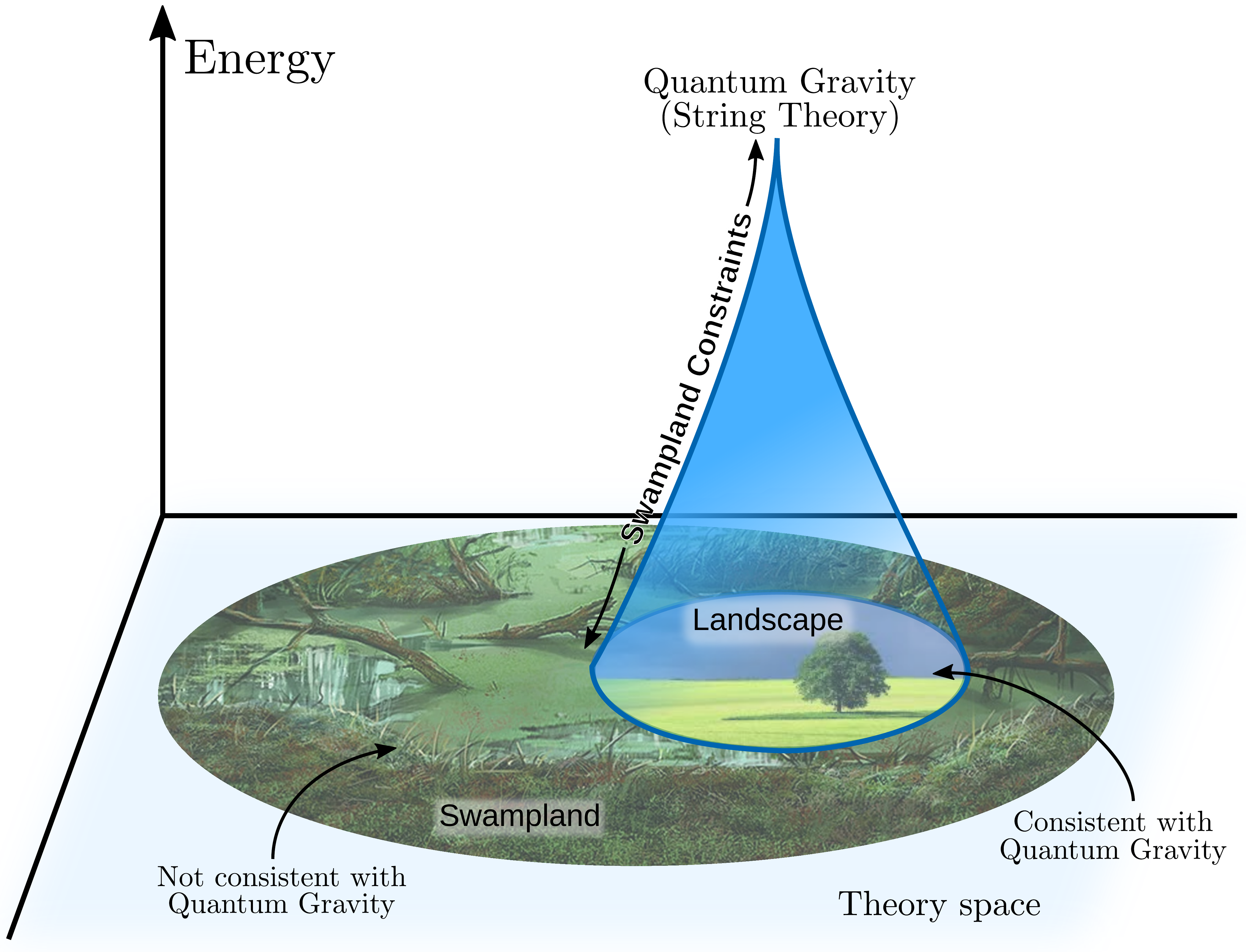}
\caption{The Swampland and Landscape of EFTs. The space of consistent EFTs forms a cone because Swampland constraints become stronger at high energies.}
\label{fig:cone}
\end{figure}

An important point to note is that these constraints, in order to be Swampland constraints, should disappear as the Planck mass goes to infinity and we decouple gravity. This also implies that they are more constraining as the energy at which the EFT should be valid increases, picking out in the end a (possibly unique) theory of QG. This is illustrated in figure \ref{fig:cone}, where the blue cone represents the set of consistent theories, which we call the landscape, and the Swampland surrounds the cone. The Swampland criteria become stronger as the energy increases and we get closer to the QG scale, as illustrated by the cone shape, and terminate in a consistent theory of QG. 
It is tantalizing to conclude, following the previous reasoning, that there is a unique QG theory in the UV, meaning that all consistent QG theories should be somehow connected to string theory. This is known as String Universality, which will be discussed in more detail in section \ref{sec:BPScomp}. As of today, it is an open question and one of the mysteries that we address in the Swampland program.

\begin{figure}[ht]
\centering
\includegraphics[width=0.35\textwidth]{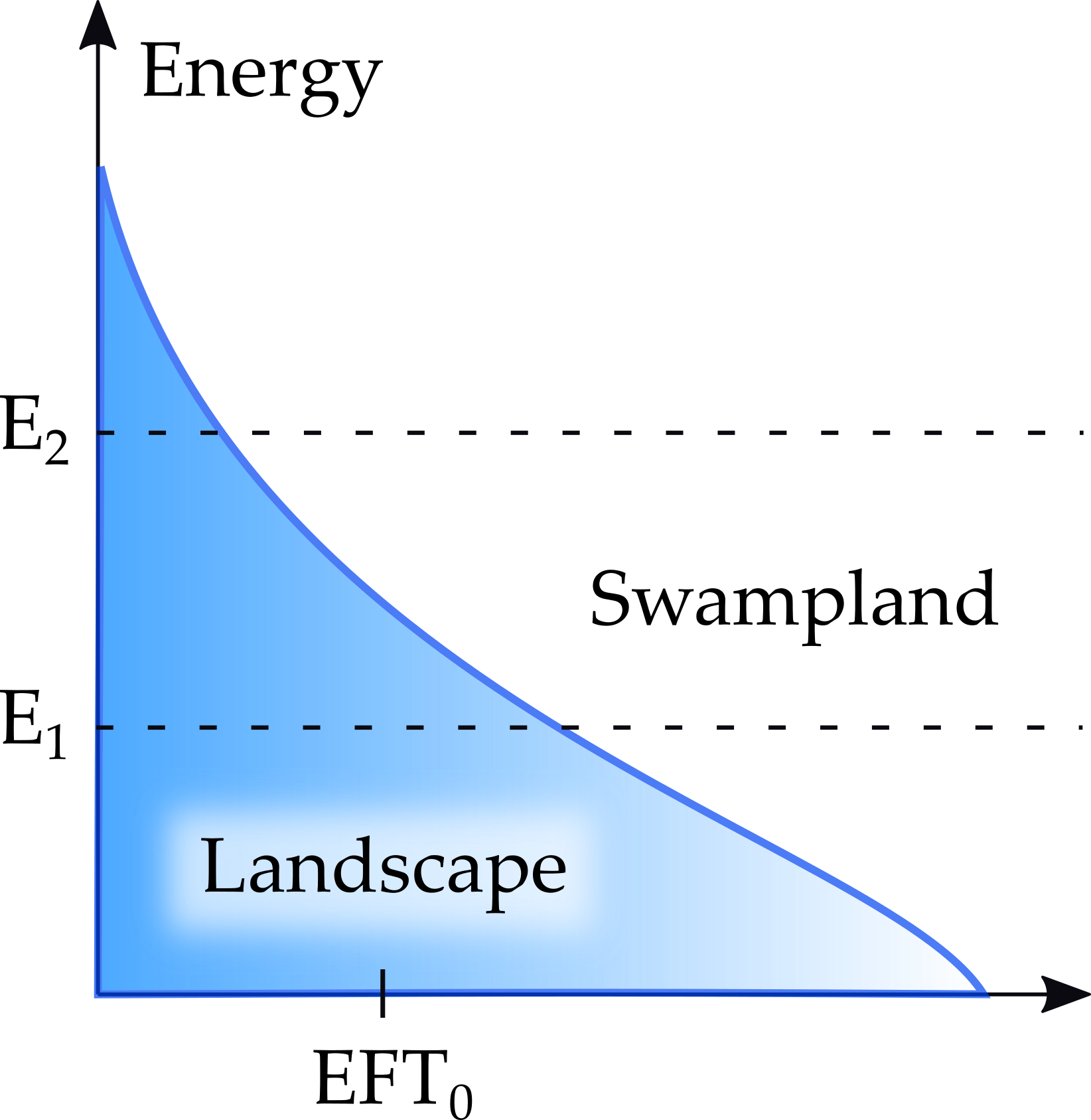}
\caption{The Swampland and Landscape of EFTs. Cross section of figure \ref{fig:cone}. The effective field theory (EFT$_0$) is in the landscape/swampland if it is used to describe a process of characteristic energy $E_1/E_2$.}
\label{fig:cone_section}
\end{figure}

As explained above, for each EFT there exists some cut-off $\Lambda_{\rm EFT}$ at which the effective description breaks down, typically because of the presence of new light degrees of freedom. Above this cut-off the EFT should be modified by ``integrating in'' these new degrees of freedom, resulting in the definition of a new EFT which is valid above that energy scale up to some new cut-off. 
However, this process cannot be repeated indefinitely: any EFT has a cut-off (dubbed the QG cut-off $\Lambda_{\rm QG}$) above which it cannot be amended to give a consistent QFT weakly coupled to Einstein gravity. In other words, it is not possible to ``integrate in'' the new light degrees of freedom while preserving the QFT description. This occurs, for example, if an infinite tower of new light states appears, they cannot simply be ``integrated in'': quantum gravitational effects become important and the EFT completely breaks down. It is then necessary to drastically change what we identify as the ``fundamental degrees of freedom" and go to a string theory description, grow an extra dimension, etc..
Therefore, an EFT belongs to the Swampland, not only if it has a feature which makes it impossible to be UV completed in QG, but also  if it is defined up to an energy scale which lies above its QG cut-off. This is represented in figure \ref{fig:cone_section}, where the same EFT used to describe a physical process at an energy $E_1$ belongs to the landscape while it belongs to the Swampland if the characteristic energy of the process is $E_2$.

To sum up, whether an EFT is in the Swampland depends on the energy scale at which we claim that it is valid. The higher this energy, the more difficult it is to be consistent with string theory (the landscape is smaller). The Swampland constraints will sometimes indicate features that the EFT must have, while others will define the QG cut-off at which the EFT should drastically break down.  


An interesting ramification of the Swampland program is that it has brought about a shift in attitude in the string phenomenology community. Instead of trying to find out, from a phenomenological point of view, which vacuum we are living in -- a daunting task given that the landscape is huge -- we can try to predict features of this vacuum by understanding what is not possible, i.e. where we do not live.
Thus, Swampland constraints can have phenomenological implications and give rise to new guiding principles for constructing Beyond Standard Model theories as well as cosmological models. Furthermore, the Swampland breaks with the logic of naturalness, which is based on scale separation, because if QG imposes non-trivial constraints in the IR, this constitutes UV/IR mixing which could explain the hierarchy problems that we observe. We will give some examples of this in the lectures, although the focus will be to understand the string theory and geometric realizations of the Swampland constraints.

\subsection{Swampland Conjectures}

\begin{figure}[ht]
\centering
\includegraphics[width=\textwidth]{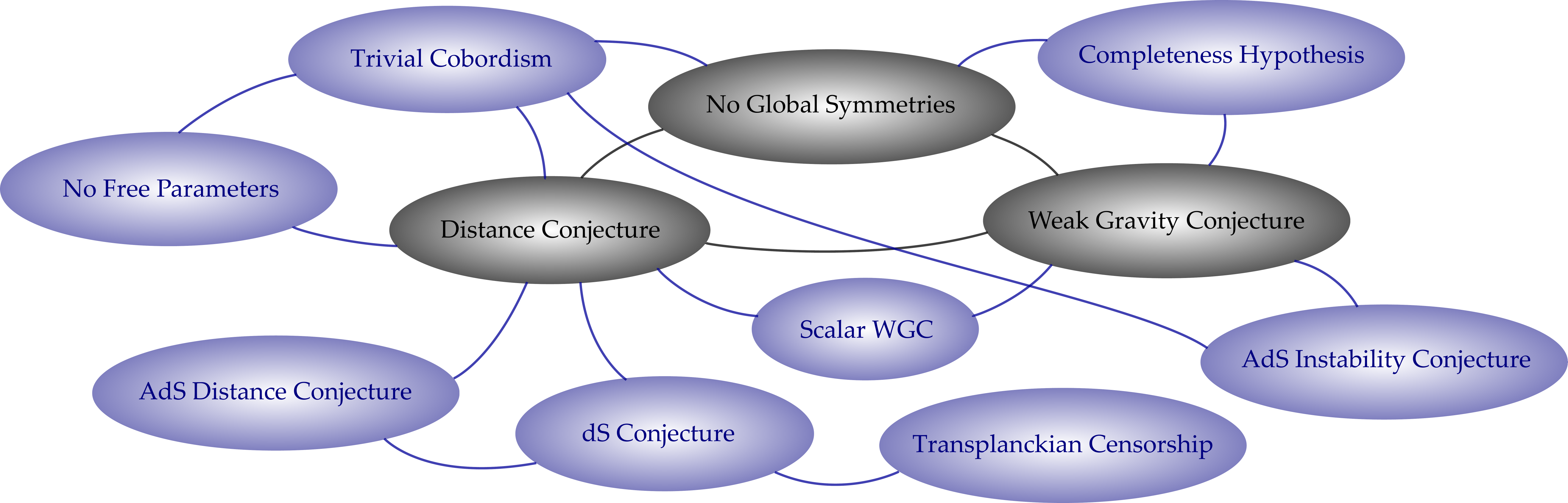}
\caption{Map of the Swampland conjectures. The conjectures in black are at the core of the Swampland program, and we will discuss them in detail in the following. The conjectures in  purple will also be discussed throughout the lectures, but sometimes in less detail.}
\label{fig:mapofconjectures}
\end{figure}

Of course, the Swampland constraints, that distinguish between what is in the landscape and what is in the Swampland, are not as well-understood as the gauge or gravitational anomalies. At the moment, we only have proposals. These proposals go under the name of Swampland conjectures because none of them are completely proven. However, in the last few years a significant amount of evidence has been gathered for several of them, which are currently widely accepted in the string theory community. The differences in the status of the evidence for each conjecture will become clear as this is discussed throughout the lectures. Figure \ref{fig:mapofconjectures} shows the Swampland conjectures as well as some connections between them.

In these lecture notes, we will go through the map and describe some of the most important conjectures. One of the most interesting things about this map are all the connections between the conjectures which are currently being discovered. This development is exciting because it might be indicating that some of the conjectures are actually different faces of the same underlying QG principles that are starting to appear. The hope is that they will all be unified in the future, improving our understanding of QG.

\begin{figure}[ht]
\centering
\includegraphics[width=\textwidth]{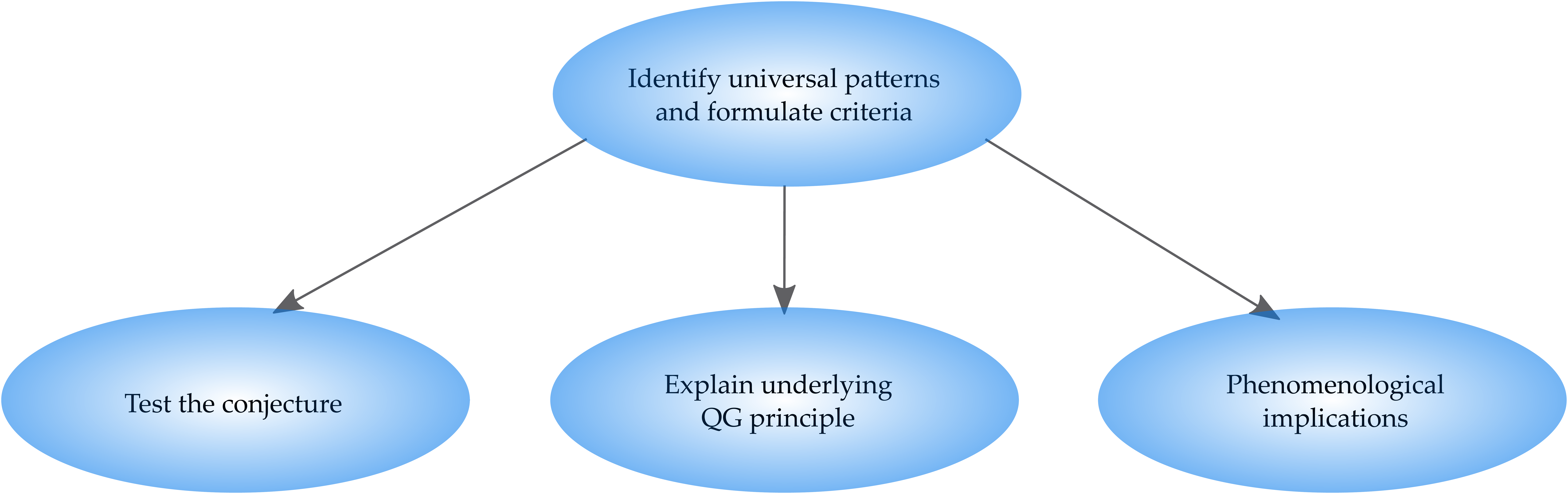}
\caption{Road map of the Swampland program.}
\label{fig:roadmap}
\end{figure}

Formulating the conjectures is only the first step in the Swampland program. Most of the work actually comes afterwards, as several tasks need to be accomplished before claiming success.  
This is illustrated in the scheme of figure \ref{fig:roadmap}.
First, one tries to identify universal patterns from the data we have from string theory or from black hole (BH) physics and formulate some criterion. Then a lot of work must go into  testing the conjectures and, if needed, refine and sharpen them. This usually implies that the conjecture is tested in string theory (or in AdS/CFT), since it is a consistent quantum gravity framework and provides a perfect playing ground to quantitatively check the conjecture. However, this is not sufficient to claim that a conjecture applies in general for QG. Although the string theory tests can be very rigorous, the proofs are typically restricted to specific classes of string compactifications so they are not completely universal. Hence, one should also be able to provide some physical rationale for the conjecture and explain what could go wrong otherwise. This is typically done using BH physics, consistency of the S-matrix, positivity constraints or some other model-independent approach. These arguments are, in general, only able to support mild versions of the conjectures, leaving several ambiguities. However, combined with the string theory tests, they can constitute strong evidence for the conjecture.  Finally, it is interesting to study the phenomenological implications of the conjectures, both for Particle Physics and Cosmology.
In these lecture notes, we will focus on explaining the conjectures themselves, their connections and how they are realized in string theory compactifications. 

\section{No Global Symmetries in Quantum Gravity}
\label{sec:globalsym}

The first conjecture, which is the oldest and the most widely accepted, is the statement that there are no global symmetries in QG. In order to discuss this conjecture, it is important to first define what we mean by a global symmetry in this context. Because, of course, this conjecture is not meant to imply that there are no symmetries in QG, just that they have to be associated to gauge degrees of freedom. A useful reference on this topic is \cite{Banks:2010zn}.
 

\subsection{No Global Symmetries Conjecture}


\begin{conj}[No Global Symmetries]{conj:noglobalsym}
There are no global symmetries in quantum gravity (i.e. any symmetry is either broken or gauged).
\end{conj}

A global symmetry can be defined as a transformation described by a unitary local operator that commutes with the Hamiltonian and acts non-trivially on the Hilbert space of physical states. The latter requires the existence of at least one charged local operator, so we say that the symmetry acts faithfully. The condition of commuting with the Hamiltonian is guaranteed by a local energy conservation condition in QFT implying that the stress-energy tensor is neutral. The local operators are required to satisfy a group law, although it is a very interesting question to understand whether this can be relaxed and whether other topological operators are likewise absent in QG. Analogously, we will also require that a global symmetry should map local operators into local operators, although it seems that this requirement might sometimes also be relaxed, as we will discuss in section \ref{sec:cob}.

{\definition[Global Symmetry] \label{def:globalsym}
$\exists$ unitary local operator $U(g)$, $g \in G$ that:
\begin{itemize}
\item Satisfies a group law: $U(g)U(g')=U(gg')$\,,
\item Acts non-trivially on the Hilbert space: $\exists$ a charged local operator $\CO(x)$ s.t. \\$U^\dagger (g) \CO(x) U(g) \neq \CO(x)$\,, 
\item Commutes with the Hamiltonian: $U^\dagger T_{\mu \nu} U=T_{\mu \nu}$\,,
\item Maps local operators into local operators.
\end{itemize}
}

While most of the evidence for this conjecture takes all these conditions into account, part of proving and understanding why global symmetries are not allowed in QG implies trying to properly define what the disallowed global symmetry is, and whether one can relax any of the above conditions. The approach is to use evidence from string theory and AdS/CFT to properly and rigorously state what does and does not occur.

Note that the global part of a gauge group is not included in the above definition of a global symmetry, i.e. this is perfectly allowed in QG. The reason is that the global part of a gauge symmetry (and also its large gauge transformations) violate the second condition above, i.e. there are no gauge invariant charged local operators $\CO(x)$. We illustrate this in the following example.\\

\textbf{Example 1.} Consider, as a typical example of a (zero-form) global symmetry, the shift symmetry of an axion
\be 
\phi \rightarrow \phi + c\,.
\ee
The current for this action is
\be 
j=\d \phi\,,
\ee
and, if the theory has a Lagrangian description, we can write the corresponding local operator describing this symmetry as
\be 
U=\exp \left( \ii \alpha \int \ast j \right)\,.
\ee
There are local operators charged under the symmetry given by
\be 
\CO(x)= \exp \left( \ii q \phi \right)\,.
\ee
This constitutes a global symmetry which is not allowed in QG. To adhere to the conjecture we must either break the symmetry or gauge it. Gauging the symmetry amounts to coupling the current to some gauge field
\be 
\label{Aj}
\int A \wedge \ast j \ \rightarrow\  \mathcal{L}  \supset  (\d \phi -A)^2\,.
\ee
Then the theory is invariant under the combined transformation of the axion and the gauge field
\be 
\phi \rightarrow \phi+c(x)\,, \qquad
A \rightarrow A+\d c(x)\,.
\ee
Thus, once the symmetry is gauged the operator $\CO (x)$ is not gauge invariant, which violates the second condition of definition \ref{def:globalsym}, so this setup is perfectly allowed in QG.\\

\textbf{Example 2.} The conjecture of No Global Symmetries also applies to $p$-form global symmetries. These are generalizations of global symmetries whose charged operators are supported on $p$-dimensional manifolds \cite{Gaiotto:2014kfa}. We can take the example of a gauge theory invariant under the transformation
\be
\label{pglobal}
A_{p}\rightarrow A_{p}+\lambda_{p} \,.
\ee
with $\lambda_p$ a closed $p$-form. This provides a $(p+1)$-form current $J=F_{p+1}$ which is conserved, $\d*J=0$, where $F_{p+1}=\d A_{p}$ is the gauge field strength. The symmetry operators read
\be 
U_g (M^{d-p-1})=\exp \left( \ii \alpha \oint_{M^{d-p-1}} \ast J_{p+1} \right)\,,
\ee 
and its topological nature follows from the conservation of the current. The charged operators are Wilson lines operators supported on $p$-dimensional manifolds $\gamma_p$ as follows,
\be
\CO(\gamma_p)=\exp\left(\ii n\oint_{\gamma_p} A_{p}\right) \,.
\label{WL}
\ee
We can either break this symmetry (e.g. by adding electrically charged states so that $\d*J\neq 0$) or gauge it by coupling the current to a $(p+1)$-form field,
\be 
\label{Stuck}
\int B_{p+1} \wedge \ast J \ \rightarrow\  \mathcal{L} \supset (\d A_{p} -B_{p+1})^2\,,
\ee
in analogy to \eqref{Aj}. In this latter case, there are no gauge invariant charged operators anymore so the gauged symmetry is consistent with QG. In string theory, the breaking and gauging of these symmetries is typically guaranteed by the presence of Chern-Simon terms.\\

\textbf{Example 3.} As a final example we can consider the case of discrete symmetries. These should also be either broken or gauged, the latter implying that the discrete symmetry can be embedded in a larger gauge group or be part of the diffeomorphisms of the geometry. Hence,  a discrete gauge symmetry  is allowed in QG as it is actually a redundancy of the theory. Consider for example the $E_8\times E_8$ heterotic string theory. The EFT has a $\mathbb{Z}_2$ symmetry interchanging the two $E_8$ gauge sectors, but this is a discrete gauge symmetry as  it has a clear geometric interpretation: the $E_8\times E_8$ heterotic arises from compactifying M-theory on an interval with the $E_8$ gauge sectors living at the endpoints, and the $\mathbb{Z}_2$ symmetry is equivalent to the geometric action of flipping the interval, so it is part of the diffeomorphism transformations. The same occurs for $SL(2,\mathbb{Z})$ of Type IIB. It is gauged, as all duality groups in string theory. The particular case of $SL(2,\mathbb{Z})$ can be understood by going to F-theory, where it becomes part of the diffeomorphisms of the torus. 

\subsection{Motivation\label{sec:motiv}}

The Swampland conjectures usually come with both evidence from string theory and some piece of motivation from a more bottom-up perspective, like black hole physics. In these notes, we will mainly emphasize the former, precisely because the arguments based on BH physics should be taken as (heuristic) motivation and not as proofs. As such, they are useful for gaining intuition for the consequences of a given conjecture, in the present case, to illustrate what would go wrong if one allowed for global symmetries. 


If a BH is charged under some global symmetry, it will evaporate loosing mass but not the global charge, since Hawking radiation is blind to the global charge; ending up as a remnant of Planckian size. Since this can occur for BHs in any representation of the symmetry group, the result is an infinite number of stable remnants. The remnants are arbitrarily long lived, since any combination of particles carrying such large representations of the global symmetry will be heavier than the remnant. Gravitational bremsstrahlung cannot carry away global charge either, as gravitons are neutral under the global symmetry. Hence, in a theory with a global symmetry, starting with a charged particle we can actually end up with an infinite number of states by constructing black holes by throwing particles together and letting them evaporate until they reach Planckian size. Having an infinite number of states below a finite energy scale sounds at the very least problematic \cite{Susskind:1995da}. However, it is difficult to rigorously prove a fundamental inconsistency as we are dealing with Planckian size objects and the semiclassical description of gravity breaks down. It has also been argued \cite{Banks:2010zn} that the existence of these remnants would violate the covariant entropy bound. According to this bound, the entropy associated to the remnant should be finite, in contradiction to the fact that there are infinitely many different states that can give rise to the same black hole geometry, suggesting instead infinite entropy.

Another argument against global symmetries makes use of the No-hair theorem for black holes \cite{Israel:1967wq}, which states that a stable black hole can be completely characterised only in terms of its mass, gauge charge and angular momentum. This implies that, given a black hole of a certain mass, there is no way to determine its global charge. This yields an infinite uncertainty to an observer outside of the black hole which can again be associated to an infinite entropy. However, this would violate the expectation that the black hole entropy should be finite and given by the Bekenstein-Hawking formula, i.e. proportional to the square of its mass.

\subsection{Evidence}

\hspace{\parindent} \underline{String perturbation theory}:\\

Much of the evidence for this conjecture comes from string theory.
In particular, it is proven in Polchinski's book \cite{Polchinski:1998rr} that there are no continuous global symmetries in the target space of perturbative string theory. In other words, any global symmetry of the worldsheet corresponds to a gauge symmetry in spacetime, and viceversa. In the following, we briefly revisit the proof for the case of bosonic string theory. Associated to any global symmetry in the worldsheet, there is a worldsheet charge given by
\be
 Q=\frac{1}{2\pi i}\oint (dzj_z - d\bar zj_{\bar z})   
\ee
by Noether's theorem, where $j_z$,$j_{\bar z}$ are the symmetry currents. This charge must be conformally invariant, which implies that  $j_z$ transforms as a (1,0) tensor and $j_{\bar z}$ as (0,1) tensor. We can then form two vertex operators of the form
\be
j_z\bar \partial X^\mu  e^{ikX} \ ,\quad  \partial X^\mu j_{\bar z}  e^{ikX}\ .
\ee
These vertex operators precisely create massless gauge vectors in the target space coupling to the left and right moving parts of the charge $Q$. Hence, any global symmetry gives rise to a gauge symmetry in spacetime, and not to a global symmetry. The same type of proof applies to superstring theory. 

As of today, there is not a single counterexample of the conjecture, even beyond perturbative string theory.\\

\underline{AdS/CFT}:\\

\begin{figure}[ht]
\centering
\includegraphics[scale=.2]{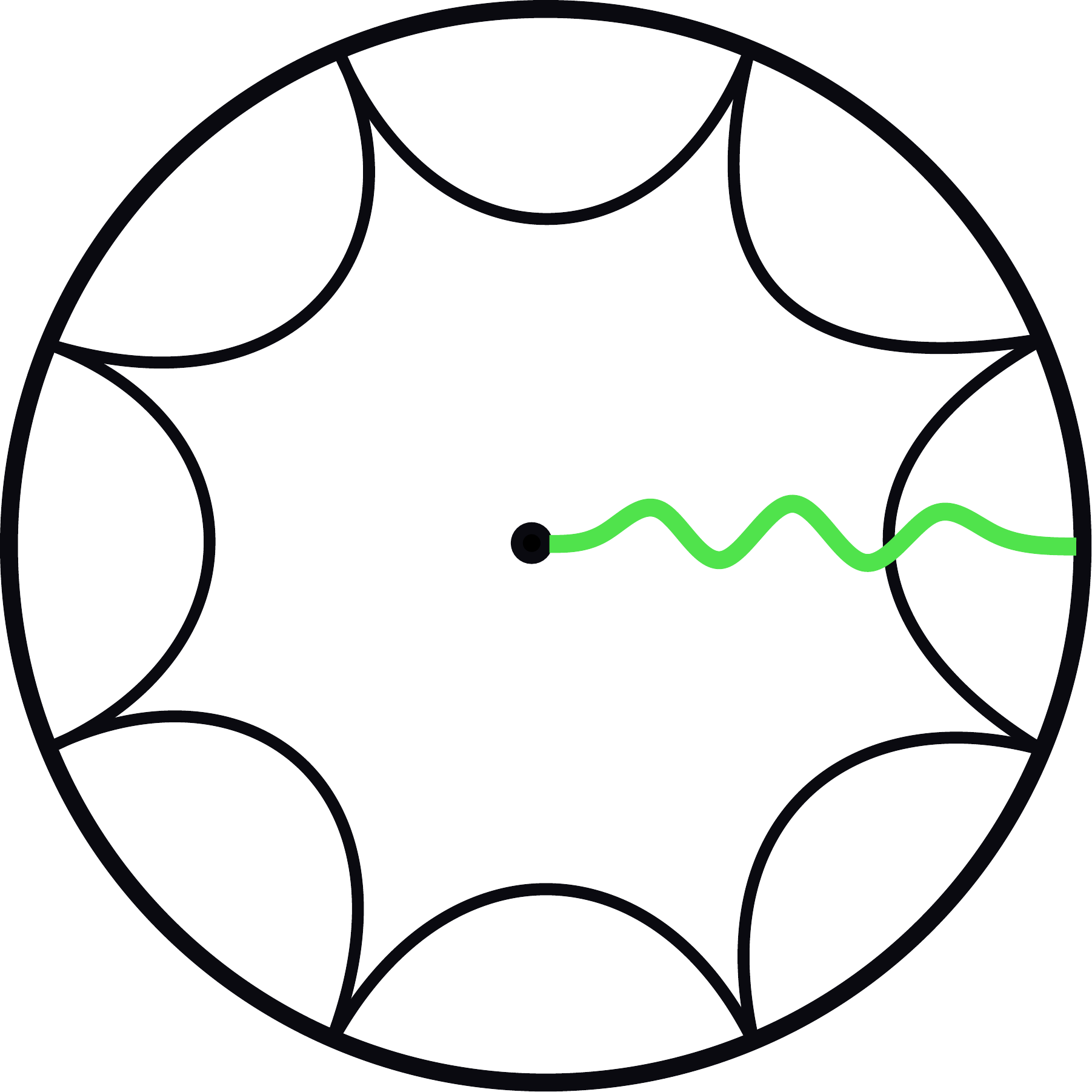}
\caption{Bulk and boundary region of AdS/CFT. In the center of the bulk sits a charged symmetry operator. The boundary is divided into subregions in which symmetry operators localize. When the symmetry is gauged a Wilson line, shown in green, can carry information of the symmetry to the boundary.}
\label{fig:AdSCFTWilsonLine}
\end{figure}

The conjecture has been proven in AdS/CFT under certain assumptions. It was shown in \cite{Harlow:2018jwu,Harlow:2018tng} that a global symmetry in the bulk would lead to a contradiction in the CFT, in the case where the symmetry is splittable on the boundary. This result is derived using entanglement wedge reconstruction, where the boundary is divided into several subregions which have access to some subregion of the bulk obtained using the Ryu-Takayanagi formula (see e.g. \cite{Harlow:2018fse}). That the symmetry is splittable implies that it splits into a product of operators, where each operator is localized in one of the different subregions of the boundary. If the subregions are small enough, they are not sensitive to a charged operator in the bulk, so none of the operators localized in these regions can be charged under the global symmetry. But without charged operators, then there is no global symmetry in the boundary, which leads to a contradiction. On the other hand, if the symmetry is gauged in the bulk, this implies that we can insert a Wilson line which can carry information of the bulk operator to the boundary. This setup is illustrated in figure \ref{fig:AdSCFTWilsonLine}. Hence, global symmetries in the boundary correspond to gauge symmetries in the bulk, and the conjecture is satisfied.

\subsection{Phenomenological Implications}

\hspace{\parindent} \underline{Approximate global symmetries in the IR}:\\

One could ask how this conjecture is useful from a bottom-up perspective, i.e. what are its phenomenological implications. In this regard, however, the conjecture that there are no global symmetries in QG does not give the strongest constraints. This is because, to satisfy the conjecture, one can either gauge the symmetry, which implies that there are degrees of freedom associated to the symmetry, or simply break it at very high energies. So global symmetries are not ruled out in the IR, so long as they are approximate global symmetries, which are broken at some high energy scale. Therefore, if this breaking is highly suppressed, the effects in the IR might be negligible and with no phenomenological consequences. 

In subsequent sections we shall discuss other related conjectures, which precisely try to quantify how approximate the global symmetries can be.\\

\underline{New defects}:\\

There are certain occasions in which the absence of global symmetries can have implications for a subsector of the theory. Sometimes this conjecture can be used to predict the existence of new defects in order to make sure that some global symmetry is broken or gauged. These defects can lead to new states or constrain some property of the theory. Examples of this will appear when discussing the triviality of cobordism classes in the next section.


\subsection{Triviality of the Cobordism Classes\label{sec:cob}}

There is a generalization of the notion that there are no global symmetries, where we include topological global charges, which states that all cobordism classes must vanish.
We will explain the notion of a cobordism class here, and will discuss it further in the context of instability of non-supersymmetric vacua in section \ref{subsec. AdS instability conjecture}, where it also plays an important role.

Cobordism is an equivalence relation on compact manifolds of the same dimensions:\\

\begin{otherbox}[Cobordism]{box:cobordism_group}
	Two given manifolds are in the same cobordism class and called ``cobordant" if their union is the boundary of another compact manifold of one dimension higher. 
\end{otherbox}

\begin{figure}[ht]
\centering
\includegraphics[width=.55\textwidth]{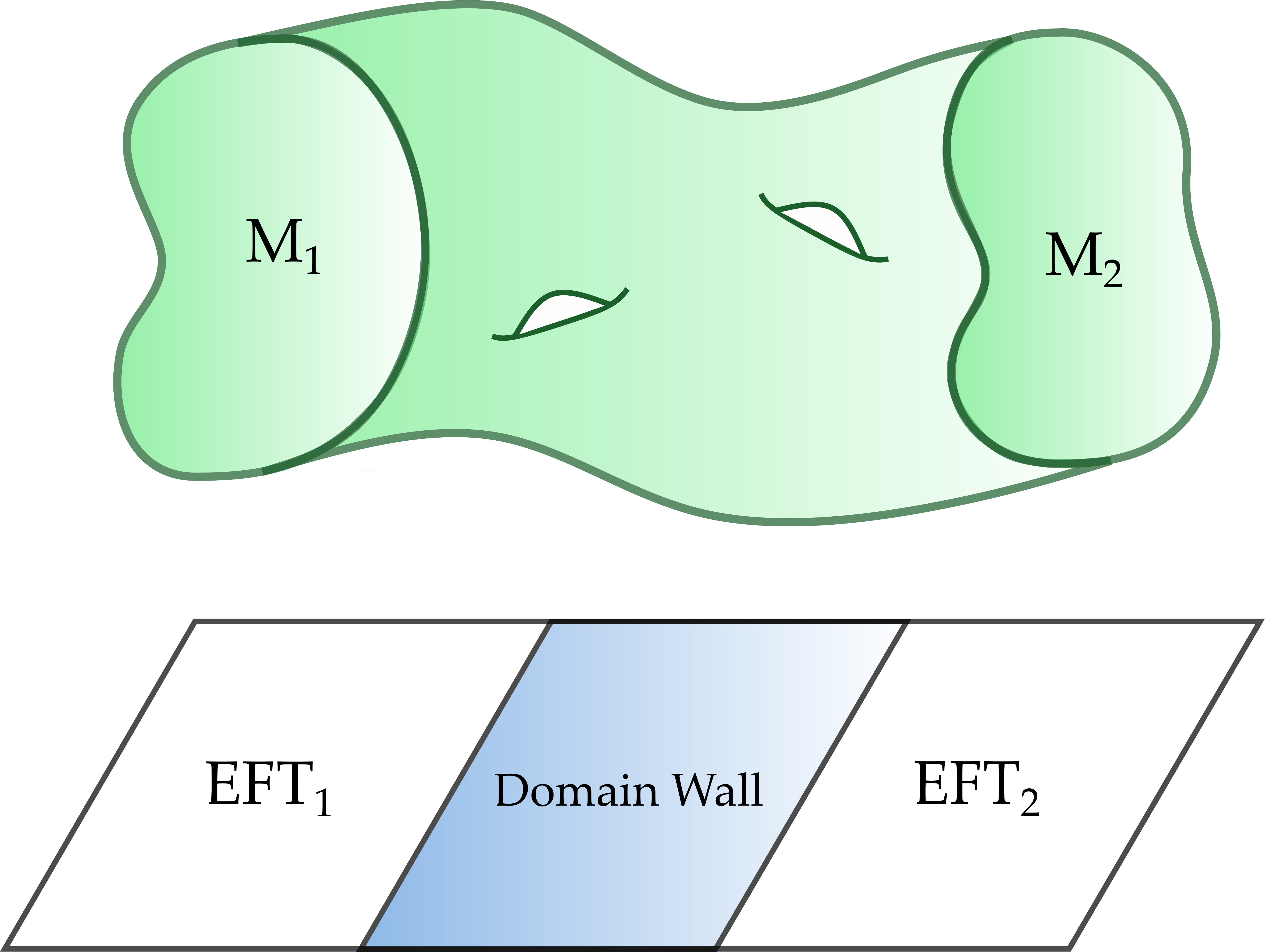}
\caption{Cobordant manifolds $M_1$ and $M_2$, and the EFTs arising from compactification on $M_1$ and $M_2$. The higher dimensional manifold of which $M_{1,2}$ are the boundary is associated to a domain wall interpolating between the two EFTs.}
\label{fig:cobordantmanifolds}
\end{figure}

This has been illustrated in figure \ref{fig:cobordantmanifolds}. It has an abelian group structure and a trivial element, which is a manifold that is a boundary by itself. The following conjecture was then proposed in  \cite{McNamara:2019rup}:

\begin{conj}[Triviality of the Cobordism Classes]{conj:cobordism}
Consider some $D$-dimensional QG theory compactified on a $d$-dimensional internal manifold. All cobordism classes must vanish \cite{McNamara:2019rup}
\be 
\Omega_d^{\text{QG}}=0\,.
\ee
Otherwise they give rise to a $(D-d-1)$-form global symmetry with charges $[M] \in \Omega_d^{\text{QG}}$.
\end{conj}

From the point of view of the lower $(D-d)$-dimensional theory, the fact that two $d$-dimensional compactification manifolds are cobordant to each other, is seen as a $(D-d-2)$-brane domain wall interpolating between two different EFTs, see figure \ref{fig:cobordantmanifolds}. If the bordism group does not vanish, this implies that there are some $d$-dimensional manifolds that cannot be connected with other ones, i.e. there are no domain walls interpolating between the associated EFTs. This in turn implies that there is some global topological charge $[M] \in \Omega_d^{\text{QG}}$, which we cannot get rid of. The global charges will give rise to some $(D-d-1)$-form global symmetry, which would contradict the statement of No Global Symmetries in QG.

The way to understand this conjecture is that a QG theory must contain the required defects to guarantee triviality of the bordism group. All these defects and any required additional structure\footnote{For instance, if the manifold admits a spin structure, the relevant cobordism group is $\Omega^{\rm spin}$. If there are fermions charged under some gauge field, one typically has to look at $\Omega^{{\rm spin}_c}$.} are encoded in the  superscript ``QG" in $\Omega^{\rm QG}$.  If at low energies the theory has a non-zero bordism group, it should be taken as an indication that the theory is missing some ingredient that should be manifest at higher energies and make the cobordism group vanish. This is why in \cite{McNamara:2019rup}, this conjecture was used to predict the existence of new non-supersymmetric defects in string theory. So there \textit{are} consequences of this conjecture, but they are not necessarily visible in the IR. 

To give an example, if the compactification includes internal fluxes, there can be non-trivial cobordism classes characterized by $\int_{X_d} F_d$ where $d$ is the dimension of the compactification manifold. According to the conjecture, this theory by itself would be inconsistent with QG, as $\Omega_d=\mathbb{Z}$. However, by adding branes acting as a source for the flux, one recovers a trivial corbordim group since $\int_{X_d} F_d$ is not an invariant anymore. Hence, consistency with QG would tell us that, in the presence of internal fluxes, the existence of charged branes is required, as occurs in string theory. In this case, triviality of the cobordism group could also be achieved by adding a Chern-Simons coupling that forces $\d F_p\neq 0$, so there is no conserved charge anymore. This is also very common in string theory.

\section{The Completeness Hypothesis}

The next conjecture we will discuss is the Completeness Hypothesis. Like the previous one, the Completeness Hypothesis is a very old and widely accepted Swampland conjecture. Sometimes it can be difficult to ascertain who was the first to propose a conjecture, because in some cases the conjectures are like common lore in the community, but it can certainly be found  both in \cite{Banks:2010zn} and in Polchinski's book \cite{Polchinski:1998rr}.\\

\begin{conj}[Completeness Hypothesis]{conj:completeness}
A gauge field theory  coupled to gravity must contain physical states with all possible gauge charges consistent with Dirac quantization.
\end{conj}


A consequence of the Completeness Hypothesis and No Global Symmetries is that  continuous gauge groups must be  compact. Consider the example where the gauge group is $\mathbb{R}$, which allows for irrational electric charges. 
Then a particle with irrational charge cannot decay to particles of rational charge (because of charge conservation), which implies that there is some global charge associated to this. This would be inconsistent with the statement that there are no global symmetries. This conjecture thus lets us understand why continuous gauge groups in string theory are always compact. Importantly, this does not apply to discrete gauge groups, which can be non-compact, as is the case with duality groups.\\

\begin{otherbox}[Consequence]{box:consequence}
All continuous gauge groups must be compact. 
\end{otherbox}

\subsection{Motivation}

\hspace{\parindent} \underline{BH physics}:\\

The conjecture states that in a theory with a gauge symmetry coupled to gravity, the full lattice of allowed gauge charges must be populated by physical states. This is not necessary in QFT because a charged particle can always be decoupled from the theory by sending the mass to infinity. This ensures that the particle in question is no longer part of the QFT. However, in QG one cannot decouple a particle in this way, because this process will simply result in a BH which is part of the spectrum in a complete theory of QG. Because BHs can have any charge, we expect every possible gauge charge to be populated by some physical state (some of which could be BHs). 
\\

\underline{Breaking global symmetries (connected gauge groups)}:\\

There are some very interesting connections between completeness of the spectrum and the absence of global symmetries in QG. Recall from section \ref{sec:globalsym} that the latter includes the case of higher form global symmetries (i.e. $p$-form global symmetries) associated to higher dimensional charge defects, with operators that live on higher dimensional manifolds. A common way of breaking these higher form global symmetries is simply by having charged states. But only if the spectrum of charged states is complete, can we break the full group. 

Consider the \emph{Example 2} given in section \ref{sec:globalsym}. The current of the $p$-form global symmetry is  given by the field strength of the gauge field $F_{p+1}=\d A_p$.
Then a very straightforward way to break this global symmetry is precisely to introduce states that are charged under the gauge group,
\be 
\d \ast J_{p+1}=\d \ast F_{p+1}=\ast j_{\text{electric}}\,.
\ee
Since the conservation equation for the current is tantamount to the equation of motion for the gauge field, this will be non-zero if there are charged states, $\d \ast J_{p+1}\neq 0$.

If the spectrum is not complete, there will be some discrete symmetry that remains intact. Consider e.g. only the existence of states with even charge; this leaves a $\mathbb{Z}_2$ discrete symmetry. Therefore, a complete spectrum is required to break any discrete remnant of a higher form global symmetry, as argued in \cite{Gaiotto:2014kfa}.
Ongoing research is trying to understand to what extend the two conjectures, i.e. No Global Symmetries and Completeness, are different or follow from each other. The answer seems to depend on whether we are talking about connected or disconnected gauge groups. In the latter case, it seems that the Completeness Hypothesis is related to the absence of more general topological operators.


\subsection{Evidence}

The evidence for the Completeness Hypothesis is very similar to that of the No Global Symmetries Conjecture. It comes from string theory and AdS/CFT. In particular, in \cite{Harlow:2015lma}, it was shown that the Completeness Hypothesis follows from solving the Factorization Problem in AdS/CFT, which pertains to a rephrasing of the Completeness Hypothesis where we require that the theory must contain enough charged states such that all Wilson lines can break. This is exactly what is happening in the gauge theory example above: there are some Wilson line charged operators defined in \eqref{WL}  which can break because of the presence of electrically charged states, which implies that the $p$-form global symmetry is broken.\\


\subsection{BPS Completeness and Implications\label{sec:BPScomp}}


There is a generalization of the Completeness Hypothesis, where we also require completeness in terms of BPS states. In particular, the BPS Completeness condition states that if a charge can support a BPS state, then there should be a BPS state populating that charge. Of course it is generally difficult to understand which BPS states can exist in a theory, and which charges can support BPS states. However, the idea is that if we were able to determine the lattice of BPS states at the same level of detail and generality that we understand it for gauge groups, then all these states should be populated. \\

\begin{otherbox}[BPS Completeness]{box:BPScompleteness}
If a charge can be populated by BPS states, then a BPS state with this charge should be part of the physical spectrum. 
\end{otherbox}

Interestingly, in the past two years there has been a lot of progress in obtaining constraints on the possible gauge groups in QG, using this simple generalization. Certain string theory compactifications admit BPS strings, and anomaly constraints in the world-volume of these strings can be utilized to obtain constraints on the gauge groups allowed by QG. 
For instance, in \cite{Kim:2019vuc,Kim:2019ths} this was used to rule out the following gauge groups:

       
      \begin{itemize}
       \item 16 supercharges: $E_8 \times U(1)^{248}, U(1)^{496}$ in 10d, rank $r_G > 26-d$ for $d<10$.
       \item 8 supercharges: $r_G> k+k' -3$ in 5d, where $k, k'$ are Chern-Simons levels.
       \end{itemize}

To get some of the constraints on the rank of the gauge group in lower dimensions it is necessary to combine the BPS Completeness with additional conjectures like the Distance Conjecture, which will be discussed in section \ref{sec:distconj}. These bounds apply e.g. to the $\mathcal{N}=4$ Super Yang-Mills theory. There also exist some bounds on the allowed number of Abelian gauge group factors, i.e. the rank of the Mordell-Weil group, in 6d $\mathcal{N}=(1,0)$ theories \cite{Lee:2019skh}. 


One of the goals of this program of constraining the gauge groups is to find out whether we can get everything from string theory that could be consistent with QG.
In principle, e.g. in a 10D theory with 16 supercharges, in addition to the $E_8 \times E_8$ and $SO(32)$ gauge groups that we observe in string theory, one could have had other anomaly-free gauge groups. However, any other possible gauge groups have turned out to be inconsistent with BPS Completeness, so this might be the explanation for why we do not observe them in string theory, i.e. that there is no known string theory construction of these gauge groups. 

This bring us to a very deep and important question: how universal is string theory? This question is coupled to the Swampland program, since much of the evidence that we have for the conjectures comes from string theory. Therefore, at some point we must ask whether our results are just an artifact of the lamppost we are looking under, or whether the conjectures are more general and are actually reflecting inconsistencies in QG. If we can show that the anomaly free theories which do not appear in string theory, do not appear for some underlying QG reason, e.g. they are inconsistent with a Swampland conjecture which is expected to be more general, this could imply that we really get everything we can get in string theory. At present, this seems to be the case for highly supersymmetric setups. The idea that everything that can possibly happen (i.e. that is not inconsistent with QG) does happen in string theory is called \emph{String Universality} or the \emph{String Lamppost Principle}. Certainly, it would be very interesting indeed if this is the case and any consistent QG theory is somehow connected to string theory! Therefore, a significant amount of work in the Swampland program goes into  studying those anomaly-free theories that we do not know how to realize in string theory, and whether there is always an additional inconsistency which prohibits them.

\section{The Weak Gravity Conjecture}
\label{sec:WGC}

We have seen that the absence of global symmetries and completeness of the charge spectra are at the core of the Swampland program. However, they lack phenomenological impact unless we can constrain \emph{how approximate} a global symmetry can be\footnote{See \cite{Fichet:2019ugl,Daus:2020vtf} for proposals to quantify how approximate a global symmetry can be.}, and whether there is any upper bound on the mass of some of the charged states. Otherwise, they only constrain the full theory but not the low energy EFT. In particular, it is phenomenologically relevant whether all charged particles can  be super heavy and even correspond to BH's, or there is some notion of completeness of the spectrum that survives at low energies. Most of the Swampland conjectures discussed in the rest of the lectures precisely deal with these questions; they aim to sharpen these statements and quantify how close we can get to the situation of recovering some global symmetries.
For example, we can a priori continuously restore a $U(1)$ global symmetry by sending the gauge coupling to zero, which should not be allowed in QG. Trying to understand how string theory forbids this, and what goes wrong if one tries to do this, can provide information about the constraints that an EFT has to satisfy to be consistent with QG. We will see that the Weak Gravity Conjecture forbids this process by signaling the presence of new light charged states invalidating the EFT description. By doing so, it also provides an upper bound on the mass of these charged states in terms of their charge.


\subsection{Definition of the WGC}
\label{sec:defWGC}

The Weak Gravity Conjecture (WGC) \cite{ArkaniHamed:2006dz} contains two parts: the electric and magnetic version.\\

\begin{conj}[Weak Gravity Conjecture -- Electric]{conj:WGCel}
Given a gauge theory, weakly coupled to Einstein gravity, there exists an electrically charged state with 
\be \label{eq:WGC}
\left. \frac{Q}{m} \geq \frac{\mathcal{\mathcal{Q}}}{M}\right\vert_{\rm extremal}= \mathcal{O}(1)
\ee
in Planck units.
Here $\mathcal{Q}$ and $M$ are the charge and mass of an extremal black hole, and
\be 
Q=q g\,,
\ee
where $q$ is the quantized charge of the state and $g$ is the gauge coupling. 
\end{conj}

The electric version of the conjecture requires the existence of an electrically charged state with a charge to mass ratio greater than the one of an extremal BH in that theory, which is typically an order one factor. The order one factor depends on the theory under consideration.
The simplest case corresponds to  a Maxwell theory coupled to Einstein gravity in which there are no massless scalar fields. In that case, we can construct Reissner-Nordstr\"{o}m BH solutions, so given a $p$-form gauge field in $d$ dimensions, the WGC implies the existence of a $(p-1)$-brane satisfying:
\be 
\label{RNBH}
\frac{p(d-p-2)}{d-2}T^2 \leq Q^2 M_p^{d-2}\,.
\ee
Here, $T$ is the tension of the brane, and instead of an unspecified $\mathcal{O}(1)$ number, we have given the precise bound for the case of a Reissner-Nordstr\"{o}m BH. Hence, for the case of particles  in four dimensions, the order one factor in \eqref{eq:WGC} is $1/\sqrt{2}$.

This is a very sharp conjecture, since it is phrased in terms of some  computable order one number, which depends on the dimension and specifics of the theory, but it does not require UV data. 

The other interpretation of this conjecture is that the bound implies that gravity acts weaker than the gauge force over this state -- hence the name the WGC. This is an equivalent formulation, since requiring that the gravitational force is weaker than the electromagnetic force
\be 
\label{eq:weakgrav}
F_{\text{grav}} \leq F_{\text{EM}}\,,
\ee
will imply that the charge is greater than the mass, so we arrive as the same condition as above. This is no longer true in the presence of massless scalar fields, as discussed in section \ref{sec:WGCscalars}.  

\medskip
We now turn to discuss the magnetic version of the conjecture, which states that the EFT cut-off $\Lambda$ is bounded from above by the gauge coupling, so that, given some EFT coupled to Einstein gravity, its cut-off is smaller than the Planck mass if the gauge coupling is small.\\

\begin{conj}[Weak Gravity Conjecture -- Magnetic]{conj:WGCmag}
The EFT cut-off $\Lambda$ is bounded from above by the gauge coupling:
\be 
\label{gMp}
\Lambda \leq g M_p^{(d-2)/2}\,.
\ee
For a $p$-form gauge field:
\be 
\Lambda \leq \left( g^2 M_p^{d-2} \right)^{\frac{1}{2(p+1)}}\,.
\ee
\end{conj}

The magnetic version of the conjecture can be derived in two ways. We can apply \eqref{eq:weakgrav} to the magnetic field (which is why it is called the magnetic version), which implies that the theory should have a monopole whose mass is smaller than its charge. Since the magnetic charge is proportional to the magnetic gauge coupling, which is the inverse of the electric gauge coupling, this means that the mass of the monopole must be smaller than the Planck mass over the electric gauge coupling:
\be 
m \leq M_p/g\,.
\ee
The mass of the monopole will typically be at least at the order of the cut-off of the theory over the gauge coupling squared (which is just the electrostatic energy integrated up to the cut-off)
\be 
m \sim \Lambda / g^2\,,
\ee
which implies that the cut-off of the theory is bounded by
\be 
\Lambda \leq g M_p\,.
\ee
We arrive at the same conclusion if we require that the theory has some monopole that is not a BH, so that the mass of the unit charge monopole must be smaller than the Schwarzschild radius
\be 
 m \leq M_p^2 R\,.
\ee
Since this radius provides a cut-off for the EFT
\be 
R \sim \Lambda^{-1}\,,
\ee
this again implies that $\Lambda$ is smaller than $g M_p$. Note that the gauge coupling is always evaluated at the given energy scale $g=g(\Lambda)$.


These are intuitive arguments for why this cut-off should be imposed. The EFT is breaking down, because at this energy scale it becomes sensitive to the monopole degrees of freedom, so these degrees of freedom can no longer be treated as solitonic objects. However, this is not necessarily a drastic breakdown of the EFT due to some QG effect. Below, we shall see a genuine QG cut-off appear in some stronger versions of the WGC, which really hints at a drastic breakdown of the EFT.

\subsection{Motivation}

The motivation for the WGC is twofold. First, it provides a QG obstruction to restoring a U(1) global symmetry by sending $g\rightarrow 0$. If a gauge coupling goes to zero, according to the WGC, this results in new light particles; and the cut-off of the theory goes to zero according to \eqref{gMp}, invalidating the EFT. 


Given some EFT, how small the gauge coupling can be depends on how high the energy of the process you want to describe with that EFT is. The smaller the energy of the process, the smaller the gauge coupling can be, as dictated by \eqref{gMp}. Conversely, if you want to keep the EFT valid up to a very high cut-off, then the gauge coupling cannot be too small. This is an example of a Swampland constraint becoming stronger for higher energies (as illustrated by the cone in figure \ref{fig:cone}). Needless to say, a theory with vanishing gauge coupling (i.e. a global symmetry) is inconsistent as the EFT cut-off is then also zero.

\begin{figure}[ht]
\centering
\includegraphics[width=.7\textwidth]{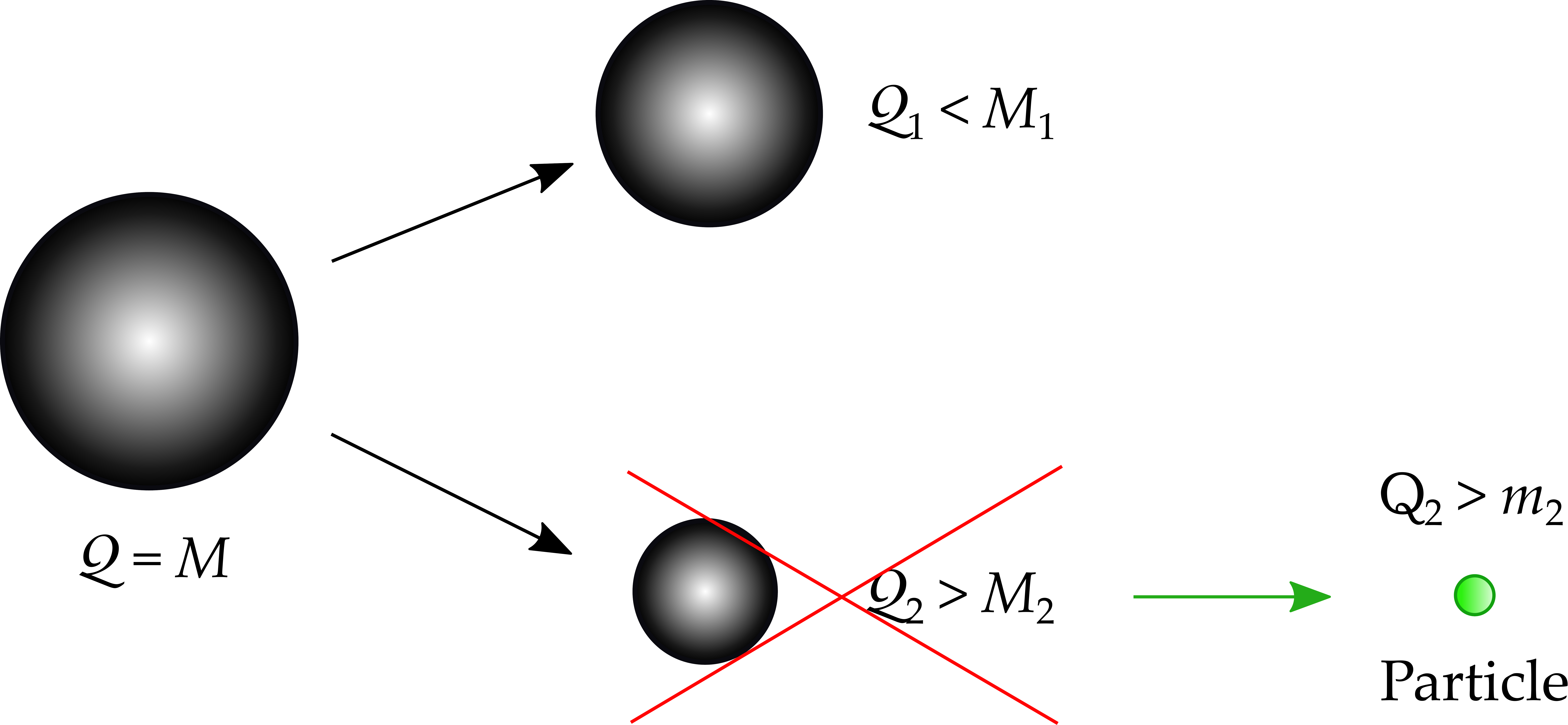}
\caption{Decay of an extremal BH. One decay product can have a charge to mass ratio $\mathcal{Q}_1/M_1 <1$ to preserve the extremality bound, but then the other must have a charge to mass ratio $Q_2/m_2 >1$. Therefore, the latter cannot be a BH but must be a particle.}
\label{fig:BHdecay}
\end{figure}

The other main motivation for the WGC is that it is a kinematic requirement that allows extremal BHs to decay. Charged BHs must satisfy an extremality bound in order to avoid the presence of naked singularities, as implied by Weak Cosmic Censhorship. For a given charge $\mathcal{Q}$, this extremality bound implies that the mass of the BH $M$ must be greater than the charge 
\be 
M \geq \mathcal{Q}\,,
\ee
for the BH to have a regular horizon. Here, we are setting the extremality $\mathcal{O}(1)$ factor in \eqref{eq:WGC} to one for simplicity. For an extremal BH the charge is equal to the mass, so the only way it can decay is if there exists a particle whose charge to mass ratio is at least $1$. To see this, consider the decay of an extremal BH, where one of the decay products has a charge smaller than its mass or saturates the extremality bound, so $\mathcal{Q}_1 \leq M_1$. Then the other decay product cannot have a charge smaller than the mass, i.e. $Q_2 \geq m_2$. This is simply a kinematic requirement. Since the second decay product violates Weak Cosmic Censorship it cannot be a BH, so it must be a particle. The system is illustrated in figure \ref{fig:BHdecay}.

The above kinematic requirement can be derived by simply imposing mass/energy conservation and charge conservation as follows. 
The mass of the initial black hole state must be greater than the sum of the decay product masses $M_i$,
and the charge of the initial black hole $\mathcal{Q}$ must equal that of the decay products $\mathcal{Q}_i$,
\be 
M \geq \sum_i M_i\,, \qquad \mathcal{Q}= \sum_i \mathcal{Q}_i\,.
\ee
Then
\be 
\frac{M}{\mathcal{Q}} \geq \frac{1}{\mathcal{Q}} \sum_i M_i=\frac{1}{\mathcal{Q}} \sum_i \frac{M_i}{\mathcal{Q}_i}\mathcal{Q}_i \geq \frac{1}{\mathcal{Q}} \left( \frac{M}{\mathcal{Q}}\right)_{\text{min}} \sum_i \mathcal{Q}_i=\left( \frac{M}{\mathcal{Q}}\right)_{\text{min}}\,.
\ee
Hence, we have shown that the mass to charge ratio of the original BH must be grater than the ratio of the decay product with minimal mass to charge ratio, which is why we need some particle satisfying the WGC.

To sum up, if a theory does not satisfy the WGC, then extremal black holes cannot decay. The question one could now ask is: what is the problem with having stable extremal BHs? The WGC is a kinematic requirement that allows extremal BHs to decay, but why is this actually necessary?
There is a heuristic argument, which is very similar to the argument for the absence of global symmetries, which suggests that we run into trouble with BH remnants. Starting with large BHs of all possible charges, they will evaporate until reaching the extremality bound. If the gauge coupling is very small, they will all reach a very similar mass, but still having very different charges. This will result in a large number of remnants (how large depends on how small the gauge coupling is)
\be 
N \sim 1/g\,,
\ee
that are almost degenerate in mass at weak coupling. The difference in mass is set by the gauge coupling 
\be 
\Delta M \sim gM_p\,.
\ee
So $N\rightarrow \infty$ and $\Delta M\rightarrow 0$ when the gauge coupling is small, and we run into trouble with entropy bounds as in section \ref{sec:motiv}. However, this heuristic argument is even less satisfactory than for the case of global symmetries because it does not imply an infinite number of remnants below a fixed energy scale, but just a large number of them. As such, the argument should be understood only as supplying some intuition of what might go wrong for an EFT violating the WGC from a bottom-up perspective.\\

\subsection{Evidence}

%
%

The motivations from BH physics are interesting because they constitute more general arguments that allow us to at least understand some difference in the physics of an EFT that satisfies the WGC and one that does not. If we only base the conjecture on evidence from examples in string theory, we risk being fooled by the lamppost we look under. However, these motivations are very far from proving anything. The reason why the WGC is one of the most important Swampland conjectures nowadays is not because of these BH arguments, but because of all the evidence that we have gathered for it in the past years. To go through all the evidence in detail is beyond the capacity of these notes, instead we will give a summary of the different works in this direction, and some useful references discussing these arguments. One of the nice things about the Swampland program in general, and in particular with this conjecture, is that there are many people working from a wide range of areas of research  trying to prove the conjecture. Hence, the assumptions that go into a particular piece of evidence coming from one corner of research is different from what another person assumes in a different corner. This suggests that there may indeed be something behind it all, and increases our confidence that some statement along the lines of the WGC is indeed a universal criterium of QG.\\

\underline{Higher derivative corrections to BHs}:\\

\begin{figure}[ht]
\centering
\includegraphics[width=.35\textwidth]{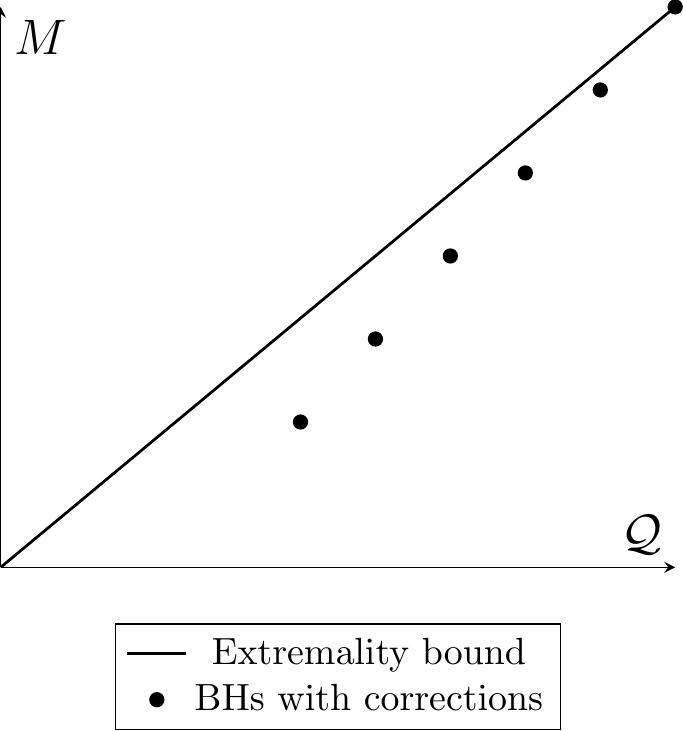}
\caption{Graph of charge $\mathcal{Q}$ versus mass $M$, where the line indicates the extremality bound for a large BH, and the marks represent smaller BHs.}
\label{fig:graphWCG}
\end{figure}

One piece of evidence comes from studying higher derivative corrections to the BH charge and mass. It turns out that these higher derivative corrections could be enough to show that the charge to mass ratio of small black holes is greater than the classical extremality bound. The latter is defined as the charge to mass ratio of large extremal black holes of large charge, for which quantum or higher derivative corrections become negligible,
\be 
\label{hd}
\frac{\mathcal{Q}}{M} \geq \left. \frac{\mathcal{Q}}{M}\right\vert_{\rm ext}\,.
\ee
If that is the case, then the WGC could already be satisfied by small BHs. The pattern found in certain setups is the following. If we plot the mass to charge ratio, as in figure \ref{fig:graphWCG}, with the straight line indicating the extremality bound for a large BH of large charge, then small BHs happen to lie below this curve upon taking into account higher derivative corrections. If this holds in general, it would be a way to prove a very mild version of the WGC, where the WGC-satisfying states are BHs instead of light particles. One could expect, though, that if we extrapolate the pattern to lower masses, small black holes will eventually turn into particles, so there will also be some particle with a charge to mass ratio bigger than one. In fact, it can be interpreted as motivation for having an infinite tower of particles satisfying the WGC, where the heavier states are actually black holes. However, this extrapolation to lower energies goes beyond the regime of validity of the computations.

For this story to work out, there is an assumption about the possible higher derivative terms that can occur built into this argument. Interestingly, this pattern of higher derivative corrections have been found to hold in certain string theory setups (see e.g. \cite{Kats:2006xp}), and there is a significant amount of research dedicated to investigate whether such assumptions on higher derivative terms can be related to other IR consistency criteria, as explained next. 
\\

\underline{Positivity constraints from unitarity and causality}:\\

The above evidence relies to some extent on specifics of the theory in which the BHs live. But there are several works trying to derive model independent results, based on the concepts of unitarity and causality, to check if positivity constraints can provide any definite answer as to the direction of the corrections to the BHs, see e.g. \cite{Cheung:2018cwt,Hamada:2018dde}. Interestingly, it seems that positivity constraints indeed imply \eqref{hd} under mild assumptions of the UV theory which are realized in string theory.
Along a different but related avenue, interesting relations between the WGC and analyticity and causality appear when studying the difference in the low energy theory between integrating out charged states that satisfy or do not satisfy the WGC \cite{Andriolo:2018lvp}.\\ 

\underline{Relation to Weak Cosmic Censorship}:\\

The WGC has interesting relations to Weak Cosmic Censorship, which forbids the presence of naked singularities not hidden by a horizon. The latter has been extensively tested in the context of general relativity in four dimensions, although there are some known possible counterexamples involving an electric field that grows in time without bound. A few years ago, it was shown in \cite{Crisford:2017gsb} that if that theory also includes states that are charged under the gauge field and satisfy the WGC, then the theory does not violate Weak Cosmic Censorship any more. Some work is currently going into trying to understand the connections between the WGC and Weak Cosmic Censorship in more detail by generalizing the above setup. It would certainly be interesting if it turns out that we always need to satisfy the WGC or some other Swampland conjecture to preserve Weak Cosmic Censorship. \\

\underline{AdS/CFT}:\\

Evidence for the WGC has also been obtained in the framework of AdS/CFT. Specifically, the WGC has been derived from imposing modular invariance of a 2d CFT \cite{Montero:2016tif,Heidenreich:2016aqi}, so it applies in the context of AdS$_3$/CFT$_2$ but also for any gauge field with a worldsheet description in string theory. It would be very interesting to find similar arguments for higher dimensional CFTs. Further evidence has been found in AdS/CFT by studying thermalization properties of the CFT \cite{Hod:2017uqc,Urbano:2018kax}, as well as from quantum information theorems related to  entanglement entropy \cite{Montero:2018fns}.

\subsubsection{Evidence from String Compactifications\label{sec:evidenceWGC}}

Finally, and very importantly, there is a plethora of examples and works in the context of string theory compactifications, which is the main focus of these lectures. Some of these examples will be discussed in more detail in section \ref{sec:SDC-ST-evidence}, since they provide evidence both for the WGC and the Distance Conjecture. However, we will already outline the basic features here in an attempt to give some intuition for how the WGC is usually realized in string theory. 
A summary of the string theory evidence can also be found in \cite{Palti:2019pca}.\\

\hspace{\parindent} \underline{$SO(32)$ heterotic string on $T^6$:}\\

\begin{figure}[ht]
\centering
\includegraphics[width=.35\textwidth]{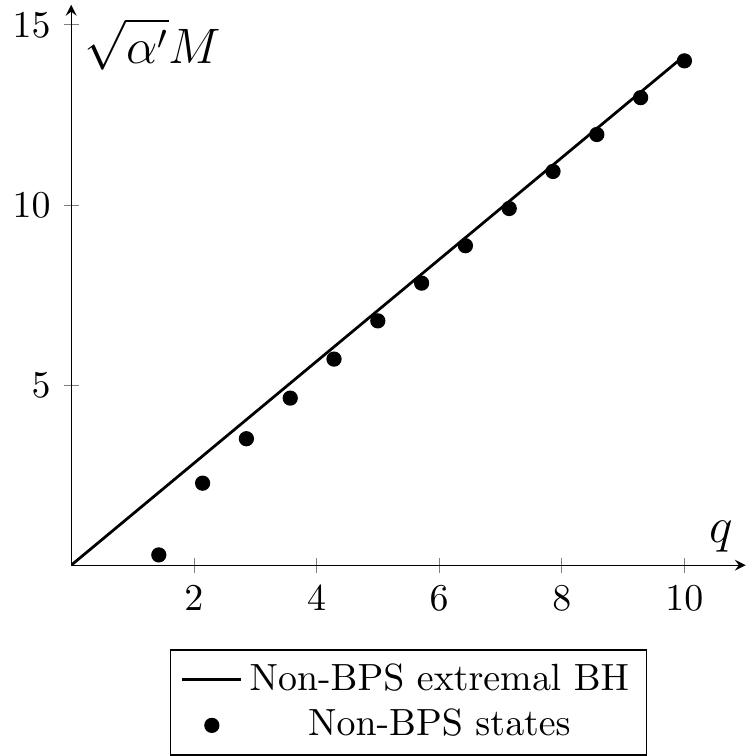}
\caption{Graph of charge $q$ versus mass $M$, where the line indicates the extremality bound for non-BPS BHs, and the marks represent non-BPS states of the theory.}
\label{fig:graphWCGhetstring}
\end{figure}

The most typical example, that already appeared in the original paper \cite{ArkaniHamed:2006dz}, is a toroidal compactification of heterotic string theory. Compactifying the $SO(32)$ heterotic string on $T^6$ yields a theory with 16 supercharges, i.e. $\mathcal{N}=4$ supergravity in 4d. Since we have a perturbative description of this theory, the full perturbative spectrum is readily available, and it is known to form a self-dual lattice, charged under the $U(1)^{28}$ gauge group. The extremal BH solutions of this theory are also known. The non-BPS extremal BHs satisfy
\be 
M^2=\frac{2}{\alpha'} \vert q \vert^2\,.
\ee
If one studies the non-BPS states in that theory, they do not saturate this extremality bound, rather they have
\be 
\label{mhet}
m^2=\frac{2}{\alpha'}\left( \vert q \vert^2-2 \right)\,.
\ee
So if we plot the mass versus the charge, as in figure \ref{fig:graphWCGhetstring}, with the straight line indicating the extremality bound, i.e. where the extremal BHs live, the states (i.e. the particles) indeed lie below this curve. 
Hence, the theory satisfies the WGC, and the states approach the extremal value for very large charges. This piece of evidence is particularly nice, because it concerns non-BPS states and not just supersymmetric particles.\\

\underline{Closed string gauge fields in Type II:}\\

In Type II compactifications, there are abelian gauge fields arising from dimensionally reducing the RR and NS 10D gauge potentials. They always come together with charged states coming from wrapping D-branes and NS5-branes. In supersymmetric compactifications, sometimes it is somewhat automatic that the WGC is satisfied because the BPS states themselves saturate the WGC. In this way, the WGC can be thought of as an anti-BPS bound, although this will be explained in more detail in section \ref{sec:Sharpening the WGC}. For example, the supersymmetric mass $M_{D_p}$ of a brane wrapping the cycle $\gamma_p$ is given, according to the DBI action, by the string scale over the string coupling times the volume that it is wrapping $\mathcal{V}_\gamma$
\be 
M_{D_p}(\gamma_p)=2\pi \frac{M_s}{g_s} \mathcal{V}_\gamma\,.
\ee
In 4d Planck units this becomes
\be 
M_{D_p}(\gamma_p)=\sqrt{\pi} M_p \frac{\mathcal{V}_\gamma}{\mathcal{V}}\,,
\ee
where $\mathcal{V}$ is the overall volume of the compactification space. Notice that any time we are checking a Swampland conjecture, it is very important to go to the Einstein frame, and write every mass in Planck units. Hence, when talking about light states in the Swampland program, we always refer to light with respect to the Planck mass.
One might think that, since the mass is proportional to the volume of the cycle, maybe we can violate the WGC by making this cycle very large, so the state becomes very massive. However, this will also change the gauge coupling as the gauge field comes from dimensionally reducing $C_{p+1}$ on the same cycle $\gamma_p$. In fact, if the state is BPS, in the asymptotic weak coupling limit the gauge coupling has the same dependence than the mass in Planck units,
\be 
\frac{M_{D_p}(\gamma_p)}{M_p} \sim g\,
\ee
saturating the WGC. A proper analysis requires the inclusion of scalars in the extremality bound, which is postponed to section \ref{sec:WGCscalars}. It will also be discussed in detail in section \ref{sec:typeII-CY}. \\

\underline{Open string gauge fields:}\\

\begin{figure}[ht]
\centering
\includegraphics[width=.7\textwidth]{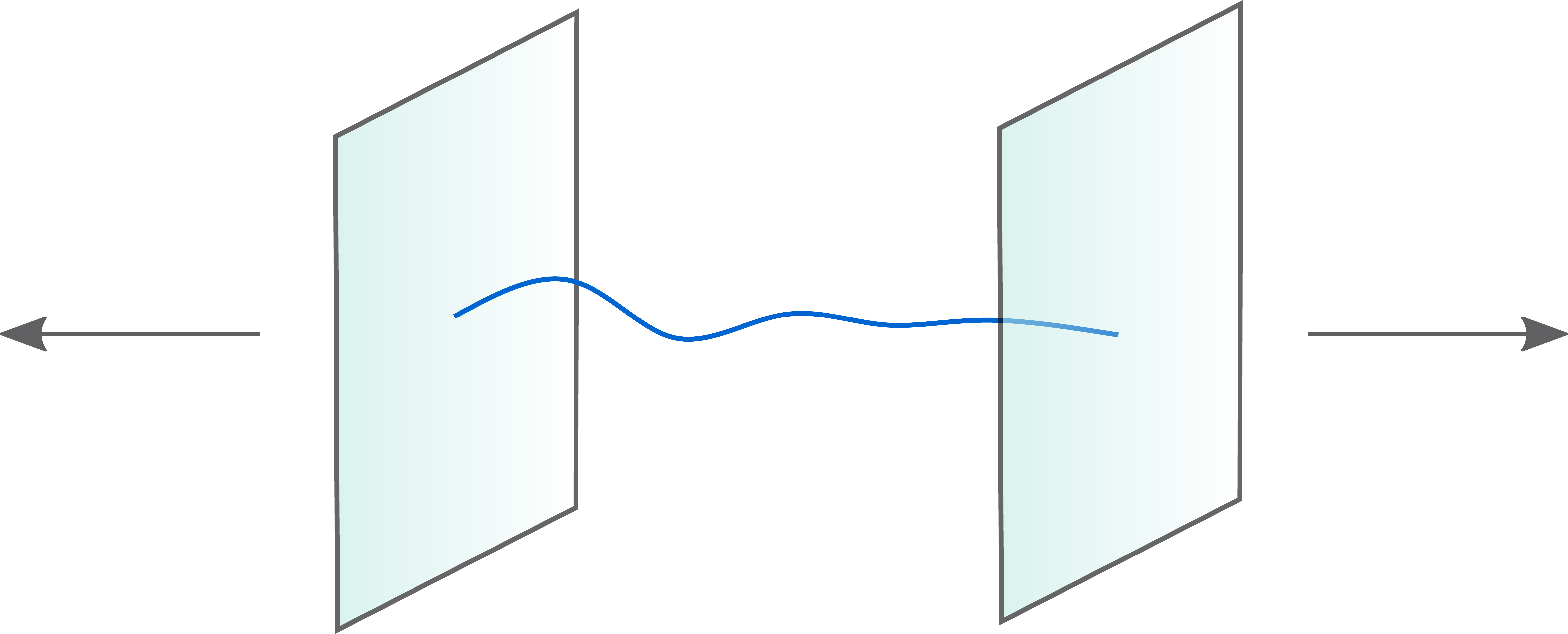}
\caption{
Open string suspended between a set of D-branes. 
}
\label{fig:openstring1}
\end{figure}

If we instead focus on open string gauge fields, one could try to play the same game. So, given some branes with a gauge field living on them, we can have strings suspended between the branes, giving rise to charged states, see figure \ref{fig:openstring1}. One could try to violate the WGC in this setup by increasing the mass of these charged states. This can be done by increasing the distance between the branes as the mass of the charged strings is proportional to the distance. Unlike in the previous setup with closed string gauge fields, here the gauge coupling remains constant, as it is not sensitive to the distance between branes. However, in a compact space, it is not possible to increase the distance between the branes as much as we want without also  increasing the overall volume of the compactification space. Increasing the overall volume decreases the gravitational strength, since it results in a larger value for the Planck mass. For this reason, there is a maximum value of the mass measured in Planck units that can be attained by increasing the distance between branes, implying that  beyond that point, the charge to mass ratio in Planck units remains constant
\be 
\frac{Q M_p}{m} \sim \text{const.}\,.
\ee

\begin{figure}[ht]
\centering
\includegraphics[width=.7\textwidth]{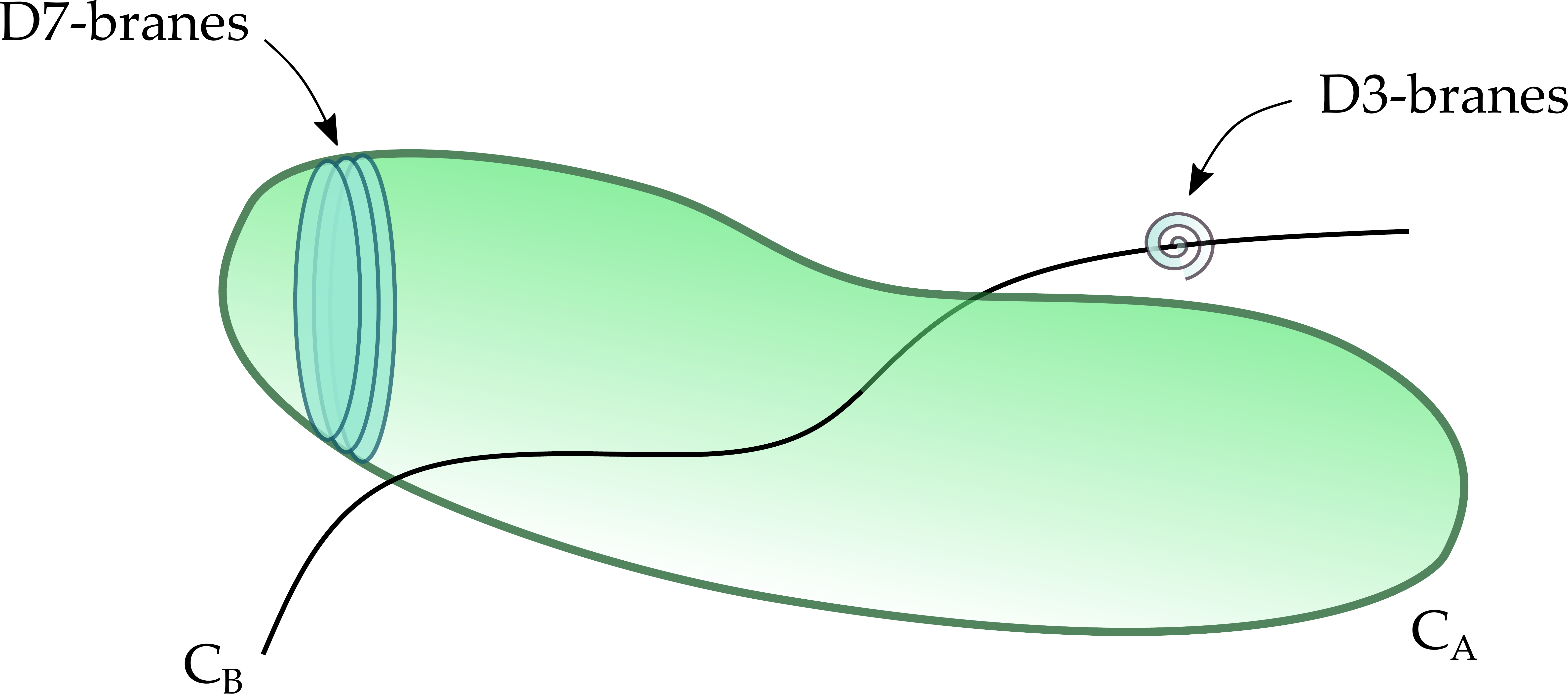}
\caption{
Equi-dimensional limit. In order for the overall volume to remain finite while the volume of the divisor $C_A$ goes to infinity, there must exist another curve $C_B$ which intersects $C_A$ and whose volume goes to zero. The gauge coupling of the D7-brane gauge sector wrapping $C_A$ will go to zero while the tension of a D3-brane wrapping $C_B$ will also go to zero.
}
\label{fig:openstring2}
\end{figure}

Upon compactification to lower dimensions, there is a richer set of possible limits that one can take varying both the mass and the charge. We could think of violating the WGC by sending a gauge coupling to zero and parametrically decreasing the charge of any state. Consider for example the open string gauge fields from a stack of D7-branes wrapping some divisor (co-dimension 1 subvariety) $C_A$ in an F-theory compactification. We can send the gauge coupling to zero by sending the volume of this divisor to infinity,
\be  
\frac{1}{g_{D7}} \sim \text{vol}(C_A) \rightarrow \infty\,.
\ee
However, we are going to run into the same above problems of increasing the overall volume and decreasing $M_p$, unless we are able to engineer a so-called equi-dimensional limit, in which the overall volume does not change.
This is possible by simultaneously decreasing the volume of another cycle $C_B$ while increasing the volume of $C_A$, so that the overall volume $\mathcal{V}$ is kept fixed, i.e.
\be 
\text{vol}(C_A) \rightarrow \infty\,, \qquad \text{vol}(C_B) \rightarrow 0\,, \qquad \text{s.t.} \qquad \mathcal{V} \sim \text{const.}
\ee
However, as shown in \cite{Lee:2018urn}, the curve $C_B$ will have a non-zero intersection with $C_A$, and a D3-brane wrapping $C_B$ will give rise to a string charged under the D7-brane gauge sector whose tension is proportional to the volume of $C_B$. The mass of the string excitation modes will then go to zero,
\be 
M_{D3}(C_B) \sim \text{vol}(C_B) \rightarrow 0\,
\ee
as $g_{D7}\rightarrow 0$, in a way consistent with the WGC.  See figure \ref{fig:openstring2} for an illustration. Hence, once again, the WGC is satisfied; in this case, thanks to the presence of the string modes arising from a wrapping D3-brane. This example will be discussed in more detail in  section \ref{sec:SDC-5d-N1}.

These examples are intended to give some intuition for how string theory typically defeats our attempts to violate the WGC. We will get deeper into these examples and the geometric structures underlying the relations between the charges and masses in section \ref{sec:SDC-ST-evidence}.

\subsection{Multiple Gauge Fields and the Convex Hull Condition} \label{sec:convex-hull-WGC}

So far we have discussed the WGC in the presence of a single gauge field, but how does one properly define the conjecture in the presence of several gauge fields? One could have guessed that it is enough to require the existence of a particle for each gauge field with a charge to mass ratio  bigger than the extremality bound, but this turns out not to be sufficient. Instead, the correct generalization is a requirement known as the Convex Hull Condition.\\

\begin{otherbox}[Convex Hull WGC]{box:convexhull}
For multiple $U(1)$ gauge fields, the WGC is satisfied if the convex hull of the charge to mass ratio $\vec{z}=\vec{Q}/m$ of all the states contains the extremal region \cite{Cheung:2014vva}.
\end{otherbox}

\begin{figure}[ht]
\centering
\includegraphics[width=.45\textwidth]{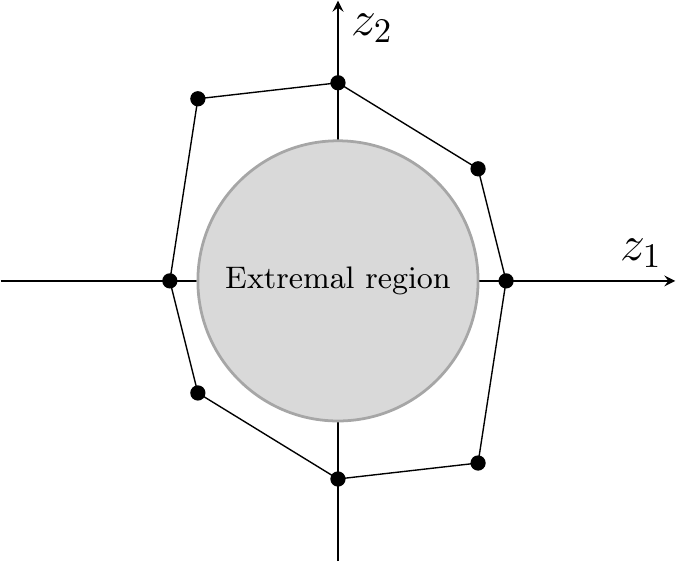}
\caption{Plot of charge to mass ratios $\vec{z}=(z_1,z_2)$ of a set of states for two $U(1)$ gauge fields. The extremal region is indicated in gray. It is contained in the convex hull of the charge to mass ratios of the states, indicated by the black lines. The WGC is satisfied.}
\label{fig:convexhull}
\end{figure}

Consider the example of two $U(1)$s, shown in the plot in figure \ref{fig:convexhull}. If we determine the extremal region, i.e. the region in this plane that allows for BHs to exist, and plot the charge to mass ratio $\vec{z}$ of all the states, then the requirement is that the convex hull of the charge to mass ratio of the states in the theory includes the extremal region. This ensures that there is always some particle with a charge to mass ratio bigger than the extremality factor along \emph{any} rational direction of the charge space.

\begin{figure}[ht]
\centering
\includegraphics[width=.35\textwidth]{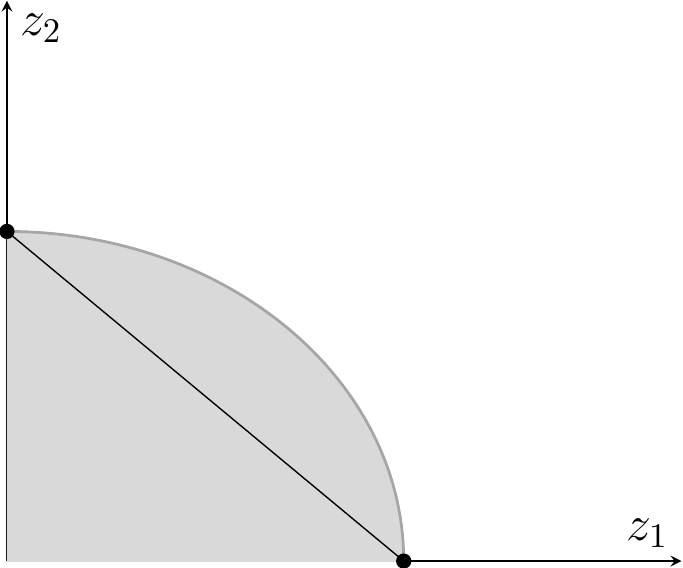}
\caption{Plot of charge to mass ratios $\vec{z}=(z_1,z_2)$ of two states saturating the WGC for the $U(1)^{(1)}$ and $U(1)^{(2)}$ gauge fields, separately. The extremal region is shown in gray, and the convex hull of the charge to mass ratios of the states is indicated by the black line. The WGC convex hull condition is not satisfied.}
\label{fig:convexhullvssep}
\end{figure}

Here we see that having a set of states saturating the WGC separately for each gauge field would cut into the extremal region, see figure \ref{fig:convexhullvssep}, so it is not enough to satisfy the Convex Hull Condition. Therefore, the Convex Hull Condition is clearly a stronger statement than requiring the WGC to be satisfied for each $U(1)$. Essentially this condition allows BHs with {\it any charge} under both gauge fields to decay. This puts non-trivial constraints on the EFTs.

The extremal region is sometimes called a unit ball, since for simplicity we tend to set the charge to mass ratio of an extremal black hole to 1. This terminology can be somewhat confusing because, first, the extremal region is not necessarily of radius one (it is some $\mathcal{O}(1)$ constant), and, second, when scalar fields are present it is not necessarily even a ball. The extremal region  can then be an ellipse or have straight lines, as discussed in section \ref{sec:WGCscalars}.


\subsection{Strong Versions of the WGC}

One of the main open questions in the WGC is: how do we identify \textit{which} states satisfy the WGC? This question is very important in order to be able to derive phenomenological implications of the conjecture. The formulation of the conjecture given so far contains no statement, or even indication, of whether there should be one or several states satisfying the WGC, if they need to be very light, have a very small charge, or the biggest charge to mass ratio, etc.. In this sense, there is a mild version and many strong versions of the WGC. The mild version simply says that there must be \textit{some} state (any state) that satisfies the WGC. It can even be above the EFT cut-off, so in this form the conjecture does not have any implications at low energies. Contrary to this, the strong versions specify which state(s) should satisfy the WGC. Such refinements typically imply particles below the cut-off, leading to potential phenomenological constraints.

For this purpose, a lot of the recent research has precisely been dedicated to understanding this question, by investigating the structure of the states satisfying the WGC in string theory setups.  In fact, there was an attempt in the original paper to get a strong version of the conjecture, by identifying the WGC-particle with the lightest one, the one with smallest charge, or the one with the biggest charge to mass ratio. However, these stronger versions are not well-defined in the presence of several gauge fields. At the moment the available strong versions that are consistent with everything we know and any string theory example are the Sublattice WGC  \cite{Montero:2016tif,Heidenreich:2016aqi,Heidenreich:2015nta} or the Tower WGC \cite{Andriolo:2018lvp}, which require the states satisfying the WGC to form a (full dimensional) sublattice of the charge lattice or a tower, respectively. The two proposals are very similar, however it is slightly stronger to require a sublattice of superextremal states than requiring a tower. It is also expected that the conjectures might be satisfied by unstable resonances but not by multiparticle states.\\

\begin{otherbox}[Sublattice/Tower WGC]{box:strongWGC}
For every site $\vec{q}$ of the charge lattice, there is a positive integer $n$ such that there is a superextremal state with charge $n\vec{q}$ satisfying the WGC. This integer can depend on $\vec{q}$ for the Tower version, while it is universal for the Sublattice.
\end{otherbox}

The main motivation for these proposals comes from demanding that the WGC is consistent under dimensional reduction \cite{Heidenreich:2015nta}. Starting from a theory with a single particle satisfying the WGC, when this theory is dimensionally reduced it can result in an EFT that does not satisfy the WGC anymore.
On the other hand, starting with a sublattice (or an infinite tower) of states satisfying the WGC ensures that the dimensionally reduced theory will also satisfy the WGC. So it is exactly this consistency under dimensional reduction that requires the existence of infinitely many states satisfying the WGC in the original theory.
Arguments from unitarity and causality \cite{Andriolo:2018lvp} motivate the weaker version in which an infinite tower (instead of a sublattice) suffices, which also relates to the Distance Conjecture as discussed in section \ref{sec:relation-WGC}. However, up to now, there is no known string theory counterexample to the Sublattice WGC (see though \cite{Lee:2019tst} for a potential candidate). 
In the following, we will see two key examples that were important in motivating these stronger versions, based on consistency under dimensional reduction. \\

\underline{Kaluza-Klein circle compactification:}\\

Consider a theory with a $U(1)$ gauge field and some particle that satisfies the WGC, so that it has charge to mass ratio
\be 
z_0=\frac{q_0}{m_0}\geq 1\,.
\ee
Dimensionally reducing the theory by compactifying on a circle $S^1_R$ of radius $R$, will give rise to a lower dimensional $U(1)$ gauge field, as well as an extra Kaluza-Klein (KK) $U(1)$ gauge field. The zero mode of the original particle with charge to mass ratio $z_0$ will still satisfy the WGC for the U(1) gauge field in lower dimensions.
There will also be states charged under the extra $U(1)$, namely the KK modes with 
\be 
m_{KK}^2=m_0^2+\frac{q_{KK}^2}{R^2}\,, \qquad g_{KK}=\frac{1}{R}\,
\ee
and $q_{KK}=n$ with $n$ being an integer.
So here we see that, if there are massless states in the original theory (i.e. $m_0 =0$), the KK modes of these massless states are going to saturate the WGC for the KK gauge field, since their mass will be equal to their charge. Thus, it is automatic in a KK compactification, that the KK modes of e.g. the graviton saturate the WGC. 

\begin{figure}[ht]
\centering
\includegraphics[width=.35\textwidth]{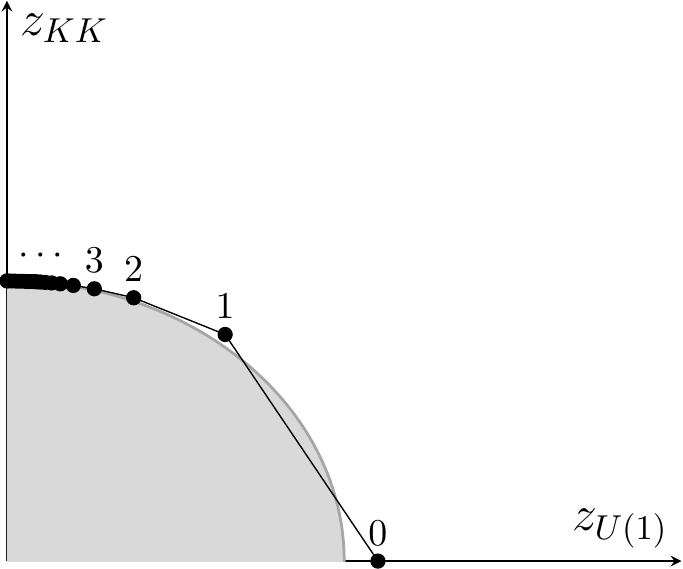}
\caption{Plot of charge to mass ratios $\vec{z}=(z_{U(1)},z_{KK})$ of KK modes of a particle charged under a higher dimensional $U(1)$ gauge field and originally satifying the WGC $z_0>1$ in higher dimensions. The extremal region is indicated in gray. The states are labelled by $q_{KK}=n \in \mathbb{Z}$. The WGC is not satisfied in the lower dimensional theory.}
\label{fig:convexhullKK}
\end{figure}

However, we have learned that in the presence of multiple $U(1)$ gauge fields, we must impose a stronger condition than satisfying the WGC for each $U(1)$. We must require that the Convex Hull Condition is satisfied. In the lower dimensional theory, the charge to mass ratio vector of the KK modes of the original WGC-satisfying particle is given by
\be
\vec z=(z_{U(1)},z_{KK})=\frac{(q_0,q_{KK}/R)}{\sqrt{m_0^2+q^2_{KK}/R^2}}\,,
\ee
as they are charged both under the original $U(1)$ gauge field and the KK photon.
If we plot this charge to mass ratio in the two directions spanned by the KK photon and the original gauge field, we get figure \ref{fig:convexhullKK}. We can see that the states lie on an ellipsoid satisfying 
\be
z_{U(1)}^2/z_0^2 + z_{KK}^2=1\,, 
\ee
outside the extremal region. However, since they only populate a finite region of the ellipsoid, they are not always enough to satisfy the WGC as their convex hull can cut the extremal region as represented in figure \ref{fig:convexhullKK}. It can be checked that the convex hull WGC will only be satisfied if the following relation holds,
\be 
\left( m_0 R^2 \right) \geq \frac{1}{4 z_0^2} \left( z_0^2-1 \right)\,.
\ee 
This implies that the Convex Hull Condition is violated unless the radius is very large or we started with a very superextremal particle. Otherwise, the dimensionally reduced theory will violate the WGC, even if it was satisfied in higher dimensions.

The solution that was proposed in \cite{Heidenreich:2015nta} was to start with a tower of states satisfying the WGC already the original theory in higher dimensions instead of a single particle.  This tower will then give rise to lower dimensional states that will exactly lie on the extremal region, filling the entire ball and satisfying the Convex Hull Condition in lower dimensions.

It is an interesting proposal, because this is in fact the way in which the WGC is satisfied in typical string theory compactifications. Take e.g. the case of Type IIA string theory on a circle, which has the $C_1$ gauge field as well as a KK photon. To satisfy the Convex Hull WGC we need to have not just one particle but a charged tower under $C_1$ in the original 10d theory. This tower of particles is precisely the tower of D0-branes, which are the KK modes from M-theory. So precisely because Type IIA already has this tower of charged states (which signals the extra M-theory direction when taking the strong coupling limit), the WGC continues to be satisfied under dimensional reduction to 9d. There are many more examples of this phenomenon in string theory, where not only one state but towers of them satisfy the WGC. Furthermore, if one focuses on toroidal compactifications, one finds that there is a superextremal state for each point of the entire charge lattice. This provides a stronger version of the Completeness Hypothesis.  However, in more general situations, only a sublattice is populated by superextremal states, as explained next.\\

\underline{Toroidal Orbifold Compactifications:}\\

Consistency under dimensional reduction was first used to formulate a Lattice WGC (since toroidal compactifications always give rise to a lattice), but in \cite{Heidenreich:2016aqi} it was shown, by studying toroidal orbifold compactifications, that only a sublattice of states satisfies the WGC. Consider a toroidal orbifold e.g. 
\be 
T^3/\mathbb{Z}_2 \times \mathbb{Z}_2'\,,
\ee
where each $\mathbb{Z}_2$ acts freely, so we mod out by a ``rototranslation'' symmetry. Compactification on the three-torus gives rise to a set of KK $U(1)$ gauge fields
\be 
\text{KK photons of } T^3 : W_\mu\,, Y_\mu\,, Z_\mu\,.
\ee
However, due to the orbifold, one of the KK photons, say $W_\mu$, is projected out. The corresponding KK modes stay, which has non-trivial implications. In fact, the KK modes associated to odd charge $n_Y$ under $Y_\mu$ receive a contribution from the KK modes of $W_\mu$, so that these odd charge states are actually subextremal. Hence, it is only a sublattice of states (with even $n_Y$ charge) that satisfy the WGC. This was the first example to suggest the idea of the sublattice, confirmed by the results using modular invariance of the CFT \cite{Montero:2016tif}.

\subsubsection{Species Bound Cut-Off}\label{Species Bound Cut-Off}

When there is not just one but infinitely many weakly coupled states, e.g. a sublattice or tower of states, becoming light, this implies a drastic breakdown of the EFT by quantum gravitational effects. Therefore, there is a natural QG cut-off associated to a tower of states known as the Species Bound \cite{ArkaniHamed:2005yv,Dvali:2007wp,Dvali:2007hz,Dvali:2010vm}.

The basic idea is that gravity becomes strongly coupled if there are too many light fields. Assuming that perturbative techniques work, one can check that the graviton propagator gets renormalized due to the presence of light degrees of freedom, so the QG cut-off is actually given by
\be 
\label{species}
\L_{\rm QG}=\frac{M_{p}}{N^{1/d-2}}\,,
\ee
where $d$ is the spacetime dimension and  $N$ is the number of light species.
Hence, the QG cut-off is smaller than the Planck scale, so quantum gravitational effects become important earlier than one would naively expect. The same species bound has been motivated using non-perturbative arguments based on black hole physics \cite{Dvali:2007wp,Dvali:2007hz}.

For a tower of states, the number of light species, i.e. how many states there are below $\L_{\rm QG}$, is related to the mass gap of the tower as follows,
\be 
\label{Nsepcies}
N=\frac{\L_{\rm QG}}{\Delta m}\,.
\ee
Plugging this into \eqref{species} one finds that the species scale  behaves as
\be 
\label{speciesm}
\L_{\rm QG}=M_{p}^{\frac{d-2}{d-1}} (\Delta m)^{\frac{1}{d-1}}\,.
\ee
When the full tower is getting light, they are also getting more degenerate in mass, so the QG cut-off goes to zero. If the states in the tower saturate the WGC, the mass gap is proportional to the gauge coupling
\be 
\Delta m \sim g\,,
\ee
so the QG cut-off scales with a power of the gauge coupling itself,
\be 
\L_{\rm QG} \sim g^{\frac{1}{d-1}} M_p^{\frac{d-2}{d-1}}\,.
\ee
Therefore, we see that this QG cut-off goes to zero as the gauge coupling goes to zero.
This situation is similar to what we encountered in the magnetic WGC, but here the cut-off is not motivated by the presence of monopoles, but by the infinite tower of states that are becoming light. Unlike for the magnetic cut-off, this truly signals a drastic break down of the effective field theory by quantum gravitational effects. 

\begin{figure}[ht]
\centering
\includegraphics[width=.23\textwidth]{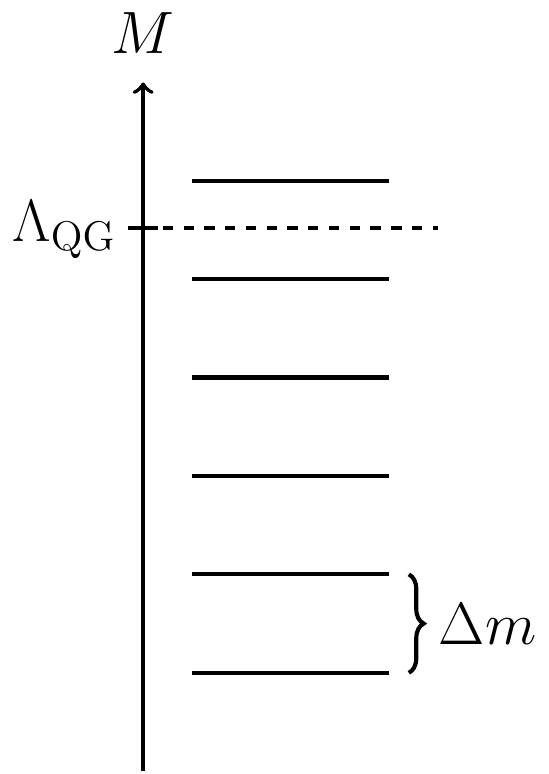}
\caption{Tower of states separated by a mass gap $\Delta m$ in a theory with cut-off $\L_{\rm QG}$.}
\label{fig:massgaptower}
\end{figure}

\subsection{Phenomenological Implications}

Among all the possible phenomenological implications of the Weak Gravity
Conjecture and its refined versions, let us discuss one that triggered the renewed great deal of activity on the Swampland program in the past five years. We are talking
about the implications for large field inflation, more specifically in the
context of natural inflation. This was triggered by the following works \cite{Rudelius:2015xta,Montero:2015ofa,Brown:2015iha} in response to the increased amount of research on large field inflationary models originated from the BICEP2 announcement about primordial gravitational waves.

Recall that the WGC should apply not only to particles, but to any kind of state
that is charged under a $p$-form gauge field. According to this, for any $p$-form
gauge field there should exist a $(p-1)$-dimensional state whose charge to mass
ratio is greater than or equal to the one of the extremal $(p-1)$ black brane.
One can apply this for axions, that are 0-form gauge fields. The WGC for
axions implies that there must exist an instanton with quantized charge $q$ whose
action satisfies
\begin{equation}
\label{WGCinstanton}
	S \lesssim q \frac{M_{p}}{f} \, .
\end{equation}
Here, we have used that the equivalent of the mass for an instanton is its
action, and that the gauge coupling of an axion is the inverse of its decay
constant, $f$. The twiddle in the inequality is due to the lack of extremal solutions for instantons. Since there is no known way to define an extremality bound in this case, the previous bound actually contains an undetermined order one factor.

In natural inflation scenarios, an axion slowly rolls down a non-perturbative potential generated by instantons. It takes the form
\begin{equation} \label{eq:axion-potential}
	V(\phi) \sim e^{-S} \cos\left( \frac{2\pi \phi}{f}\right) \, ,
\end{equation}
where we have assumed that the leading contribution comes from a charge one
instanton and we have neglected the higher harmonics contributions, which are
suppressed with respect to the fundamental one. Looking at this expression, we
see that the field range available for inflation is of the order of the decay
constant. Notice that, if the leading contribution comes from an instanton with
quantized charge $q$ greater than one, this field range is smaller, so in general
$\Delta\phi \lesssim f$.

\begin{figure}[ht]
\begin{center}
\includegraphics[width=\textwidth]{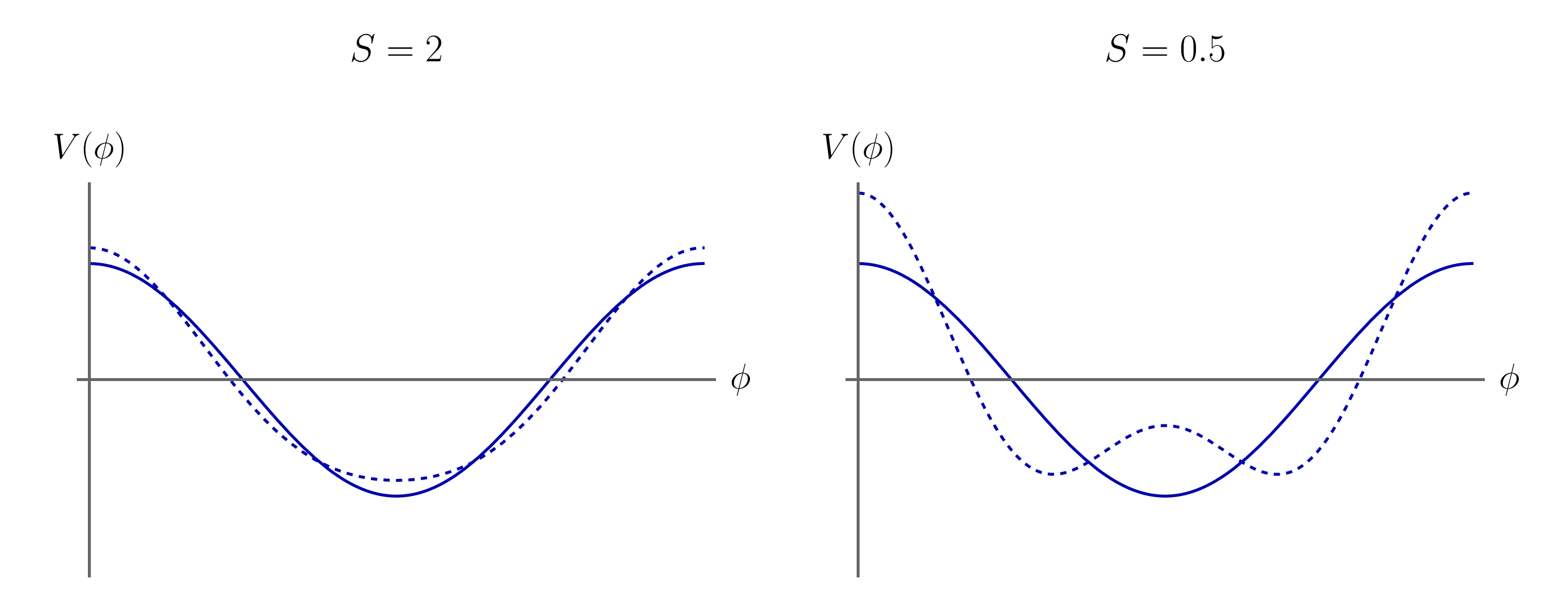}
\caption{Non-perturbative potential for the axion. The plain line is the
fundamental harmonic contribution while the dashed one includes the first higher
harmonic correction. Left: The action is larger than one so the correction does
not modify the field range available for inflation. Right: For small action the
corrections spoil the monotonicity of the potential, shortening the field range
for inflation.}
\label{fig:inst-potential}
\end{center}
\end{figure}

It is important to take into account that \eqref{eq:axion-potential} applies in
the diluted instanton gas approximation, which requires that $S\gtrsim
1$ to keep control of the instanton expansion. Moreover, if this condition is relaxed in some model, the higher harmonics
that we previously neglected will not be suppressed enough anymore and will reduce the
field range available for inflation (see figure \ref{fig:inst-potential}). So
the maximum field range is attainable by having the non-perturbative potential generated
by a charge one instanton with $S\gtrsim 1$.

If this instanton satisfies the WGC, we can combine this condition on the action $S\gtrsim 1$ with the WGC bound \eqref{WGCinstanton},
obtaining
\begin{equation}
	\Delta \phi \lesssim f \lesssim M_{p} \, ,
\end{equation}
which means that transplanckian axionic excursions, and consequently, natural (large field) inflation, is disfavoured by the WGC. Of
course what we are considering here is only the simplest case which includes a single axion.
One can use the convex hull WGC to place bounds on models with several axions, including N-flation and alignment models.

A loophole in this argumentation is to be noted. It could be the case that the instanton satisfying the WGC is not the one contributing the dominant piece of the potential for inflation. This could happen if both the action $S$ and the quantized charge $q$ of the WGC-satisfying instanton are larger than the ones corresponding to the instanton generating the inflationary potential. In this way, the exponential suppression in \eqref{eq:axion-potential} is greater for the WGC instanton, but the field range is instead bounded by
\begin{equation}
	\Delta \phi \lesssim f \lesssim q\, M_{p} \, ,
\end{equation}
where we have considered the best case scenario in which the instanton generating the potential has unit charge.
So a transplanckian field range is allowed if the charge of the WGC-satisfying instanton is $q>1$. However, this requires a somewhat unnatural scenario in which the charge to mass ratio of the WGC instanton is bigger than for the \emph{light} instanton generating the potential, which goes against the monotonicity found in string theory examples and which is explained in figure \ref{fig:graphWCGhetstring}. Nonetheless, we cannot discard this scenario (see e.g. \cite{Hebecker:2015rya}).

Thus we have seen that the mild version of the WGC, in which it is not specified which instanton should satisfy the WGC, is not enough to rule out large field inflation. 
In order to determine meaningful constraints, some strong version of the WGC is
needed. For instance, if the full lattice
of WGC states is present, the unit charge one will imply an important constraint
for large field inflation. But if only a sublattice is required, as the index of the lattice grows, the quantized
charge of the first state satisfying the WGC grows and the constraint gets
weaker. If this index is very large, the implications stop being relevant for
constraining large field inflation. This is why constraining the index of the sublattice is one of the most important open questions of research in the WGC. All  known string theory examples have an index of order one, but we are lacking a fundamental universal bound. This issue is highly linked to the question of the maximum rank for discrete symmetries; a large sublattice index hints at a large discrete symmetry. However, this rank is expected to be bounded in order to avoid a global symmetry.

\subsection{The Weak Gravity Conjecture with Scalar Fields\label{sec:WGCscalars}}

The presence of massless scalar fields strongly affects the WGC bounds. 
As discussed in section \ref{sec:defWGC}, without scalar fields, there are two different but equivalent interpretations of the WGC. We could either think of it as a super-extremality condition or a repulsive force condition in the sense that the gauge repulsion is  stronger than the gravity attraction over two equal states. However, these two conditions will not be equivalent any more in the presence of massless scalar fields. Clearly, the second condition gets modified due to the presence of additional scalar forces, and the extremality bound is also sensitive to the contribution from massless scalars. The current convention is to still identify the WGC with a superextremality condition, while the the second interpretation receives the name of the Repulsive Force Condition.
\nl
\begin{itemize}
    \item \underline{\textbf{WGC (superextremality condition):}}\\

Given some p-form gauge field, $\exists$ a superextremal state satisfying
\begin{equation}
Q^{2}\geq \g(\f_{i})M^{2}~,
\end{equation}
where the massless scalars $ \{\f_{i}\} $ can also contribute to the extremality factor $ \g(\f_{i})$. Recall that this factor is the charge to mass ratio of an extremal black hole in that theory. For instance, a famous example is a dilatonic black hole, which is an extremal black hole solution in an Einstein-Maxwell-dilaton theory in which the gauge kinetic function of the gauge fields is parametrized by (canonically normalised) massless scalars as follows,
\begin{equation}
\label{gaugekinetic}
\mathcal{L}\supset\frac{1}{2g^{2}_{0}} e^{\a_{i}\f_{i}}F^{2}_{p+1}~.
\end{equation}
The scalars induce an additional contribution to the charge to mass ratio of the extremal black holes, so the WGC becomes
\begin{equation}\label{extremality bound with scalars}
Q^{2}M^{d-2}_{p}\geq \Big( \frac{\lvert\vec{\a}\rvert^{2}}{2}+\frac{p(d-p-2)}{d-2} \Big)T^{2}
\end{equation}
with $Q=qg_0$. By comparing to \eqref{RNBH} we can see that there is an extra term proportional to the scalar couplings  $ \vec{\a}:=\{\a_{i}\} $ in \eqref{gaugekinetic}.

The generalization to multi-fields will again involve a convex hull condition as in section \ref{sec:convex-hull-WGC}. However, notice that the extremal region will \emph{no longer} be a \emph{unit ball} but will depend on the scalar contribution. For states that are mutually BPS, the extremal region becomes an ellipse, while for states that are non-mutually BPS it develops straight lines. The take-home message is that, in order to check the WGC, it is essential to first know the charge to mass ratio of extremal black holes in that theory, which can vary extremely from one theory to another if massless scalar fields are present.\\


\item \underline{\textbf{Repulsive Force Condition (RFC):}}\\

Given some p-form gauge field, $\exists$ a state which is self-repulsive, namely the gauge repulsion between two copies of the state acts stronger than the sum of the gravitational and scalar interactions \cite{Palti:2017elp,Heidenreich:2019zkl}, i.e.
\begin{equation}\label{repulsive force condition}
F_{\text{gauge}}\geq F_{\text{gravity}}+F_{\text{scalar}}~,
\end{equation}
In a d-dimensional theory, each of these forces for a particle of mass $m$ and gauge charge $Q$ take the following form
\begin{eqnarray}
F_{\text{gauge}} = \frac{  Q^2}{r^{d-2}}~,\quad
F_{\te{gravity}} = \frac{m^2}{M_p^{d-2}r^{d-2}}\frac{d-3}{d-2}~,\quad
F_{\te{scalar}} = \frac{\m^2}{M_p^{d-2}r^{d-2}}~,
\label{forces}
\end{eqnarray}
where $ \m $ is the scalar Yukawa charge.
The scalar Yukawa force emerges whenever the mass $ m $ of the state is parametrized by a massless scalar $\phi$, as we can always expand the mass term as follows,
\begin{equation}
\mathcal{L}\supset m^{2}(\f)\chi^{2}=\Big( m^{2}_{0}+2m_{0}(\partial_{\f}m)\f \Big)\chi^{2}+\dots~,
\end{equation}
where the scalar Yukawa charge is then given by $\mu= \partial_{\f}m $. 
 For the case of a particle in four dimensions, the RFC in (\ref{repulsive force condition}) reads
\begin{equation}\label{extremality bound from repulsive force condition}
Q^{2}M^{2}_{p}\geq \frac12 m^{2}+g^{ij}(\partial_{\f^{i}}m)(\partial_{\f^{j}}m)M^{2}_{p}~,
\end{equation}
where we have allowed for the presence of multiple scalars with inverse field metric $ g^{ij}$.  This condition was first proposed in \cite{Palti:2017elp} as the proper interpretation of the WGC in the presence of scalar fields, and later named as the RFC in \cite{Heidenreich:2019zkl}.

For a p-form gauge field, we can generalize \eqref{forces} and derive the existence of a $ (p-1) $-brane of tension $T$ satisfying
\begin{equation}\label{eq:RFC-general}
f^{ab}q_{a}q_{b}\geq g^{ij}(\partial_{\f^{i}}T)(\partial_{\f^{j}}T)+\frac{p(d-p-2)}{d-2}T^{2}\ ,
\end{equation}
where we have written explicitly the gauge charge in terms of the inverse gauge kinetic function $f^{ab}$ and gauge quantized charges $q_a$.

\end{itemize}

Note that \eqref{extremality bound with scalars} and \eqref{eq:RFC-general} are identical apart from the contribution from the massless scalars. In the WGC, the scalars contribute through the dependence in the gauge kinetic function; while in the RFC, they enter through the behaviour of the mass/tension. Consequently, it seems we now have two different bounds, so which one is realised in quantum gravity? This is an open question, although there is no known counterexample for any of the two conditions in string theory, so it might be that both are realised.
In fact, although seemingly different, they coincide in many cases. For instance, string theory evidence shows that they are fulfilled by the same states at the weak coupling limits $ g\rightarrow 0 $ \cite{Lee:2018spm,Gendler:2020dfp}. Furthermore, extremal BPS states saturate both conditions, as by definition they satisfy a no-force condition. In these cases, the charges and masses are related such that the scalar contributions in \eqref{extremality bound with scalars} and \eqref{eq:RFC-general} become equal.
This is in fact related to the sharpening of the WGC discussed in the next section.



\subsection{Sharpening the WGC and BPS states}\label{sec:Sharpening the WGC}

A sharpening of  the WGC was proposed in \cite{Ooguri:2016pdq}:\\

\begin{otherbox}[Sharpening of the WGC]{box:sharpWGC}
The WGC can only be saturated by BPS states in a supersymmetric theory.
\end{otherbox}

This has important implications for the stability of non-susy vacua, as discussed in section \ref{sec:motivation-from-WGC}. The idea is that only supersymmetry can guarantee an exact equality relating the mass and charge of a state. Otherwise, one should expect quantum corrections preventing these physical quantities from saturating the WGC bound.
In the following, we will summarize some examples from string theory compactifications that support this sharpening. Unfortunately, the evidence is scarce.  This  is due to the difficulties in computing the mass of non-BPS states.
\begin{itemize}
	\item The first typical example is the toroidal compactification of heterotic string theory, discussed in section \ref{sec:evidenceWGC}. In this setup, we can compute the whole spectrum perturbatively including non-BPS states. These states do not saturate the WGC as their mass is strictly smaller than the charge,
	\begin{equation}
	m^{2}_{\text{non-BPS}}=\frac{2}{\a^{'}}\big(\lvert q^{2} \rvert-2\big)<\frac{2}{\a^{'}}\lvert q^{2}\rvert =M^2_{\rm BH} ~,
	\end{equation}
	as given in \eqref{mhet}. 
	 Indeed, by comparing to the charge to mass ratio of the extremal non-BPS black hole solutions of the theory, these states are superextremal. This is why they lie below the extremal curve in figure \ref{fig:graphWCGhetstring}.
	 
	\item Another example are the exictation modes of strings becoming tensionless at the weak coupling limit of gauge field theories in F-theory CY$_3$ compactifications. The string arises from wrapping a D3-brane on a shrinking curve. This example was briefly discussed in section \ref{sec:evidenceWGC} and we will have a more detailed explanation in section \ref{sec:SDC-5d-N1}. The string spectrum is given by non-BPS states satisfying
	\begin{equation}
	g^{2}q^{2}=m^{2}+4jg^{2}>m^{2}~,
	\end{equation}
	where $m$ and $q$ are the mass and quantized charge of the state, $g$ the gauge coupling and $j$ some integer depending on some intersection number defined below \eqref{eq:string-excitations}. Again, these states are superextremal with respect to the black hole extremality bound, which in the weak coupling limit coincides with a no-force condition.

    \item Higher derivative corrections to non-BPS black holes in heterotic string theory also motivate the sharpening of the WGC. The corrections go in the direction of increasing the charge to mass ratio of small black holes, so they satisfy the strict inequality of the WGC.

    \item Consider some supersymmetric theory with massless scalar fields, so they contribute to the extremality bound as given in \eqref{extremality bound with scalars}. 
    Upon compactification, if supersymmetry is preserved, the scalar contribution $|\alpha|^2$ changes in such a way as to compensate the variation of the second term in  \eqref{extremality bound with scalars} due to the change in the space-time dimension. This way, a particle saturating the WGC in higher dimensions will also saturate it in lower dimensions. However, if supersymmetry is broken while compactifying, the scalars will typically get a mass, so they will not contribute to the extremality bound anymore, since $\alpha=0$. Hence, the charge to mass ratio of extremal black holes decreases, and a previously extremal particle will become superextremal with respect to the new extremality bound in the lower dimensional non-supersymmetric theory. More generally, since the scalar contribution to the extremality bound is positive, this helps to make the states satify the strict inequality when supersymmetry gets broken and the states are no longer BPS.
\end{itemize}

All the previous items are examples of non-BPS states satisfying the strict inequality of the WGC, and not saturating it, in agreement with the sharpening of the WGC above, according to which, only BPS states could exactly saturate the WGC. If that happens, then the BPS state is saturating both the WGC and the RFC (as it feels no force), so the scalar contributions in \eqref{extremality bound with scalars} and \eqref{eq:RFC-general} coincide. One should keep in mind, though, that it is not necessary for a BPS state to saturate the WGC; there are examples of non-extremal BPS states near finite distance singularities of the moduli space, e.g. conifold singularities. Contrary, BPS states becoming light at the infinite field distance singularities are typically extremal \cite{Gendler:2020dfp}.

If the sharpening of the WGC is true, it predicts interesting conditions on the geometry of string theory compactifications. For instance, take M-theory compactified on $ K3 $  where M2-branes can wrap non-holomorphic curves, so the resulting states are non-BPS. Then, validity of the sharpened version of the conjecture implies that the mass associated to the area of two-cycles has to be smaller than the charge which can be obtained by integrating the K\"{a}hler class over the two-cycle. A check of this geometric statement would provide a robust piece of evidence for the sharpened WGC.
Another example would be to consider M-theory on $ G2 $ manifolds. Since the resulting 4d theory is $ \mathcal{N}=1 $, the particles are non-BPS. According to the sharpened WGC, these states should satisfy only the strict inequality of the WGC.
\nl

\subsection{Open Questions}

Here, to conclude the WGC, we mention some of the key open questions as well as some generalizations and other current avenues of research that we did not cover in these lectures:

\begin{itemize}
	\item WGC version for discrete gauge symmetries. Some proposals have appeared, see e.g. \cite{Craig:2018yvw} and $Z_K$ Weak coupling conjecture in \cite{Buratti:2020kda}.
	\item Bottom-up motivation of the WGC for axions and low codimension objects. There are no extremal black hole solutions of such dimensionalities, so the motivation for the WGC based on black hole physics is missing in these cases. 
	\item Very little evidence in non-susy setups.
	\item Better understanding of the interplay between the extremality bound, the BPS bound and the repulsive force condition.
	\item What is the exact strong version of the WGC realized in string theory? If it is the sublattice WGC, is there any upper bound on the sublattice index?
	\item Possible loophole: WGC under Higgsing, see e.g. \cite{Saraswat:2016eaz}. The WGC might not be satisfied in the IR upon Higgsing even if satisfied in the UV, unless the amount of Higgsing is also constrained by quantum gravity.
	\item Can we rigorously prove any fundamental inconsistency with having stable non-BPS black holes?
	\item WGC in AdS$ _{d+1}  $/CFT$ _{d} $ for $ d>2 $. The evidence for the WGC using modular invariance only applies for $d=2$.
	\item WGC in de Sitter space, see e.g. \cite{Montero:2019ekk}.
\end{itemize}

\section{Swampland Distance Conjecture}
\label{sec:distconj}

We now turn to discuss the Distance Conjecture, which plays a central role in the Swampland program as it comes with many interesting relations to the other Swampland conjectures.

\subsection{Asymptotic Limits in Moduli Space} \label{sec:asymptotic-limits}

In string theory every coupling, mass, etc., is controlled by the vacuum
expectation value (vev) of some scalar fields. They are called moduli as they are massless before adding fluxes or other ingredients to the compactification manifold. From the
10d perspective, they control the size and shape of the extra dimensions. This motivates  the common lore that there
are no free parameters\footnote{This  is actually related to the absence of $(-1)$-form global symmetries, where the role of the symmetry current is played by the parameter.} in string theory, but all of them are \emph{dynamical}.  

Different effective field theories can be explored by moving in this moduli
space. For example, one may try to move towards a point in which a global
symmetry is restored, by e.g. sending some gauge coupling to zero. This case is
particularly interesting, since global symmetries are forbidden in Quantum
Gravity, so something dramatic is expected to happen when approaching such a limit.
Leaving its definition for later, let us recall that the moduli space is
naturally equipped with a Riemannian metric. The obvious way in which the theory
is protected against restoring a global symmetry in this way is that,
according to this metric, weak coupling limits are at infinite field distance
in moduli space. But this is not the end of the story. One would expect the effective field theory description to continuously break down as the approximate global symmetry looks more and more exact as we move towards the infinite distance loci.
This is precisely the behaviour predicted e.g. by the magnetic version of the WGC or the species bound for the WGC tower, since the
EFT cut-off will fall down to zero as the gauge coupling goes to zero.

It is worth noting that the cut-off decreasing is purely a quantum gravity
mechanism. From the Quantum Field Theory perspective, even if the point with vanishing gauge coupling is unreachable since it is at
infinite distance, there seems to be nothing wrong with being as close as possible.
In fact, it is precisely in these weak coupling limits where we have good control of the EFT, so a tiny gauge coupling is very appealing  from a bottom-up perspective. However, restoring
a global symmetry is strictly forbidden in quantum gravity and, as a consequence, the effective
field theory description should break down in a continuous way by quantum gravitational effects when taking these
infinite distance limits. 

We could try to run a similar argument when taking other limits restoring other types of global symmetries. Or more generally, we could wonder what happens when
approaching any infinite distance boundary in moduli space. Does the EFT always break down? And if so, what is the mechanism introduced by QG so that it happens? This is quantified by
the Swampland Distance Conjecture (SDC), to which this section is
devoted.

\subsection{Swampland Distance Conjecture}

Consider a $D$-dimensional effective field theory coupled to Einstein gravity
and with some moduli space $\mathcal{M}$ parametrized by massless scalar fields.
The action includes the term
\begin{equation}
    S \supset M_{p}^{D-2} \int d^{D}x \sqrt{-h} \left( \frac{R}{2} - \frac{1}{2} g_{ij} \partial_{\mu} \phi^{i} \partial^{\mu} \phi^{j} \right) \, ,
\end{equation}
where $g_{ij}$ is identified as the metric in moduli space.

The first statement of this conjecture is that the moduli space is non-compact.
Starting from a point $P\in\mathcal{M}$, there always exist another point
$Q\in\mathcal{M}$ at infinite geodesic distance $d(P,Q)$. Note that a point is at infinite distance if every trajectory approaching the point has infinite length. One typical example in
string theory compactifications is the decompactification limit. The second and more important statement describes what happens if we try to approach some point at infinite field distance:
\\
\begin{conj}[Swampland Distance Conjecture]{conj:SDC}
There is an infinite tower of states that becomes exponentially light at any
infinite field distance limit as  
\begin{equation}
	M(Q) \sim M(P) \, e^{-\lambda \Delta\phi} \quad \text{when } \Delta\phi\to\infty,
\end{equation}
in terms of the geodesic field distance $\Delta\phi \equiv d(P,Q)$ \cite{Ooguri:2006in}.
\end{conj}

\begin{figure}[ht]
\begin{center}
\includegraphics[width=0.9\textwidth]{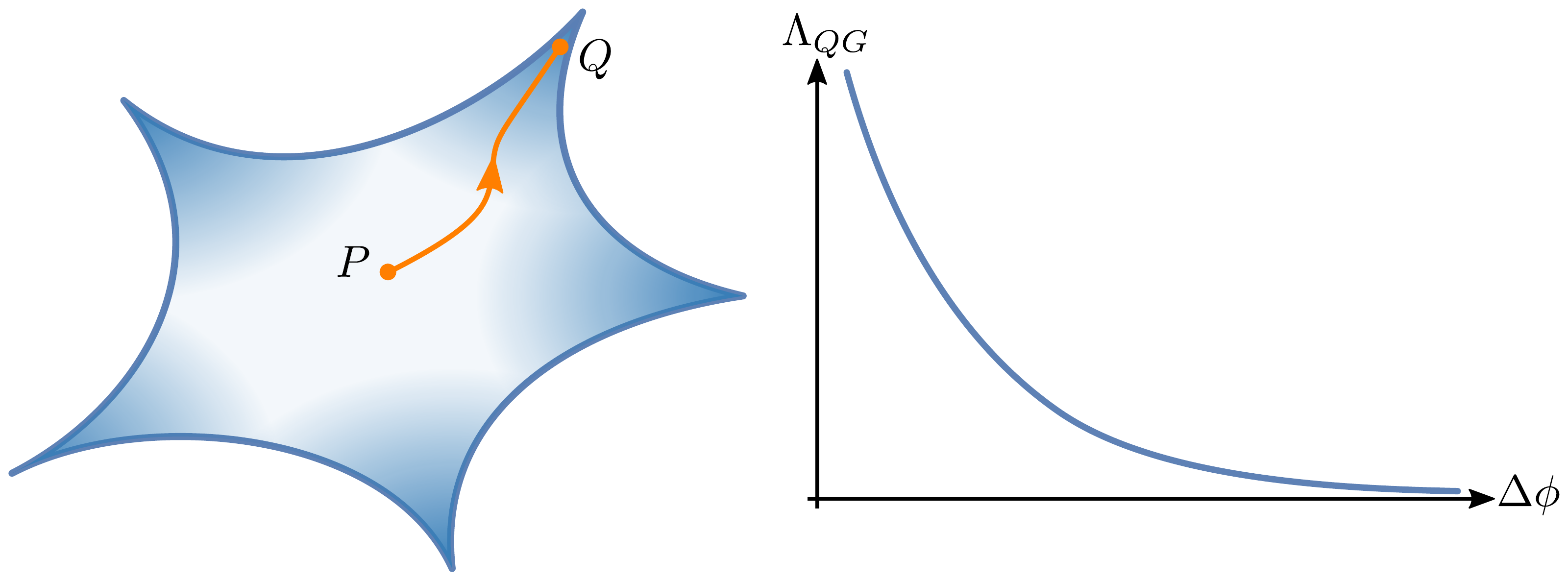}
\caption{Left: Pictorial representation of quantum gravity moduli space with several asymptotic limits at infinite distance. Right: Quantum gravity cut-off falling exponentially with the geodesic distance as we approach an infinite distance point, due to the tower of states predicted by the SDC.}
\label{fig:moduli-space}
\end{center}
\end{figure}

The exponential rate $\lambda$, apart from being positive, is not specified by
the conjecture.
 It is expected to be an $\mathcal{O} \left( 1\right)$ constant, as otherwise it could spoil the exponential behaviour,
but its origin is not known. 
In fact, determining how small this parameter can be is the main open question
about the SDC, and we will see that significant progress on this has been achieved in certain string theory setups. Concrete lower bounds have also been proposed in the literature \cite{Gendler:2020dfp,Bedroya:2019snp,Andriot:2020lea,Lanza:2020qmt} and will be discussed later. Clearly, having an unspecified factor in the conjecture is not
satisfactory; we would like to be able to compute it from first principles or
effective field theory data. This was the case for the WGC, in which the order
one factor is specified in terms of the extremality bound for black holes.

The infinite tower of states is weakly coupled and signals the breakdown of the effective field
theory, as it is impossible to have an effective field theory description weakly coupled to Einstein gravity with infinitely many light degrees of freedom.  Hence, there is a quantum gravity cut-off
associated to the infinite tower of states, 
which decreases exponentially in terms of the proper field distance,
\begin{equation} \label{eq:SDC-cut-off}
	\Lambda_{QG} = \Lambda_{0}\, e^{-\lambda \Delta\phi}\, ,
\end{equation}
as represented in figure
\ref{fig:moduli-space}. 
For simplicity, here we are taking this cut-off
to coincide with the first state of the tower. But a more accurate way of defining
it is via the species bound cut-off given in \eqref{speciesm}, whose
exponential rate will differ from $\lambda$ by an order one factor depending on the space-time dimension.

 
An immediate consequence is that effective field theories are only valid for
finite scalar field variations. From \eqref{eq:SDC-cut-off} and taking into account that $\Lambda_{0} \leq M_{p}$ one gets
\begin{equation} \label{eq:SDC-bound}
	\Delta\phi \leq \frac{1}{\lambda} \log \frac{M_{p}}{\Lambda} \, ,
\end{equation}
which is telling us that the maximum field variation actually depends on the
cut-off of the effective field theory. This means that the higher the cut-off or the process changing the vev of the scalar, the smaller is the maximum field
distance that can be described within the effective field theory. This statement
has direct implications for inflation that will be discussed later. Notice that this correlation between the cut-off and the maximum field range is intrinsically quantum gravitational, as these two quantities are a priori unrelated from a QFT perspective.

\subsection{Examples and Connection to String Dualities\label{sec:SDCdual}}

The prototypical example of how the SDC is realized is a KK circle
compactification of string theory to $d$ space-time dimensions. Taking $r$ to be
the modulus controlling the radius of the circle one finds:
\begin{equation}
\label{SKK}
	S \supset M_{p}^{d-2} \int d^{d}x \sqrt{-h} \left( \frac{R}{2} - \frac{1}{2} \frac{d-1}{d-2} \frac{(\partial r)^{2}}{r^{2}} \right) \, 
\end{equation}
where we show only the Einstein-Hilbert term and the kinetic term for the modulus $r$.
There are thus two limits at infinite distance, small radius $r\to0$ and large
radius $r\to\infty$. The proper field distance is given by the canonically
normalized field
\begin{equation}
	\Delta R = \sqrt{\frac{d-1}{d-2}} \, \log r \, .
\end{equation}
Taking the decompactification limit $r\to\infty$ we know that the KK tower
becomes light. The mass of these modes is given by
\begin{equation} \label{eq:KK-mass}
	m_{KK} = \frac{q}{r^{\frac{d-1}{d-2}}} = q \, e^{-\sqrt{\frac{d-1}{d-2}} \, \Delta R} \, ,
\end{equation}
so we indeed find an infinite tower of states becoming exponentially light with
the distance, as required by the SDC. Moreover, in this case we have been able
to compute the exponential rate in terms of the space-time dimension of the
effective field theory as
\begin{equation}
\label{lambdaKK}
	\lambda = \sqrt{\frac{d-1}{d-2}} \, .
\end{equation}
As expected, it turns out to be an order one constant.

\begin{figure}[ht]
\begin{center}
\includegraphics[width=0.8\textwidth]{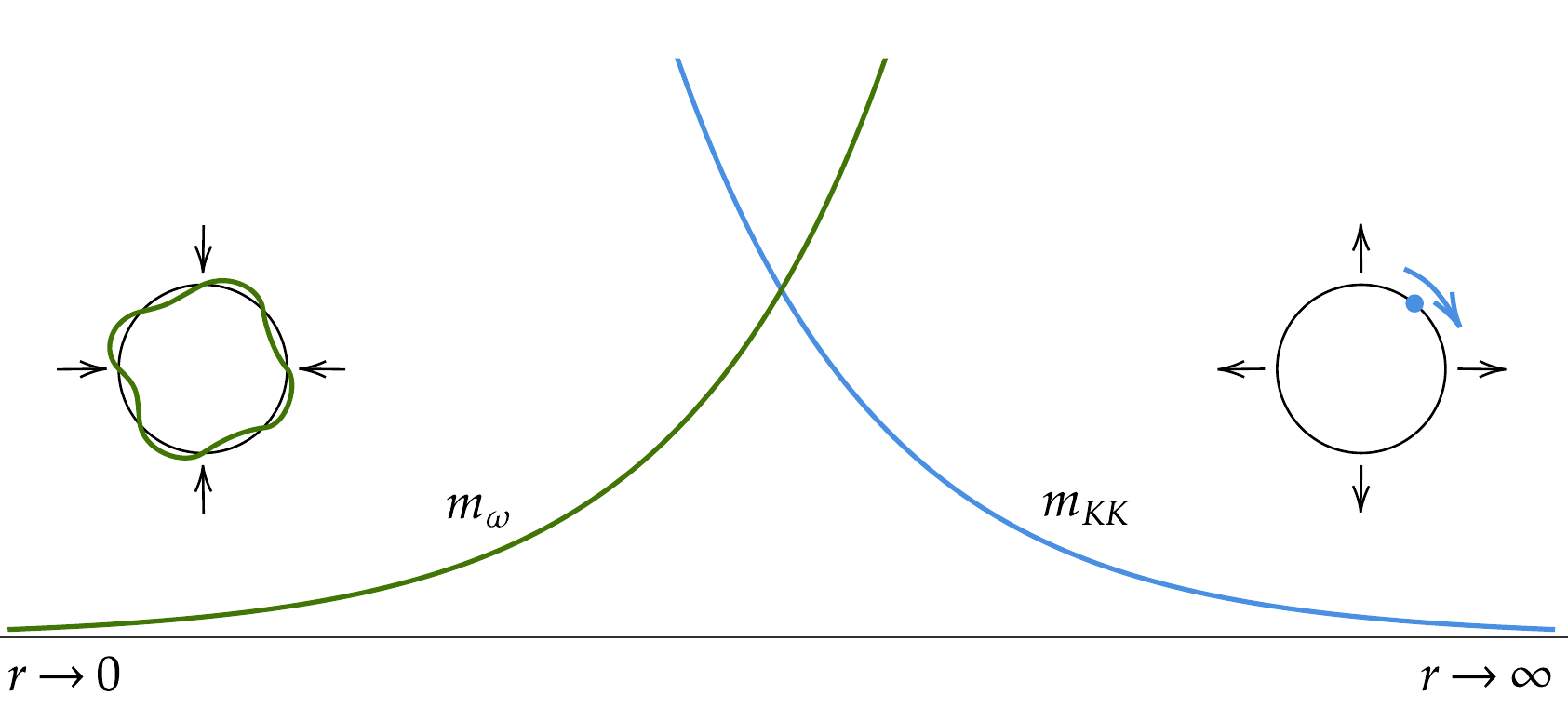}
\caption{Illustration of the SDC being realized in circle compactifications of string theory. KK (winding) modes become exponentially light at large (small) radius.}
\label{fig:SDC-KKcircle}
\end{center}
\end{figure}

In the opposite limit at infinite distance $r\to0$ the KK tower cannot be the
one satisfying the SDC. In this limit it is another
infinite tower that is becoming light. These are the winding modes (wrapping strings) and they
present the same behaviour as the KK modes in the decompactification limit. This
leads to the striking observation that in order to satisfy the SDC in KK
compactifications we need extended objects that can wrap some compact directions
and become light in the limit in which they shrink to zero size. For an illustration of the SDC in this setup see figure \ref{fig:SDC-KKcircle}.

Another interesting feature of this example is that the realization of the SDC
is inexorably linked to the existence of a T-duality involving both limits and their
tower of states. When the radius becomes small we can go to the dual description
in which the radius is again large and the winding modes are exchanged by a KK tower becoming light. Notice
that going to the dual description does not help when it comes to saving the
effective field theory, but it can tell us how to resolve it in the full Quantum Gravity theory.
In this case it is telling us that we have to grow a new extra dimension to resolve the singular limit.

It is very common in string theory that a dual description arises in these
asymptotic limits. Hence, the Distance conjecture can also be understood as a \emph{Duality conjecture}, in the sense that it is predicting the existence of a duality at every infinite distance limit such that the infinite tower provides the new fundamental (weakly coupled) degrees of freedom of the dual description. In this fashion, trying to understand why the SDC is true from
first principles in Quantum Gravity is closely related to giving a reason why
dualities exist in the first place.  
A bottom-up explanation of this feature or of
the SDC is missing. Currently, the SDC just points out a universal feature observed in string theory compactifications, but there is no clear intuition why it should hold in general in quantum gravity. It is fair to say, though, that there is so much evidence in support of the conjecture in string theory, that the SDC is widely accepted in the community. 
It has also been noted that a possible explanation arises
by connecting the Distance conjecture with the WGC and the absence of global symmetries. This connection is natural
when some global symmetry is restored in the limit at infinite distance, as was
discussed in section \ref{sec:asymptotic-limits}. In the next section we will
elaborate more in this direction.

Before moving on, it is worth mentioning a refinement of the SDC regarding the nature of the tower of states, which also connects with known dualities. This is the Emergent String Conjecture
\cite{Lee:2019wij}:\\

\begin{otherbox}[Emergent String Conjecture]{conj:ESC}
    Any infinite distance limit is either a decompactification limit or a limit in which there is a weakly coupled string becoming
    tensionless.
\end{otherbox}
This means that the leading tower becoming light is either a KK tower (in some appropiate dual frame) or some string excitation modes, such that the resolution in Quantum Gravity of the EFT breaking down is given by growing an extra dimension
or by considering a string perturbation theory. The first case is associated to a T-duality while the second one fits with an S-duality.

Let us show the typical example in which these two possibilities are realised in the
same moduli space. This is the case of type IIA in ten dimensions, whose action in the Einstein frame
contains
\begin{equation}
	S \supset M_{p}^{8} \int d^{10}x \sqrt{-h} \left( \frac{R}{2} - \frac{(\partial s)^{2}}{4s^2} \right) \, ,
\end{equation}
where $s$ is related to the string coupling $g_s$ and the 10d dilaton $\Phi$ via
\begin{equation}
	s = \frac{1}{g_s} = e^{-\Phi} \, .
\end{equation}
We again find two infinite distance limits that correspond to strong ($g_s\rightarrow \infty$) or weak
($g_s\rightarrow 0$) string coupling. The proper field distance grows logarithmically with the string coupling in both limits, $\Delta\phi=\frac1{\sqrt{2}}\log s$.

In the strong coupling limit, a tower of D0-branes
becomes exponentially light with the distance. Indeed, a D0-brane becomes
massless polynomially with the string coupling as
\begin{equation}
	m_{D0} \sim \frac{M_{s}}{g_s} \sim \frac{M_{p}}{g_{s}^{3/4}}\sim M_{p} \, e^{-\frac{3\sqrt{2}}{4} \Delta\phi} \, ,
\end{equation}
which means that they do so exponentially with the proper field distance.
Importantly, these D0-branes are the KK modes of M-theory on $S^1$, and thus we see that
this corresponds to a decompactification limit.

Contrary, in the weak coupling limit, the string
excitations are the modes becoming exponentially light. The mass of these states
falls polynomially to $0$ as $g_s\to 0$,
\begin{equation}
	M_{s} \sim M_{p} \, g_{s}^{1/4}\sim M_{p} \, e^{-\frac{\sqrt{2}}{4} \Delta\phi}
\end{equation}
which again decreases exponentially with the proper field distance. We see
that in this case there is no extra dimension growing up, rather this is the perturbative string limit of Type IIA.
In both cases, as dictated by the SDC, the ten dimensional quantum field theory description breaks down, and we need to either grow extra dimensions or consider a string theoretical description of the theory.

The motivation for this refinement of the SDC is example based. So, as
with the SDC itself, there is no known quantum gravity reason why the leading
towers should always be KK modes or a weakly coupled string. For instance, one
could a priori expect a limit in which the tower corresponds to a tensionless
membrane. The absence of this type of limits is, though, in agreement with the lack of a perturbative description for a membrane theory, which has been extensively searched for in the string theory community. 

\subsection{Relation to the WGC} \label{sec:relation-WGC}

As already mentioned in section \ref{sec:asymptotic-limits}, all limits in
which a global symmetry is restored because some gauge coupling vanishes are at
infinite field distance in moduli space. The opposite direction has also been proposed to be true \cite{Gendler:2020dfp} although  not proven in
general. Interestingly, in all string theory examples studied so far, there seems to be a global symmetry restored at infinite distance coming from a $p$-form gauge coupling going to zero (sometimes 
also an infinite order discrete global symmetry appears). The SDC, like the WGC, can then be understood as a quantum gravity obstruction to restore a global symmetry.

If there is a vanishing $p$-form gauge coupling in a certain infinite distance
limit, charged states satisfying some tower or sublattice version of the WGC
would be the natural candidates for the SDC tower of states becoming light. And in that case,  the exponential rate $\lambda$ will be fixed by the black hole extremality
bound \cite{Gendler:2020dfp,Lee:2018spm}. Note that even if this charged tower is not the leading one becoming light, it can be used to put a lower bound on the SDC exponential rate, as any other fast-decaying tower will have a bigger $\lambda$. Thus,
this is a very appealing situation in which the WGC and the SDC merges in one
single statement in which no undetermined order one factors are left.

To give an example, this situation is realized in the previously discussed KK
compactification on a circle of section \ref{sec:SDCdual}. In this case we have a KK photon whose kinetic
gauge function can be read from
\begin{equation}
	S \supset - M_{p}^{d-2} \int d^{d}x \sqrt{-h} \, \frac{1}{2} \, e^{-2\sqrt{\frac{d-1}{d-2}} \phi} \, F_{KK}^{2} \, ,
\end{equation}
which supplements the Einstein-Hilbert term and the scalar kinetic term given in \eqref{SKK}. Here, $\phi=\log r$ with $r$ the radius of the circle. 
This implies that the scalar contribution to the black hole extremality bound in \eqref{extremality bound 
with scalars} is given by
\begin{equation}
\label{alpha}
	\alpha = 2 \sqrt{\frac{d-1}{d-2}} \, .
\end{equation}
By plugging this into \eqref{extremality bound 
with scalars}  one
finds that the WGC requires the existence of some charge states satisfying 
\begin{equation}
\label{massKK}
	m^{2} = \frac{q^2}{r^2} \, ,
\end{equation} 
where $q$ is the quantized charge and we have used that the gauge coupling for the KK photon is given by $g_0=\frac{2}{r^2M_p^{d-2}}$.
KK modes are charged under the KK photon and precisely satisfy \eqref{massKK}. They also become exponentially light with the proper field distance, as shown in
 \eqref{eq:KK-mass}. Therefore, a KK tower 
satisfies simultaneously the WGC and the SDC with an exponential rate given by $\lambda=\alpha/2$, obtained from comparing \eqref{lambdaKK} and \eqref{alpha}.

This matching is actually quite general and will happen whenever the extremality
condition \eqref{extremality bound with scalars} and the RFC
\eqref{eq:RFC-general} coincide, in the sense that extremal states feel no force. In that case, the second term on the left hand
side of \eqref{extremality bound with scalars} is equal to the second term in
\eqref{eq:RFC-general}, which is  related to the exponential decay rate of the tower.
Then, by equating both bounds for the case of particles, one gets
\begin{equation}
	\lambda = \frac{\alpha}{2} \, ,
\end{equation}
thus fixing the exponential decay rate in terms of the extremality bound. This
merging of the conjectures has been seen to occur at weak coupling limits of several string theory setups \cite{Lee:2018spm,Gendler:2020dfp} and it is expected to hold whenever there is a gauge coupling vanishing at infinite distance.
If the tower of states of the SDC comes from the excitations of a weakly coupled
string, the exponential rate can be fixed in a similar way by using the WGC for
the two-form gauge field to which it couples.

Hence, the generality of the lower bound for the SDC exponential factor in terms of the extremality factor in the WGC depends on whether there is always a gauge coupling vanishing
at any infinite distance limit. This has been proposed to be the case in \cite{Gendler:2020dfp} and it is supported by the string theory evidence  and the
Emergent String Conjecture. The latter suggests that we can always use either the extremality bound associated to a KK photon or to a two-form gauge field for the case 
of a weakly coupled string, since in both cases their gauge coupling
is going to zero in the asymptotic limit.
\medskip

There is an alternative connection between the WGC and the SDC which does not require a gauge
coupling vanishing in each infinite distance limit. More specifically, this
requires a version of the WGC dubbed the Scalar WGC that does not even need the
presence of a gauge theory at all to be formulated. It was proposed in
\cite{Palti:2017elp} and poses that gravity should be weaker than any scalar
force, even if both of them are attractive, so they are not actually competing with
each other. For a particle with mass $m$, controlled by the vev of some moduli,
this translates into the following condition:\\

\begin{otherbox}[Scalar WGC]{box:scalarWGC}
Given some EFT weakly coupled to Einstein gravity with some massless scalar fields $\phi_i$, there must exist a state with mass $m$ satisfying:
\begin{equation} \label{eq:Scalar-WGC}
	g^{ij} \, \partial_{i}m \, \partial_{j}m > \frac{d-3}{d-2}\, m^{2}\, ,
\end{equation}
where $g_{ij}$ is the field metric and $d$ the space-time dimension.
\end{otherbox}

The above condition can be obtained from simply setting the gauge charge to zero in \eqref{extremality bound from repulsive force condition}. Unlike the usual WGC, it is not related to any black hole extremality bound, as there are no extremal black holes with only scalar charge by the no-hair theorem. Therefore, the motivation in terms of black hole decay is missing, although one could run a similar argument in terms of gravitational bound states instead of black hole remnants. 

Notice that, in order to satisfy this condition at large field distance, the
mass of the state has to decrease exponentially as required for the SDC. Hence, using the scalar WGC one can also
argue for an exponential rate  of order one. We would like to remark, though, that the evidence behind this scalar WGC is very scarce, so it is not on the same level of rigor as the usual WGC or the SDC. In fact, refinements involving second derivatives of the mass \cite{Gonzalo:2019gjp,Freivogel:2019mtr,DallAgata:2020ino,Benakli:2020pkm,Gonzalo:2020kke} have appeared, which suggests that the simplest form in \eqref{eq:Scalar-WGC} might not be the end of the story, and a proper formulation of the conjecture (if any) is still a subject for further research.

\subsection{String Theory Evidence} \label{sec:SDC-ST-evidence}

The evidence for the SDC comprises a plethora of examples in string theory.
Those discussed in section \ref{sec:SDCdual} are so simple that the conjecture might look trivial from a string theoretical perspective. However, in general there are many different
infinite distance limits that one can engineer in more involved compactifications, and how string theory manages to
always have such light towers of states is non-trivial. In fact, testing the SDC
against different classes of models has triggered research in several corners of
string theory compactifications, uncovering very interesting connections to cutting-edge 
mathematics, such as algebraic geometry and modular forms.

In the following we will be discussing the setups in which most of the
progress testing the SDC has been made, providing an introduction to the underlying mathematical structures uncovered on the way. These are supersymmetric compactifications of the
following types:
\begin{itemize}
    \item 4d $\mathcal{N}=2$ from Type II.
    \item 5d $\mathcal{N}=1$ from M-theory and F-theory duals.
    \item 4d $\mathcal{N}=1$ from Type II.
\end{itemize}
In all these examples the towers have been shown to satisfy both the WGC and the
SDC.

\subsubsection{4d \texorpdfstring{$\mathcal{N}=2$}{N=2} %
Type II Calabi-Yau Compactifications}
\label{sec:typeII-CY}

The moduli space in this class of compactifications is known to be the product
of the K\"{a}hler and the complex structure moduli spaces. As a consequence we
can study them separately to leading order. Let us first focus on the vector multiplet moduli
space, which is the complex structure in Type IIB or K\"{a}hler in Type IIA.

Infinite (geodesic) distance can only occur when approaching some singularity. Hence, testing  the SDC reduces to studying the physics near singularities in the moduli space. In fact, it is known that there are always \emph{some} BPS
states becoming massless at the singularities of the
vector multiplet moduli space, so they are perfect candidates for the SDC tower. Note that one can have different kinds of singularities
in moduli space, and not all of them are necessarily at infinite distance. For
instance, the well-known conifold points are indeed singular but at finite
distance. Interestingly, in this case one does not find an infinite number of
BPS states becoming light, but a finite amount of them. Then, for the singularities at infinite distance the SDC
is predicting that the number of BPS states becoming massless has to actually be
infinite, and this is not trivial at all.

Following \cite{Grimm:2018ohb,Grimm:2018cpv}, let us work with the complex
structure moduli space of Type IIB CY$_3$ compactifications. These works constitute a non-trivial test of the SDC, as they show the existence of an infinite tower of BPS states becoming exponentially light with the distance at any infinite distance singularity of any Calabi-Yau manifold. Notice that, upon
mirror symmetry, the analysis translates to the K\"{a}hler moduli space in Type IIA
\cite{Corvilain:2018lgw}. First let us introduce some background about the
metric and the set of BPS states which will be relevant for realizing the SDC.

The vector multiplet moduli space of a Calabi-Yau three-fold is known to be a special K\"{a}hler manifold.
This means that the metric is determined by the K\"{a}hler potential, which can
be written in terms of a holomorphic $(D,0)$-form $\Omega$ as follows,
\begin{equation}
	g_{I\bar{J}} = \partial_{z^{I}} \partial_{\bar{z}^{J}} K \, , \quad K = - \log \left( i^{D} \, \int_{CY_D} \Omega\wedge \bar \Omega \right) ,
\end{equation}
with $D=3$ for a $CY_3$.
Here, we have chosen a set of complex coordinates $z^{I}$, known as the complex structure deformation moduli, that locally parametrise the moduli space. It is always possible to find an appropiate symplectic basis of three-cycles, such that the K\"ahler potential can be written as
\be
\label{KIIB}
    K = - \log \left(- i^{D} \, \Pi^{T} \eta \, \bar{\Pi} \right) \,,
\ee
where $\eta$ is the symplectic intersection product and $\Pi$ is a vector containing the periods of $\Omega$ over the basis of
three-cycles,
\begin{equation}
	\Pi^{\mathcal{I}} = \int_{\Gamma_{\mathcal{I}}} \Omega \, .
\end{equation}
BPS particles can be obtained by wrapping D3-branes on special Lagrangian
three-cycles $\Gamma_\mathcal{I}$.\footnote{In the IIA mirror, they correspond to bound states of D0
and D2-branes wrapping two-cycles.} Their mass is given by its central charge, which is
again determined by the periods as
\begin{equation}
\label{Z}
	M = |Z| = e^{K/2} \, |q^{T} \, \eta \, \Pi | .
\end{equation}
Here, $q$ is the vector of quantized charges of the BPS particle with respect to
the gauge fields coming from the dimensional reduction of the Type IIB RR 4-form
field,
\begin{equation}
\label{Afield}
	A_{1}^{\mathcal{I}} = \int_{\Gamma_{\mathcal{I}}} C_{4} \, .
\end{equation}

One could now start testing the SDC case by case, computing the periods $\Pi$ in different
Calabi-Yau manifolds, but given the incredible amount of them this does not seem
like the best way of gathering evidence for this conjecture. Fortunately, we can
do better and give a systematic approach which is valid for any Calabi-Yau, and
may be even more general than that. In order to do so, we will be using some
powerful tools of algebraic geometry and BPS counting. Namely, we will use theorems of limiting Hodge theory to describe
the behaviour of the periods near any infinite distance loci, and the study of walls of
marginal stability to count the number of BPS states becoming light.

The key ingredient of this analysis are
the monodromy transformations. As mentioned before, we are interested in approaching
singularities in moduli space that are at infinite distance. It turns out that
the periods undergo a monodromy transformation when circling around the
singularity. That is, choosing some special complex coordinates, $z^{I}$, for
which the singularity is located at $z^{i}\to i\infty$, the periods transform as
\begin{equation}
	\Pi(...,z^{i}+1,...) = T_{i}\, \Pi(...,z^{i}+1,...)\, ,
\end{equation}
while the K\"ahler potential remains invariant.
From the point of view of the effective field theory, $z^{i}\to z^{i}+1$ is a discrete shift of an axion, so the monodromies correspond to  axionic shift symmetries in the EFT.

The monodromy around a singularity can be of finite or infinite order. That is,
after going around the singularity several times we either get back to the same point, so $T^n=\mathbb{I}$ for some $n>0$;
or this is not possible, so $T^n\neq \mathbb{I}$ for any $n>0$. We will see that infinite distance singularities
are necessarily related to infinite order monodromies.

If the monodromy is of infinite order, one can define a non-trivial nilpotent operator by taking $N_{i} = \log
T_{i}$. In this situation, the Nilpotent Orbit Theorem of Schmid \cite{schmid} gives an asymptotic expansion of the periods near the
singular locus such that they are well approximated by the following nilpotent orbit
\begin{equation}
\label{nil}
	\Pi (z) = e^{z^{i}N_{i}} \Pi_{0} + \mathcal{O} (e^{2\pi i z^{i}}) \, ,
\end{equation}
up to exponentially suppressed corrections,
where $\Pi_{0}$ only depends on the complex coordinates which are not sent
to infinity.

For simplicity, let us now stick to the case of a single modulus $z\to\infty$. By plugging \eqref{nil} into \eqref{KIIB} and using that $N^T\eta =-\eta N$, we can see that $K=-\log(p_d(\text{Im}z)+\mathcal{O}(e^{2\pi iz}))$ near the infinite distance locus, where $p_d(\text{Im}z)$ is a polynomial of degree $d$. This further implies the following leading term for the
metric,
\begin{equation}
\label{metric}
	g_{z\bar{z}} = \frac{d}{4(\text{Im} z)^2}\, .
\end{equation}
Here, $d$ is the effective nilpotency order defined as
\begin{equation}
	N^{d}\Pi_{0}\neq0 \, , \quad N^{d+1}\Pi_{0}=0 \, ,
\end{equation}
which is bounded by the complex dimension of the Calabi-Yau, so in the case at hand it can take values $d=0,1,2,3$. Clearly, only for $d\neq 0$  the singularity will be at infinite distance, so an infinite order monodromy is a necessary condition for infinite distance.  
This nilpotency order can be used to
classify different asymptotic limits and, interestingly, it also fixes
the exponential rate of the tower for the SDC. Indeed, one can show that there
is always an infinite tower of BPS states behaving as required by the SDC and
whose exponential rate is bounded by
\begin{equation} \label{eq:bound-exponential-rate}
	\frac{1}{\sqrt{2d}} < \lambda < \sqrt{2d} \, .
\end{equation}
Since $d\leq 3$ for a Calabi-Yau three-fold, this yields the following lower bound for the exponential rate:
\begin{equation}
    \lambda\geq \frac{1}{\sqrt{6}}\,.
\end{equation}
The exponential behaviour of the mass is a consequence of the asymptotic behaviour of the metric. By plugging \eqref{nil} into \eqref{Z} the BPS mass behaves as a polynomial in the coordinates $z^i$ which, by using \eqref{metric} to canonically normalise the scalars, depends exponentially on the proper field distance
\begin{equation}
    \Delta\phi= \sqrt{\frac{d}{2}}\log \text{Im}z \,.
\end{equation}

The role of the monodromy transformations does not end here. The existence of an
infinite tower of states (and not only a finite number of them) can be intuitively explained by it. The monodromy
transformation acts on the charge of the BPS states, generating an orbit of
charges $q_{n}=T^nq_0$. Monodromies are redundancies of the theory, which implies that even if the charge of a single state varies, the full tower must reorder itself and remain invariant under the monodromy so that the physics does not change. In fact, these monodromies become actual global symmetries at infinite distance. Hence, the existence of a single state implies that the entire orbit must be populated by physical BPS states, which yields an infinite tower if the monodromy is of infinite order. In this way, the fact that
there is an infinite tower of BPS states tracks back to the monodromy
transformations around infinite distance singularities being of infinite order. This intuitive reasoning must be accompanied by the analysis of walls of marginal stability and a careful check of the existence of the seed charges $q_0$, but we refer the student interested in the details to the original paper \cite{Grimm:2018ohb}. The generalization to multiple moduli limits can be found in \cite{Grimm:2018cpv}.

In this scenario we can relate the SDC with the WGC as already discussed in
section \ref{sec:relation-WGC}. We find that in these limits there is always a
gauge coupling going to zero for some gauge field $A^\mathcal{I}$ of \eqref{Afield} and that the bound in
\eqref{eq:bound-exponential-rate} can be related to the black hole extremality bound. Hence, the SDC tower also saturates the WGC. These conjectures can then be understood as quantum gravity obstructions to restore some global symmetries at infinite distance, coming both from the vanishing gauge couplings and from the monodromies, which can be promoted to continuous global
symmetries of the theory at infinite distance. 

After having studied the vector multiplet moduli space, we are left with the
hypermultiplet moduli space. In what follows, let us give a very brief review of the results
that have been obtained, referring to
\cite{Marchesano:2019ifh,Baume:2019sry} for more details. In this case, one does not find a tower of BPS states but KK towers, charged strings from wrapping branes and instantons becoming light. Importantly, the fact that
some instantons become light (in the sense that their action goes to zero) means
that there could be relevant quantum corrections to the field metric, which could even obstruct certain infinite distance limits from
being taken. What typically happens then is that the trajectories approaching such
limits get deviated towards another one in which there is a weakly-coupled string becoming
light at the same rate as a KK tower,
\begin{equation}
	\frac{T_{\text{string}}}{M_{p}^{2}} \sim \frac{M_{KK}^{2}}{M_{p}^{2}} \to 0 \, .
\end{equation}
Since the KK modes have a less dense spectrum than the string excitation modes, i.e. $m^2_{KK}\sim k^2 M^2_{KK}$ vs $m^2_{\rm str}\sim k T_{\rm str}$ for $k\in \mathbb{N}$, the leading tower fulfilling the SDC is given by string excitation modes in these limits.

\subsubsection{5d \texorpdfstring{$\mathcal{N}=1$}{N=1} %
 M-theory and 6d F-theory Duals} \label{sec:SDC-5d-N1}

We consider the K\"{a}hler moduli space of M-theory compactified on a Calabi-Yau three-fold, and their 6d F-theory duals.
In this setup, the SDC is realized by a tower of particles coming from wrapping M2-branes on two-cycles, which corresponds to strings becoming tensionless from
the F-theory perspective. 

There have been two approaches to analyse and classify the possible infinite distance limits in this setup. One approach consists in borrowing the classification of singular limits on Calabi-Yau three-folds
used to study 4d $\mathcal{N}=2$ theories in section \ref{sec:typeII-CY} and
applying it to compactifications of M-theory to 5d on these spaces. Since the geometry is the same, the classification based on the properties of the monodromies still applies,\footnote{In the K\"ahler moduli space, different properties of the monodromies, like the effective nilpotency order, translates into different patterns of vanishing and non-vanihising intersection numbers \cite{Corvilain:2018lgw}.} and only the microscopic interpretation changes. Now the
monodromies will not be related to shift symmetries of the axions but to large gauge transformations of the one-form gauge
fields obtained from reducing $C_{3}$ on the two-cycles. They will act on the
M2-brane charges, populating an infinite tower of particles. 

The other approach involves an equivalent classification of these singularities in
terms of the fibration structure that the Calabi-Yau three-fold develops in the limit.
This was worked out in
\cite{Lee:2018urn,Lee:2019tst,Lee:2019xtm,Lee:2019wij}, where it was
shown that the possible fibrations are $T^{2}$, $K3$ and $T^{4}$. The limits studied in this series of works involve shrinking some of these fibers. Interestingly, this classification translates into certain properties of the intersection numbers which can be mapped one-to-one to the properties of the monodromy transformations and the previous classification of singularities. In the following, we will focus on this second approach and explain the appearance of the infinite tower from the F-theory perspective. To illustrate the main features we will restrict to the case of a K3 fibration.

Consider F-theory compactified on an elliptically fibered $CY_3$. The 6d Planck mass is given by
\begin{equation}
	M_{p}^{4} = 4 \pi \text{Vol}(B_{2}) \, ,
\end{equation} 
where $\text{Vol}(B_{2})$ is the volume of the base of the fibration. The theory
contains a gauge theory sector coming from 7-branes wrapping some divisor of the
base, $C$. The associated gauge coupling is given by
\begin{equation}
	\frac{1}{g_{YM}^{2}} = \frac{1}{2\pi} \text{Vol}(C).
\end{equation}
The limit $\text{Vol}(C)\to\infty$ is a weak coupling limit at
infinite distance in moduli space. There are now two possibilities. It could be that, when taking the volume of the divisor to be large, the overall volume of the base also goes to infinity, leading to  a decompactification limit
and the corresponding tower of KK modes. The second possibility is to engineer an equi-dimensional limit, in which $\text{Vol}(B_{2})$ remains
finite even if $\text{Vol}(C)\to\infty$. To make this geometrically feasible, it can be proven that there
is always another curve $C_{0}$ that intersects $C$ and whose volume is going to
zero. D3-branes wrapping this shrinking curve will therefore give rise to strings that become tensionless in the weak coupling limit,
\begin{equation}
	T \sim \text{Vol}(C_{0}) \to 0 \, .
\end{equation}

For the SDC to hold it is still necessary to show that the string spectrum contains
infinitely many excitation modes in this limit. In this present case it is easy to show this
by using dualities and thus identifying this string as the critical heterotic
string on a K3 surface. More generally, it has been shown in \cite{Lee:2020gvu} that the modular
properties of quasi-Jacobi forms of the elliptic genus of the string can be used to
determine the spectrum and yield infinitely many states. 
For the present case,
this results in an infinite tower of excitations whose
mass is given by \cite{Lee:2018urn}
\begin{equation} \label{eq:string-excitations}
	m_{n}^{2} = 8\pi T (n-1) 
\end{equation}
with $j=\frac{1}{2}C\cdotp C_{0}$ and $n$ the excitation level of the string. 
In fact, one obtains a sublattice of charges with index $2j$ such that for each charge
\be
q_k=2jk\ ,\quad k\in \mathbb{Z} \, ,
\ee
there exists a state at excitation level $n(k)=jk^2$.
These states satisfy
\begin{equation}
	g_{YM}^{2} q_{k}^{2} = m_{k}^{2} + 4 j g_{YM}^{2} > m_{k}^{2} \, ,
\end{equation}
in agreement with the WGC. 
Therefore, one gets a sublattice of non-BPS states
satisfying the SDC and the strict inequality of the WGC at the same time. This
provides evidence for both conjectures in the case in which the states are not
necessarily supersymmetric, and it is also consistent with the sharpened WGC in section \ref{sec:Sharpening the WGC}. 
The analysis can also be  extended to 4d $\mathcal{N}=1$ theories obtained from F-theory on Calabi-Yau
four-folds \cite{Lee:2019tst}.

These works are again an example in which the swampland conjectures have served to uncover the underlying geometric structures of string compactifications and new relations to mathematics.
In particular, this has triggered research on modular forms, yielding interesting non-trivial connections
between algebraic geometry of Calabi Yau manifolds and modular properties of
quasi-Jacobi forms that allow the WGC and the SDC to hold.

\subsubsection{4d \texorpdfstring{$\mathcal{N}=1$}{N=1} %
 Effective Field Theories}

In this case there are no BPS particles in the spectrum, and that makes it more
difficult to keep track of the states that could satisfy the SDC at the
boundaries of moduli space. However, there are BPS charged strings and membranes
that can be shown to become tensionless in these limits \cite{Font:2019cxq}. Even though they are not
necessarily the leading towers,\footnote{In the string theory setups analysed, the leading tower  either correspond to KK modes (in some appropiate dual frame) or excitation modes of these BPS strings, in agreement with the Emergent String Conjecture.} they can still give us valuable information about
the physics at these boundaries \cite{Herraez:2020tih,Lanza:2020qmt}.

The strings are charged under two-form gauge fields that are dual to
axions in 4d. The gauge kinetic function for the two-forms is equal to the inverse of the axionic field metric, which in $\mathcal{N}=1$ can be given in terms of the Kähler potential,
\begin{equation}
	g_{ij} = \frac{\partial^{2}K}{\partial s^{i} \partial s^{j}} \, .
\end{equation}
Here, $s^i$ are the saxions which, together with the axions $\phi^i$, make up the complex scalars of the chiral multiplets. 
Thus, information about the gauge kinetic terms
of the two-forms under which the strings are charged gets translated into information
about the metric in moduli space.

In the same way, membranes are charged under three-forms and their gauge kinetic
function $T_{ab}$ is related to the scalar potential as follows,
\begin{equation} \label{eq:CY-flux-potential}
	V =\frac12 T^{ab}(s,\phi) f_af_b= \frac{1}{2} Z^{ab}(s) \rho_{a}(f,\phi) \rho_{b}(f,\phi) \, ,
\end{equation}
where $f_a$ are the discrete internal fluxes dual to the three-form gauge fields. 
Interestingly, the flux potential of string compactifications can always be factorized between a saxionic an axionic part, where the $\rho_{a}$ functions are shift invariant axion polynomials including the fluxes, see e.g.\cite{Bielleman:2015ina,Herraez:2018vae}. Therefore, studying the behaviour of
the three-form gauge couplings and charges of BPS membranes in the asymptotic limits provides information about the
scalar potential.

In conclusion, the asymptotic behaviour of the field metric and the scalar potential can be translated into properties of BPS strings and membranes charged under two-form and three-form gauge fields, respectively, in 4d $\mathcal{N}=1$ EFTs. This yields interesting relations between different Swampland conjectures  \cite{Lanza:2020qmt}. For instance, the exponential behaviour of the SDC tower becomes a consequence of having a BPS string satisfying the WGC. This allows us to provide a lower bound for the SDC exponential rate in 4d $\mathcal{N}=1$ EFTs in terms of the extremality factor for the strings. Similarly, a potential described by a WGC-saturating membrane satisfies the de Sitter conjecture that will be explained in section \ref{sec:dS}.

\subsection{SDC with Potential and Implications for Inflation}

Up to now, the SDC has been a statement about the moduli space
of the theory. This means that it regards a set of scalars with an exactly flat
potential (typically protected by extended supersymmetry) that parametrizes
different vacua of the theory. However, it is phenomenologically relevant, e.g. for inflation, to understand what happens
when a potential is added so that this moduli space is lifted. Based on physical grounds, one would expect that the SDC should also apply to the valleys of the potential, i.e. to directions along which the potential may not be exactly flat but the
relevant energies are smaller than a given cut-off.

The refined SDC \cite{Klaewer:2016kiy} proposes that:
\begin{enumerate}

	\item The exponential behaviour with $\lambda \sim \mathcal{O}(1)$ should be
	manifest when $\Delta\phi \gtrsim M_{p}$.

    \item The conjecture should also hold for scalars with nearly flat
    potential.
	
\end{enumerate}
This can have important implications for inflationary models.\footnote{The SDC also places important constraints on relaxation models of the EW scale, since they typically require transplanckian field ranges.} Consider the bound 
for the field distance in terms of the cut-off in \eqref{eq:SDC-bound}. In order 
to accommodate inflation in the effective field theory we need the cut-off to be
above the Hubble scale, $\Lambda > H$. These two conditions together yield
\begin{equation}\label{eq:SDC-inflation}
	\Delta \phi \leq \frac{1}{\lambda} \log \left(\frac{M_{p}}{H}\right) \, .
\end{equation}
Thus, the SDC gives an upper bound on the field range of inflation in terms of 
the Hubble scale. In particular, when $H$ is close to the Planck scale one finds 
that the field range is bounded by an order one number in Planck units.

\begin{figure}[ht]
\begin{center}
\includegraphics[width=\textwidth]{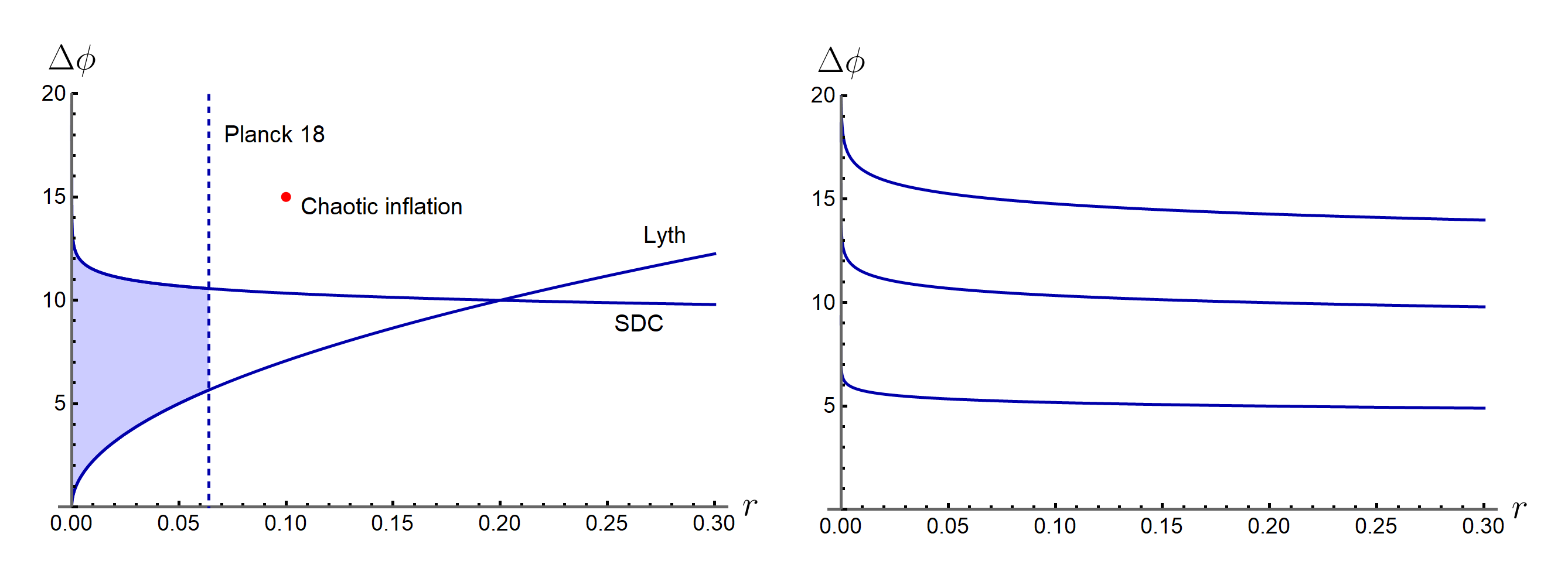}
\caption{Left: Allowed (blue) region of scalar field excursion during inflation
against tensor-to-scalar ratio obtained by combining Lyth bound and SDC
constraint for $\lambda=1$, together with the experimental bound put by Planck
18 \cite{Akrami:2018odb}. The red point corresponds to chaotic inflation \cite{linde_chaotic_1983}. Right: SDC bound for $\lambda=0.7,1,2$ starting from
the upper one. We note that the constraint is very sensitive to the precise
value of $\lambda$.}
\label{fig:SDC-Lyth-bounds}
\end{center}
\end{figure}

In slow roll inflation, the field range  provides an upper bound on the tensor-to-scalar ratio $r$ via the Lyth bound,
\begin{equation}
	\Delta \phi \geq \left( \frac{r}{0.002}\right)^{1/2} \, .
\end{equation}
The tensor-to-scalar ratio can be related
to the Hubble scale as 
\begin{equation}
\label{Hr}
	\frac{M_p}{H} = \sqrt{\frac{2}{\pi^2 A_s r}} \, ,
\end{equation}
where $A_s$ is the amplitude of scalar perturbations whose value has been 
measured experimentally to $\log(10^{10} A_s)=3.047\pm 0.015$ \cite{Aghanim:2018eyx}.  Using \eqref{Hr} we can write the SDC bound  in \eqref{eq:SDC-inflation} in terms of $r$ as follows,
\begin{equation}
	\Delta \phi \leq - \frac{1}{2\lambda} \left( \log \left(\frac{\pi^2 A_s}{2} \right) + \log r \right) \, ,
\end{equation}
so it provides a lower bound on the tensor-to-scalar ratio.
When written in this way, we note that the Lyth and the SDC bounds are
complementary and can constrain different models of inflation
\cite{Scalisi:2018eaz}. To give an example, chaotic inflation \cite{linde_chaotic_1983} would be ruled out. This is shown in figure \ref{fig:SDC-Lyth-bounds}.

In conclusion, we see that the SDC can constrain some models of large field
inflation (including axion monodromy) but it does not rule it out, as a moderate transplanckian field range might suffice for some models. Notice that
the SDC constraint on large field inflation highly depends on the exact value of
$\lambda$. For this reason, it is really important to learn about this
exponential decay rate and how it can be computed from first principles or
effective field theory data. It should also be noted that replacing $H$ as the cut-off in \eqref{eq:SDC-inflation} gives a very conservative bound, in the sense that the EFT will likely break down (or at least get sensitive to the infinite tower) before the mass of the first state becomes of order Hubble, so the constraints might be stronger than represented here.
\medskip

The implications of the SDC do not end here.
Remember that the SDC applies
to geodesics in moduli space but, with the inclusion of a potential, the
trajectory followed during inflation is not necessarily a
geodesic from the perspective of the UV moduli space. However, it still makes sense to apply the SDC if it corresponds to a geodesic from the perspective of the IR pseudo-moduli space generated by the valleys of the potential. 
In such a case, for the SDC to hold, it should be impossible to
generate a potential such that the trajectory is sufficiently non-geodesic so
that the exponential behaviour of the tower is violated \cite{Calderon-Infante:2020dhm}. This is to say, by requiring consistency of the SDC at any energy scale, we can put
constraints on the type of potentials that one can engineer in quantum gravity!

\begin{figure}[ht]
	\centering
	\includegraphics[width=.35\textwidth]{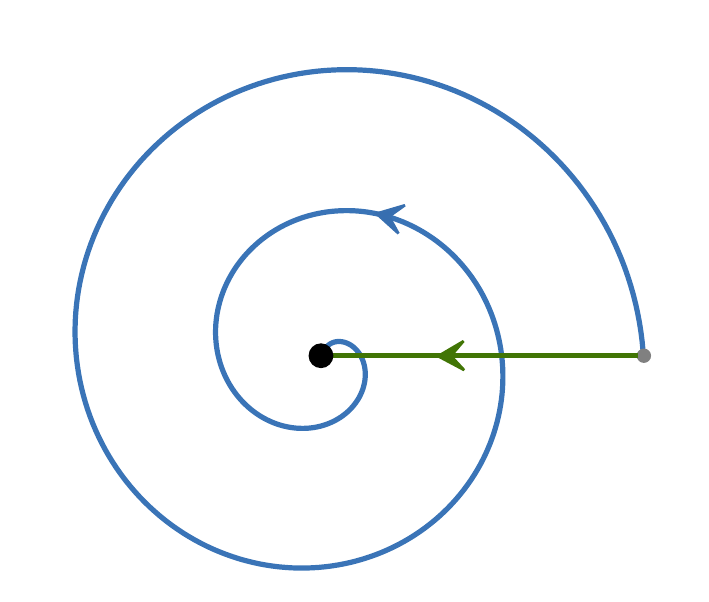}
	\caption{Green: purely saxionic (geodesic) trajectory. Blue: a highly turning trajectory, like in axion monodromy. The black dot represents a locus at infinite distance, while the radial and angular coordinates correspond to the saxion and axion scalars respectively.}
	\label{fig:non-geodesic}
\end{figure}

The main evidence that we have for this statement, and that motivated the Refined
SDC, comes from Calabi-Yau flux compactifications in string theory. Since
the theory is 4d $\mathcal{N}=1$, the scalars come in chiral multiplets including
an axion and its partner, the saxion. 
Before adding the scalar potential, purely saxionic trajectories are geodesics of the moduli space. When adding the flux potential, both saxions and axions get stabilized.
One could think that a way of avoiding any tower of states in inflation could be by moving only along an axionic trajectory, while
keeping the saxion fixed. This way, one would engineer an axion monodromy model with a highly turning trajectory and very large field range, see figure \ref{fig:non-geodesic}. However, the structure of the flux potential coming from string theory is such that the minimization of the potential in the saxionic
directions implies
\begin{equation}
	\partial_{s^{i}}V = 0 \, \Longrightarrow \, s^{i} = \tilde{\lambda} \phi^{i} + \cdots \,
\end{equation}
for large $\phi^{i}$. That is, when trying to move in an axionic direction, the
potential induces a linear backreaction on the saxion. So one cannot move along large values of the axion without displacing the saxion in the same way. This implies that if there was a tower decaying exponentially with the saxionic distance, it will also decay exponentially upon taking into account the backreaction in the axion monodromy model.

This backreaction issue was first introduced in
\cite{Baume:2016psm} and further checked in
\cite{Valenzuela:2016yny,Blumenhagen:2017cxt,Grimm:2019ixq}.
In the
context of F-theory compactifications on Calabi-Yau four-folds with fluxes, this
backreaction was proven to hold more generally \cite{Grimm:2019ixq} due to an asymptotic homogeneity
of the potential
\begin{equation}
	V(\alpha s^{i},\alpha \phi^{i}) \simeq  \alpha^{d_{i}} V(s^{i},\phi^{i}) \, 
\end{equation}
realised near the infinite distance loci.

\subsection{AdS Distance Conjecture}

Finally, we are going to very briefly mention a generalization of the Distance conjecture to other field space configurations beyond the moduli space. In particular, one can define a notion of distance between different metric configurations\footnote{Taking a background metric and considering perturbations around it, we will have a metric configuration space on which we can define distances.} of AdS spacetime. As discussed in \cite{Lust:2019zwm}, the flat spacetime limit $ \L\rightarrow 0 $ is then at infinite distance in this metric configuration space.  Assuming that the Distance conjecture also applies in this space, and taking into account that the distance grows logarithmically as $\L\rightarrow 0$, one gets the following conjecture:\\

\begin{conj}[AdS Distance Conjecture]{conj:adsDC}
Any AdS vacuum has an infinite tower of states that becomes light in the flat space limit $ \L\rightarrow 0 $, satisfying
\begin{equation}
\label{mL}
	m\sim \lvert \L \rvert^{\a}~.
\end{equation}
\end{conj}
A strong version of the conjecture implies $\alpha=\frac12$ if the vacuum is supersymmetric and $\a\geq\frac12$ for non-susy AdS and $\a\leq\frac12$ for dS space. A consequence of the strong version of the conjecture is that a supersymmetric $d$-dimensional AdS vacuum cannot exhibit scale separation between the AdS scale and the cut-off of the $d$-dimensional theory (typically given by the KK scale), as the associated scale of the tower is of order the AdS scale $\L^{1/2}$. Most of the AdS vacua constructed in string theory satisfy this conjecture, as they do not have scale separation between the internal dimensions and the AdS length. However, there is one exception associated to Type IIA compactified on a particular limit of a flux orientifold  CY$_3$ \cite{DeWolfe:2005uu}, which has captured a lot of recent attention since the vacuum is in principle at parametric control, although its validity is still under debate (see e.g. \cite{McOrist:2012yc,Junghans:2020acz,Marchesano:2020qvg}). In any case, this latter example would only be a counterexample to the strong version of the conjecture but not to \eqref{mL}.

\subsection{Open Questions}

There are a number of interesting open questions about the SDC:

\begin{itemize}

	\item All the evidence for the conjecture comes from string theory compactifications. Hence, a botom-up explanation for the conjecture is missing.

    \item The existence of the towers is linked to the manifestation of dualities at the asymptotic limits. Why are dualities ubiquitous in string theory? Can we map the geometric classification of infinite distance limits to different types of dualities?

	\item SDC in AdS/CFT, see e.g. \cite{Baume:2020dqd} and \cite{Perlmutter:2020buo}.

 \item Can the SDC always be understood as a quatum gravity obstruction to restore a global symmetry at infinite distance?

    \item More concretely, is there always some gauge coupling going to zero at infinite distance
    in moduli space? If so, there is a WGC tower satisfying the SDC, which would bound the exponential rate, see \cite{Gendler:2020dfp}, and would serve as a quantum gravity obstruction to restore a global symmetry.

    \item What is the specific value of the exponential rate? Can we compute it or bound it from first
    principles or effective field theory data?

    \item Better understand the constraints on scalar potentials coming from the Refined SDC.

    \item What is the nature of the leading tower? Could it ever originate from $p$-branes with
    $p\geq 2$? This would disprove the Emergent String Conjecture.

    \item If the scalars have a non-trivial profile in the non-compact space-time dimensions, does the EFT also break down at large distances and how? This is known as the Local SDC, see e.g. \cite{Klaewer:2016kiy,Draper:2019utz}.
    
    \item What are the cosmological signatures of the tower during inflation or late-time cosmology?
	
\end{itemize}

\section{Emergence Proposal}

The so-called Emergence proposal \cite{Harlow:2015lma,Grimm:2018cpv,Corvilain:2018lgw,Heidenreich:2017sim,Heidenreich:2018kpg,Palti:2019pca} provides a general rationale for why the Swampland conjectures, especially the WGC and the SDC, should hold true in any theory of quantum gravity.  Recall that these two conjectures relate the kinetic terms (i.e., gauge couplings, field metrics, etc.) to the mass of some new states. But it is well known that relations between the mass of heavy states that are integrated out and couplings of the low energy effective field theory Lagrangian are already guaranteed by the renormalization group equations. So could it be that the relations provided by the Weak Gravity and Distance conjectures are simply due to the renormalization group equations upon integrating out the WGC- and SDC-satifying states? The answer would be affirmative according to the Emergence proposal.
\nl


\begin{otherbox}[Emergence Proposal]{box:emergenceproposal}
All the kinetic terms in an EFT \textit{emerge} from integrating out the massive states up to some quantum gravity cut-off\cite{Grimm:2018ohb,Palti:2019pca}.
\end{otherbox}

This implies that all fields are non-dynamical at the quantum gravity scale where gravity becomes strongly coupled, hinting at some sort of UV topological description. That is, there is no kinetic term to start with. It is only upon going to the IR and
 integrating out the tower of states predicted by the Weak Gravity and/or Distance conjecture, that we get some finite kinetic terms. We will see that when the tower becomes light, it generates small gauge couplings and parametrically large field distances. In other words, in the UV there are no tiny gauge couplings or infinite distances; these, as well as the approximate global symmetries that come with them, are just artifacts of the IR description.

A weaker version of the proposal \cite{Heidenreich:2017sim,Heidenreich:2018kpg} allows for a classical piece of the kinetic terms of the same form as the quantum corrected piece, so the IR value is not purely coming from quantum corrections. In that case, at the quantum gravity scale the fields are still dynamical but a sort of unification occurs, in the sense that gravity and the other gauge and scalar forces become strongly coupled at the same quantum gravity cut-off scale.

\begin{figure}[ht]
	\centering
	\includegraphics[width=.8\textwidth]{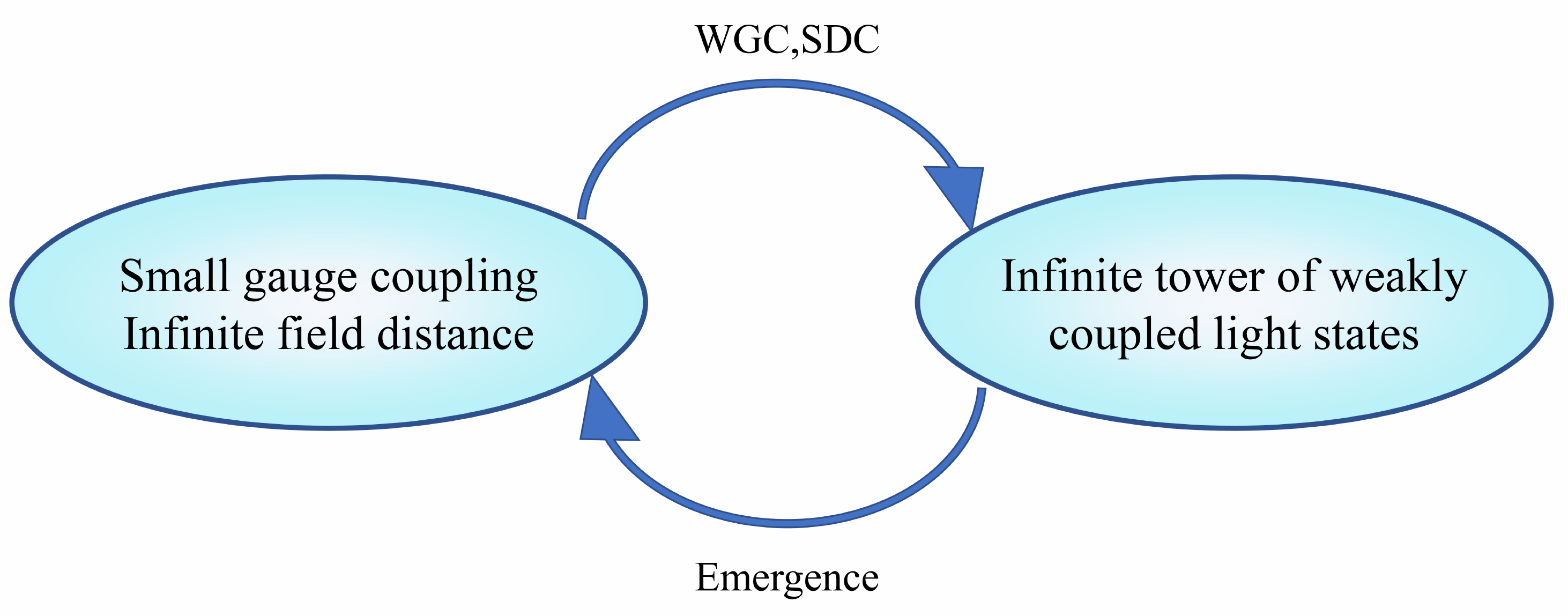}
	\caption{Comparison of the emergence proposal and the WGC/SDC.}
	\label{fig:emergence}
\end{figure}

We started the section by promising that the emergence proposal provides a bottom-up explanation for the Swampland conjectures. The WGC and the SDC imply the presence of an infinite tower of  states getting light, whenever the EFT allows for very small gauge couplings and/or very large field distance (how light depends on how small the gauge couplings or how large is the field distance). The emergence proposal goes the other way around by saying that having an infinite tower of states becoming light at some point in the field space generates the infinite field distance and the small gauge coupling there, see figure \ref{fig:emergence}.

We can perform computations in a field theory toy model to exemplify how the infinite tower can indeed generate the small gauge coupling and the infinite field distance.
Consider the IR EFT resulting from integrating out a tower of states charged under some gauge field, for simplicity we stick to the 4d case. We should integrate out the tower up to a species bound $\Lambda_{\rm QG}$, as in \eqref{species}, which acts as a QG cut-off above which quantum gravitational effects becomes important and QFT computations are not reliable any more. Using \eqref{species} and \eqref{Nsepcies}, the number of states $N$ of the tower running in the loops can be written purely in terms of the mass gap $\Delta m$ as follows,
\be \label{eq:N-species-bound}
    N = \left( \frac{\Delta m}{M_p} \right)^{- \frac{2}{3}} \,.
\ee
Computing the one-loop correction from integrating out the tower of states, the IR quantum-corrected gauge coupling is given by
\begin{eqnarray}\label{IR gauge coupling in terms of mass gap}
\dfrac{1}{g^{2}_{\text{IR}}}=\sum^{N}_{k=1} q^{2}_{k}\log \Big( \dfrac{\L^{2}_{\rm QG}}{m^{2}_{k}} \Big) \simeq \sum^{N}_{k=1} k^{2}\log\frac{N^2}{k^2} \simeq N^3 \simeq \dfrac{1}{(\D m)^{2}}~,
\end{eqnarray}
where $ m_{k}=k\D m $ and $ q_{k}=k $ are the mass and charge of the $k$-th state of the tower, and $ \L_{\rm QG} $ is the species scale in \eqref{species}.
This result implies that the mass of the states that we integrated out is proportional to the renormalized gauge coupling in the IR, i.e. 
\begin{equation}
    m_k^2\simeq k^2 g^2_{IR}\,,
\end{equation}
and this is precisely the WGC! 

It should be emphasized that, in order to obtain this result, it is essential that we are summing an infinite tower up to its species bound. 
The dependence of the cut-off and $N$ on $\Delta m$ qualitatively changes the result, transforming the logarithmic behaviour on the mass into polynomial in \eqref{IR gauge coupling in terms of mass gap}. 
In contrast, we could not reach this conclusion by integrating out only a finite number of states and having a constant cut-off. This motivates the stronger versions of the WGC in which there is not only one but an infinite number of states. 

A similar calculation can be done for the Distance conjecture. If the mass of the tower of states is parametrised by some scalar field, quantum corrections from integrating out the tower will modify the field metric of the scalar. The one-loop contribution to the scalar propagator is given by
\begin{equation} \label{eq:emergence-metric}
g_{\f\f}=\sum_{k=1}^N \big( \partial_{\f}m_{k} \big)^{2}=\big( \partial_{\f}\Delta m \big)^{2}\sum_{k=1}^N k^2\simeq \big( \partial_{\f}\Delta m \big)^{2}N^3\simeq \Big( \dfrac{\partial_{\f}\D m_{k}}{\D m_{k}} \Big)^{2}~,
\end{equation}
where $ m_{k} $ is the mass of the particle of the tower running in the loop. As before, it is important to sum up to the species bound \eqref{eq:N-species-bound} where quantum field theory computations are not reliable any more. The proper field distance measured by using this renormalized field metric reads
\begin{equation}
\D\f\simeq\int_{\phi_1}^{\phi_2}\sqrt{g_{\f\f}}\simeq\log\left( \frac{\Delta m(\phi_2)}{\Delta m(\phi_1)}\right)~.
\end{equation}
 This gives rise to the exponential behaviour of the tower predicted by the SDC in terms of the proper field distance. If the full tower becomes massless at some point $\phi_2$ of the moduli space, so $\Delta m(\phi_2)=0$, then the field distance becomes infinite. Again, this only occurs upon integrating out an infinite tower up to its species scale, since a finite number of fields will always generate a finite distance.\footnote{Quantum corrections from integrating out only a \emph{finite} number $n$ of fields would generate a finite field distance given by $\Delta\phi\simeq n\int  \partial_\phi m=n(m(\phi_2)-m(\phi_1))<\infty$.} So from this point of view it is the tower itself that is responsible for the infinite distance.

 Although these are just toy model computations, the emergence of the WGC and SDC behaviours arising from integrating out the towers of states serves as motivation for  the Emergence proposal. Unfortunately, the  estimations of the species bound are not precise enough to trust the resulting numerical factors.  
There are some string theory setups, though, in which these toy model computations actually provide a good approximation of the physics and the Emergence proposal can be checked. Consider 
the type IIB Calabi-Yau compactifications of section \ref{sec:typeII-CY}. It is known that the geometry near the conifold singularity in the complex structure moduli space can be reproduced by computing the quantum corrections of the BPS state becoming massless at the singular point \cite{Strominger:1995cz}. Thanks to supersymmetry, it is enough to compute the one-loop correction to the gauge coupling and the field metric to reproduce the geometric result. Hence, the complex structure moduli space is quantum in nature, in the sense that it incorporates the effects from integrating out the BPS states. The singularities are then generated because one is `incorrectly' integrating out some BPS states that are actually massless. Analogously, one can compute the quantum corrections from integrating out the tower of BPS states near the infinite distance singularities and check that the resulting gauge coupling and field metric in \eqref{IR gauge coupling in terms of mass gap} and \eqref{eq:emergence-metric} matches with the results obtained from the geometry. This was done in \cite{Grimm:2018cpv}, providing evidence for the Emergence proposal. The difficulty behind extending these computations beyond the complex structure moduli space is that the species bound of the tower in other scenarios is typically above the KK scale or around the string scale, so that a low energy field theory computation is not possible.

\section{Non-SUSY AdS Instability Conjecture}\label{subsec. AdS instability conjecture}

The conjecture first appeared a few years ago in \cite{Ooguri:2016pdq,Freivogel:2016qwc}. While \cite{Ooguri:2016pdq} focused on the non-trivial case of AdS spacetime, it was emphasized in \cite{Freivogel:2016qwc} that it should apply to any non-susy vacuum. 

\begin{conj}[No Non-susy Stable Vacuum]{conj:nonsusy}
Any non-supersymmetric vacuum is at best metastable and has to decay eventually.
\end{conj}
In other words, supersymmetry is the only mechanism to protect a vacuum from decaying in quantum gravity. Typically, it is quite common to generate some instabilities when breaking supersymmetry. But it is highly non-trivial to propose that this should always be the case; in particular, it is false for a QFT in the absence of gravity.

\begin{figure}[H]
	\centering
	\includegraphics[width=.4\textwidth]{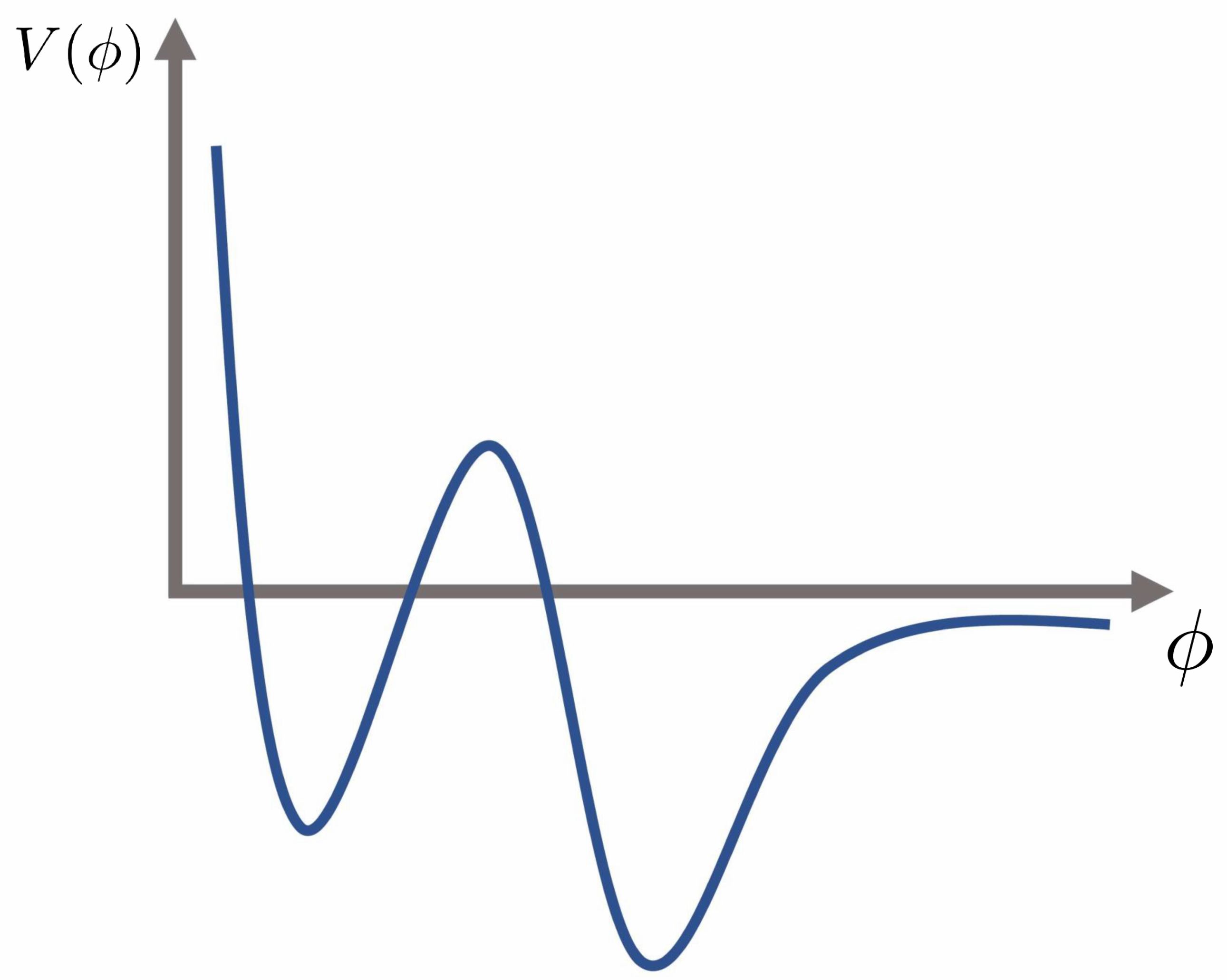}
	\caption{Potential with a metastable vacuum (the left one). It can decay non-perturbatively to the lower energy minimum on the right, which presumably should be supersymmetric if stable.}
	\label{fig:metastable}
\end{figure}

The instability can be perturbative or non-perturbative. In the latter case, it is called a metastable vacuum, as the decay rate is exponentially suppressed, see figure \ref{fig:metastable}.
The conjecture does not give any bound on the decay rate, so the vacuum can be very long-lived. Hence, the implications of this conjecture for dS could be negligible (however see the next section for stronger conjectures about dS vacua). On the other hand, the implications for AdS are huge, as the existence of the boundary implies that a mestastable AdS vacuum cannot live longer than a Hubble time. We will discuss the implications for AdS/CFT in section \ref{sec:AdS-inst-implications}. Regarding Minkowski vacua, they could in principle also be very long-lived, although it is actually not clear whether non-susy Minkowski vacua even exist as they require an extreme fine-tunning to keep a vanishing cosmological constant.


\subsection{Motivation from the WGC} \label{sec:motivation-from-WGC}

\begin{figure}[H]
	\centering
	\vspace{-10pt}
	\includegraphics[width=.55\textwidth]{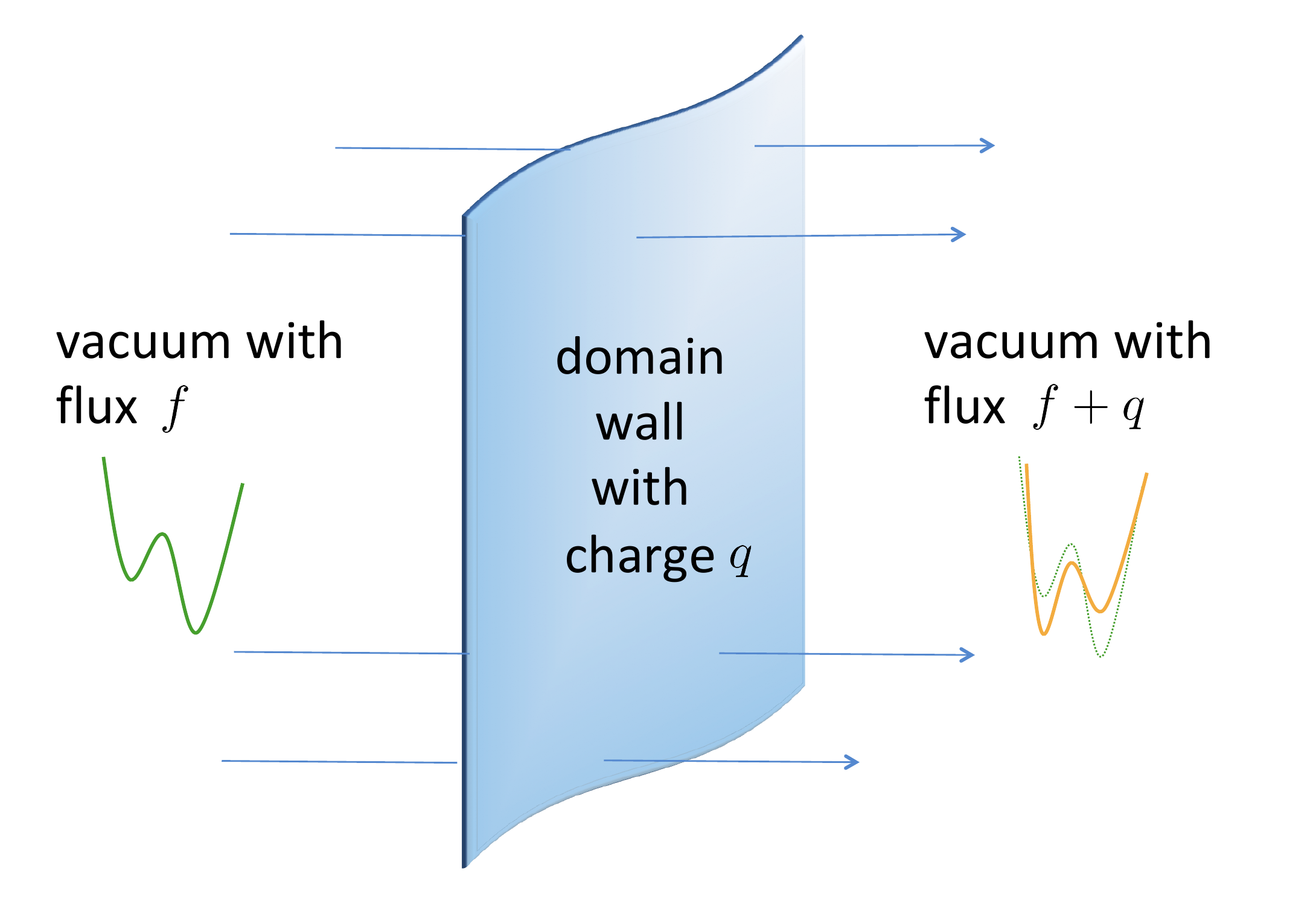}
	\caption{A charged brane interpolates between vacua with different fluxes.}
	\label{fig:domainwall}
\end{figure}

The original motivation for the conjecture comes from the WGC, as it provides a decay mode for non-susy vacua supported by fluxes. 
Consider an AdS vacuum supported by fluxes, meaning that some of the scalars are stabilized by a scalar potential, which is generated by these fluxes. Specifically, fluxes can originate from gauge fields with field strength $ F_{p}$ propagating in the extra dimensions, so the fluxes $ f $ are given by the integrals
\begin{equation}\label{Def. flux}
\int_{\S_{p}}F_{p}=f~
\end{equation}
over some internal non-trivial $p$-cycle.
These fluxes induce a scalar potential which can stabilize the scalars in a particular minimum, say an AdS vacuum.\footnote{It is not necessary that all scalars are stabilized by fluxes, but rather changing the flux changes the vacuum we are in.} The fluxes are Hodge dual to top form gauge field strengths $ \mathcal{F}_{d}=\text{d} C_{d-1} $ where $ d $ is the spacetime dimension. We can now apply the WGC to the gauge field $C_{d-1}$ which implies the existence of some electrically charged $ (d-2) $-brane that plays the role of a domain wall of codimension one.\footnote{Here we applied WGC to the non-dynamical gauge fields dual to the fluxes in the lower dimensional EFT. Equivalently, from the higher dimensional perspective, we can apply the WGC to the magnetic duals $ \star F_{p} $ of the gauge fields in the extra dimensions. It implies that there should be a charged $ (D-p-2) $-brane, where $ D $ is the total spacetime dimension. Then, dimensional reduction of these branes on the dual cycles to $ \S_{p} $ gives $ (d-2) $-branes, which are the domain walls obtained above.} The WGC requires such a brane to satisfy 
\be 
\label{inequality-tension-charge}
T\leq Q M^{2}_{p} \,,
\ee
where $ T $ denotes the brane tension. These branes are magnetically charged under the fluxes and  interpolate between different vacua with different values for the fluxes.
Suppose that the vacuum state on one side of the domain wall has flux $ f $. Then the charge conservation law determines the flux of the vacuum state on the other side of the wall to be equal to $ f+q $, where $q$ is the quantized charge of the brane (see figure \ref{fig:domainwall}). The presence of these domain walls is somewhat expected in a flux landscape, but the novelty of the WGC is the upper bound on the tension.

\medskip
Now, let us assume that the sharpened version of the WGC discussed in section \ref{sec:Sharpening the WGC} is valid.
 If the vacuum is not supersymmetric, then the sharpening of the WGC requires the strict inequality $ T<Q M_p^2 $. But a codimension one brane with a tension smaller than its charge corresponds to an instability in AdS!
 
\begin{figure}[ht]
	\centering
	\includegraphics[width=.4\textwidth]{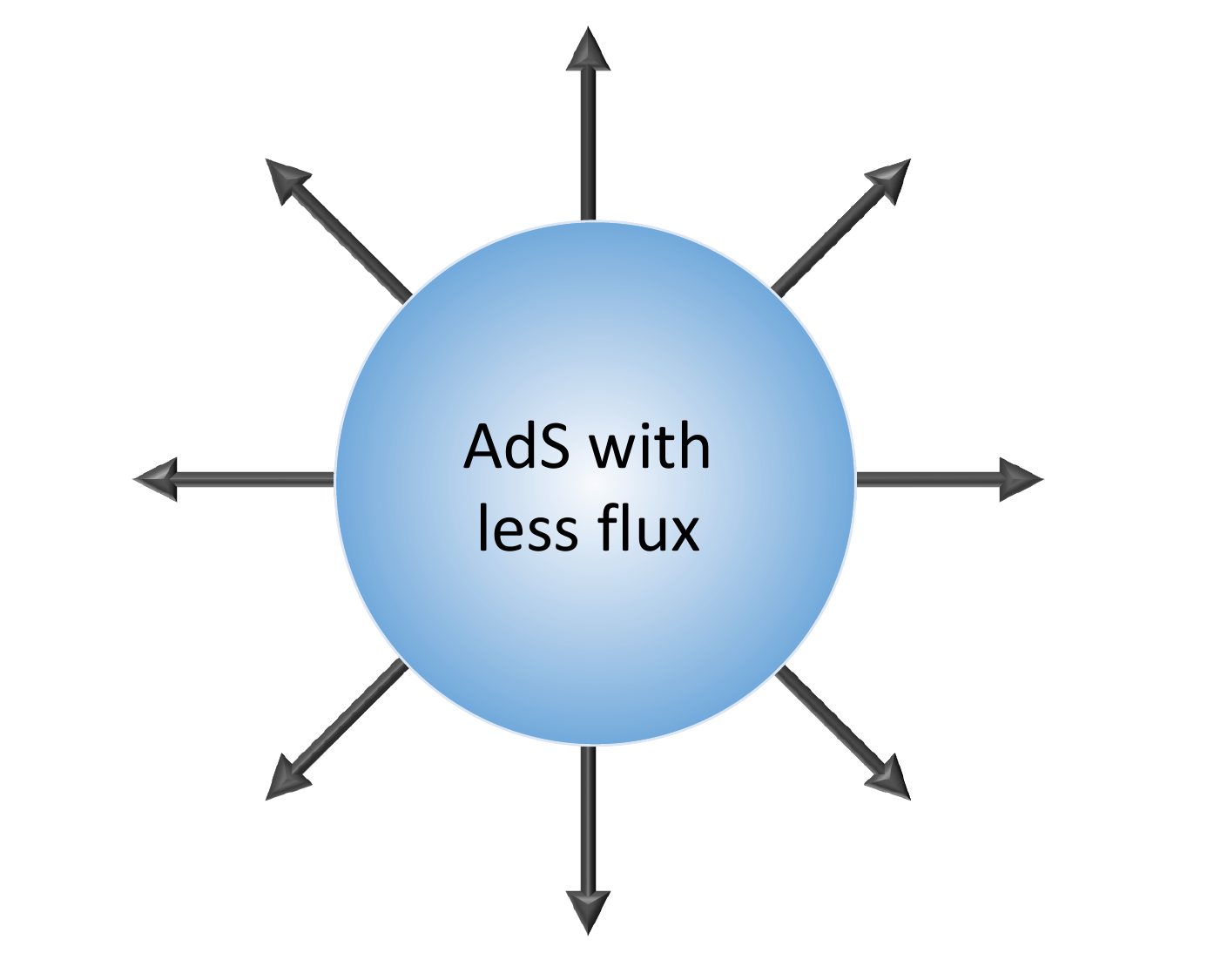}
	\caption{Bubble nucleation of an AdS vacuum with less flux inside. It will expand mediating vacuum decay.}
	\label{fig:bubble}
\end{figure}

 The intuition is simple and it is based on energy conservation. We can nucleate a bubble (whose wall is the brane) and, if the tension is smaller than the charge, the energy cost of expanding the bubble will be smaller than the energy gain from the electric repulsion between different points on the wall. So the bubble will grow indefinitely and will eventually describe a false vacuum decay to the true vacuum inside the bubble, in the sense that as the bubble comes to occupy the entire space, everything will be in the true vacuum, see figure \ref{fig:bubble}.
 
 More concretely, one can construct the instanton solution that mediates the vacuum decay. For this purpose, consider the following parametrization of the AdS$_d$ metric,
 \begin{equation}
  ds^2 = L^2(\cosh^2r\,d\tau^2 + dr^2 + \sinh^2r \,d\Omega^2_{d-2})
 \end{equation}
with $L$ the AdS radius. Following \cite{Maldacena:1998uz}, the instanton solution corresponds to a spherical brane of action
\be
S = L^{d-1}\Omega_{d-2}\int d\tau\left(T\sinh^{d-2}r \sqrt{\cosh^2r+\left(\frac{dr}{d\tau}\right)^2}-Q\sinh^{d-1}r\right) \, .
\ee
Upon solving the euclidean Einstein's equations, one gets that the  radius for the spherical brane solution is given by\footnote{This computation was performed in the probe/thin wall approximation, neglecting e.g. the backreaction on the scalar fields. Although this is a good approximation far away from the object, it would be desirable to have a more complete analysis including the scalar contribution to the WGC bound.}
 \begin{equation}
 \tanh R=\frac{T}{Q}\,,    
 \end{equation}
in Planck units. In Lorentzian signature, this will be associated with the nucleation of a bubble of radius $R$ that will continue expanding at the speed of light. The strict inequality on the WGC condition \eqref{inequality-tension-charge} implies 
 \begin{equation}
 \tanh R<1 \,,    
 \end{equation}
 which admits solutions for finite values of the radius. On the contrary, if the vacuum is supersymmetric, we could satisfy the WGC by a BPS brane saturating the inequality \eqref{inequality-tension-charge}, implying 
 \begin{equation}
 \label{infinite-radius-domain-wall}
 \tanh R=1\,,    
 \end{equation}
 whose only solution has infinite radius. Thus, we get a straight domain wall interpolating between different susy vacua which remain stable.

 

In conclusion, the sharpening of the WGC in section \ref{sec:Sharpening the WGC}, implies the presence of a codimension one brane with a tension smaller than its charge in non-susy vacua, which describes a bubble instability. According to this argument, any AdS non-susy vacuum with fluxes should be at best metastable.

\subsection{Bubbles of Nothing}

We have seen that the sharpened WGC provides a decay mode for non-susy vacua with fluxes. But what if there are no fluxes  and, therefore, we cannot use the WGC?  In principle, conjecture \ref{conj:nonsusy} should apply to any non-susy vacuum. It would be very interesting then to look for an universal instability whenever supersymmetry is broken, regardless of the ingredients of the theory.

A potential candidate for a universal instability, whenever there are extra dimensions, is a \emph{bubble of nothing}. This was first constructed by Witten in \cite{witten_instability_1982}, and we have recently learned that these instabilities are far more common than thought \cite{GarciaEtxebarria:2020xsr}.\\


\begin{otherbox}[Bubble of Nothing]{box:bubbleofnothing}
A bubble of nothing is a non-perturbative instability mediating the decay from the vacuum  to nothing, i.e. the vacuum annihilates itself. This occurs when a compact dimension collapses to zero size at the bubble wall. 
\end{otherbox}

Hence, it is literally a bubble with  nothing inside (not even a spacetime) that pops up in the vacuum and starts expanding at the speed of light, leaving \emph{nothing} behind. Expressed this way,
it may sounds very exotic. However, it is a perfectly well defined smooth solution to the equations of motion. The trick is that some extra dimension collapses to zero size at the bubble wall, allowing for the full solution to be regular and geodesically complete from the higher dimensional perspective.

\begin{figure}[h!]
	\centering
	\includegraphics[width=.75\textwidth]{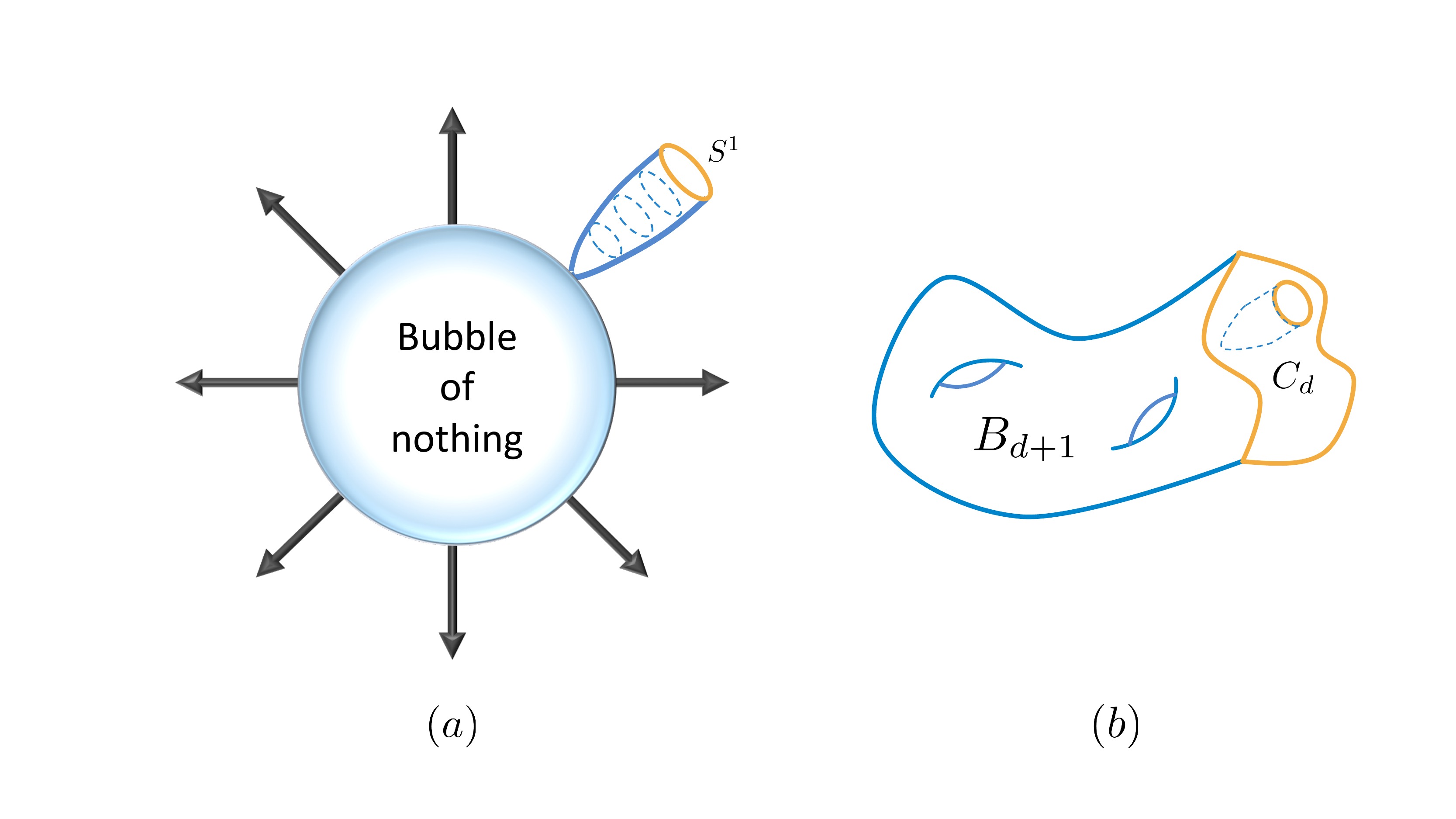}
	\caption{(a) The bubble of nothing is expanding at the speed of light. The internal compact manifold shrinks to zero size on the bubble wall. Here $ \te{S}^{1} $ is an example of internal compact manifold that belongs to the trivial class, i.e. $ \te{S}^{1}=\partial D $. (b) A compact manifold $ C_{d}=\partial B_{d+1} $ that belongs to the trivial bordism class $ \O^{\#}_{d} $, can become a bubble of nothing.} 
	\label{fig:cobordism}
\end{figure}

The existence of bubbles of nothing was first brought up by Witten \cite{witten_instability_1982} in the case of a Kaluza-Klein circle compactification of five dimensional Einstein gravity to four dimensions. The circle shrinks to zero size when approaching the bubble wall, as represented in figure  \ref{fig:cobordism}. More generally, we can have bubbles of nothing for higher dimensional compact spaces as long as the compact dimensions are allowed to collapse. Bubbles of nothing have e.g. been constructed for
$AdS _{5}\times S^{5}/\G  $ with freely acting $ \G $ and
$AdS _{4}\times \mathbb{C}P^{3}  $, where no other decay mode was previously known, providing evidence for conjecture \ref{conj:nonsusy}.

Consider some compactification $M_{D}\times C_d$ with $C_d$ the compact space and $M_D$ the space-time. Whether this compact space is topologically allowed to shrink to zero size gets translated to the question of whether it can be written as the boundary of another manifold of one dimension higher,
 i.e. $ C_{d}=\partial B_{d+1} $. 
 For Witten's case with $C_1=S^1$, the circle is the boundary of a disk (figure \ref{fig:cobordism}.a). 
 In general the compact manifold will be a boundary if it belongs to the trivial class of the cobordism group $ \O_{d}$ (figure \ref{fig:cobordism}.b).
Recall that the bordism group was defined in section \ref{sec:cob}, so we refer the reader there for details. In a nutshell, we say that two manifolds are bordant to each other if together they form a boundary of another manifold, and a manifold which is a boundary by itself belongs to the trivial class. Manifolds of this latter type admit bubbles of nothing at the topological level. This can be summarized as follows:
\begin{eqnarray*}
&&\text{Compact manifold }C_d\text{ can shrink to zero size} \Longleftrightarrow C_d=\partial B_{d+1}\Longleftrightarrow \\
&&C_d \text{ belongs to trivial class of } \O_{d} \Longleftrightarrow \text{A bubble of nothing is topologically allowed.}
\end{eqnarray*}

The  euclidean instanton solution mediating the decay is a warped product of the bordism manifold over a $(D-1)$-sphere given by
\be
\mathcal{M}_{\text{BON}}= B_{d+1}\times_W S^{D-1}\ .
\ee
The decay rate is exponentially suppressed by the euclidean action, i.e. $\Gamma\sim e^{-S_{\text{BON}}}$, where $S_{\text{BON}}$ can be computed to be proportional to the volume of the compact space $C_d$ at spatial infinity.

For the case of a KK circle compactification with fermions, $C_d=S^1$ preserves a spin structure. We then have $\Omega_1^{\rm spin}=Z_2$, implying that only the circle with antiperiodic boundary conditions for the fermions belongs to the trivial class on bordism and admits a bubble of nothing. Contrary, the circle with periodic (SUSY-compatible) boundary conditions cannot decay via a bubble of nothing, even if the vacuum breaks supersymmetry.

However, this topological obstruction can be absent in higher dimensions since e.g. $\Omega_d^{\text{spin}}=0$ for $d=3,5,6,7$. Therefore, any spin manifold of these dimensions admits a bubble of nothing! This was shown in \cite{GarciaEtxebarria:2020xsr} and used to construct new types of bubbles of nothing for a toroidal $T^3$ compactification with periodic boundary conditions. This is very interesting as it implies that a bubble of nothing can exist even if the boundary conditions are periodic and supersymmetry is restored at high energies (as long as the vacuum breaks susy spontaneously).

If the compactification includes other ingredients, like fluxes or charged fermions, these need to be added to the bordism group, which we will label in general as $\Omega_d^{\#}$. This is where the relation to other Swampland conjectures arises. Recall from section \ref{sec:cob} that a Swampland conjecture was proposed in \cite{McNamara:2019rup} stating that the cobordism group of a consistent theory of quantum gravity must be trivial, i.e $ \O^{\text{QG}}_{d}=0 $, in order to avoid the presence of global symmetries. Hence, the theory should always include the necessary defects or ingredients in order to guarantee the vanishing of all cobordism classes. As explained in \cite{GarciaEtxebarria:2020xsr}, this conjecture has an important consequence for the discussion at hand, as it implies that all compactification manifolds belong to the trivial class of the bordism group. Therefore, any internal compact manifold of a consistent QG compactification can shrink to a point, implying that there is \emph{no topological obstruction to construct bubbles of nothing in quantum gravity}. 




It is very important to remark, though, that the fact that a bubble of nothing is topologically allowed,  does not mean that it will expand after nucleation, describing an instability. For this to happen, the decay must be dynamically allowed, in the sense that it must be energetically favourable to decay. Otherwise, it will either collapse after nucleation or it will describe a flat domain wall instead of a bubble instability, as happens for a vacuum that preserves supersymmetry. In the latter case the radius of the euclidean solution will be infinite, describing an end-of-the-world brane. This dynamical obstruction is tied to whether a certain local energy condition is satisfied \cite{GarciaEtxebarria:2020xsr}. For instance, in the case of a manifold with covariantly constant spinors, the bubble of nothing will be dynamically allowed and the vacuum will decay only if the Dominant Energy Condition is violated.\footnote{The Dominant Energy Condition requires the vector $ -T^{\m}_{\n}n^{\n}>0 $ to be causal and future-pointing, for every causal and future-pointing vector $ n^{\m} $ and $T_{\m\n}$ being the stress-energy tensor.} Interestingly, this condition is generically violated when the vacuum breaks supersymmetry. For a theory with charged fermions, the energy condition is modified and resembles the WGC. By studying the relation between these energy conditions and supersymmetry in more detail, we could end up proving or disproving the conjecture about the instability of any non-susy vacuum.


\subsection{Implications of AdS Instability Conjecture} \label{sec:AdS-inst-implications}

Here, we are going to discuss two implications of the AdS instability conjecture.
\nl

\hspace{\parindent} \underline{Non-susy AdS/CFT}:\\

An instability in AdS, even if highly suppressed, has dramatic implications for the boundary CFT. Note that 
an observer at the boundary has access to an infinite volume of the bulk and hence can detect any bulk instability instantaneously. This implies that the AdS vacuum has no dual unitary CFT. From the gravity side, the instability will reach the boundary and come back in Hubble time, so this is the maximum lifetime of the vacuum. 

This can have important implications for non-susy holography. Notice, though, that the previous discussion about AdS instabilities was performed in the context of a weakly coupled Einstein gravity theory, in which a semiclassical description of bubble instabilities is possible and the WGC applies. Consequently, non-susy holography can be consistent but only if the non-susy unitary CFT is dual to a gravity theory which is not AdS weakly coupled to Einstein gravity. In other words, a non-susy unitary CFT cannot admit a large N expansion and/or a large gap to single-trace higher spin operators (spin greater than two). 
\nl

\hspace{\parindent} \underline{Compactifications of Standard Model of particle physics}:\\

This conjecture can also have implications for our universe. Even though we do not live in an AdS spacetime, non-susy AdS vacua can still be generated in lower dimensions from compactifying the Standard Model. The simplest case is to compactify on a circle to three dimensions. Assuming background independence, compactifications of a theory consistent with QG should also be consistent. Hence, according to the conjecture, these 3d AdS vacua should be unstable because otherwise the Standard Model would be inconsistent with quantum gravity! But this cannot possibly be the case, since the Standard Model is clearly a low energy theory consistent with QG describing our universe. We then need to find a way to exclude the existence of these stable 3d vacua, which can be done by placing bounds on beyond Standard Model parameters and the cosmological constant of our universe. In particular, we can guarantee the absence of stable AdS vacua by imposing a lower bound on the value of the cosmological constant $\Lambda_4$ in terms of the Dirac mass\footnote{Majorana neutrinos (only 2 degrees of freedom) would be ruled out according to the conjecture, as they always generate an AdS vacuum.} of the lightest neutrino  $m_\nu$  \cite{Ibanez:2017kvh}, which roughly goes as
\begin{equation}\label{naturalness}
\L_{4}\gtrsim m^{4}_{\nu}\ .
\end{equation}    
Interestingly, this could explain the numerical coincidence $\L_{4}\sim m^{4}_{\nu}$ observed in our universe. Furthermore, we can write the neutrino mass in terms of the Higgs vev, so that the previous bound gets translated to an upper bound on the electroweak scale
\begin{equation}
\langle H \rangle\lesssim \frac{\L^{1/4}}{Y_\nu} \,,
\end{equation}
where $ Y_\nu $ is the neutrino Yukawa coupling. Amazingly, this relates the two naturalness problems associated to the cosmological constant and the EW scale, so that if we happened to live in a universe with a small $\L^{1/4}$, the EW scale would also have to be small (of order TeV at most) by consistency of quantum gravity. Hence, there is no EW hierarchy problem to start with, as greater values of $\langle H \rangle$ would not lead to theories consistent with quantum gravity.

Let us emphasize, though, that this is  \emph{not} a solution to the EW hierarchy problem yet. Clearly, we first need to prove the validity of the Swampland conjecture. But also, we are assuming that there is no other way to destabilize the lower dimensional AdS vacua beyond those provided by the Standard Model, which can be criticized. Yet, this result exemplifies the potential of the Swampland conjectures to change our logic of naturalness, which breaks down when taking into account quantum gravity constraints. The  naturalness issues observed in our universe could simply be artifacts of not yet knowing the space of parameters that is consistent with quantum gravity. Remember, not everything goes!


\section{de Sitter Conjecture\label{sec:dS}}

For the time being, it is still an open question whether string theory admits a de Sitter vacuum or not. At the moment, there is no full-fledged top-down de Sitter construction in a controllable regime of string theory. Although there is no universal no-go theorem either. This is a challenging and very important objective in string theory, due to the obvious phenomenological implications for the cosmological expansion of our universe. Based on the relations to other Swampland conjectures, and the difficulties of constructing a dS vacuum in string theory, the following Swampland conjecture has been proposed \cite{Obied:2018sgi,Ooguri:2018wrx,Garg:2018reu} asserting that de Sitter vacua (even if metastable) are not consistent with quantum gravity:\\ 

\begin{conj}[de Sitter Conjecture]{conj:dS}
	A scalar potential of an EFT weakly coupled to Einstein gravity must satisfy
\begin{equation}
\label{dS}
M_{P}\frac{|\nabla V|}{V}\geq c\,,
\end{equation}
with $c$ some $\mathcal{O}(1)$ constant. 
This was further refined by stating that the previous bound only needs to be imposed if the following condition on the second derivative of the potential is violated,
\begin{equation}
\min \big( \nabla_{i}\nabla_{j}V \big)\leq \frac{-c^{'}V}{M^{2}_{P}}~\,,
\end{equation}
with $c'$ another $\mathcal{O}(1)$ constant. This way, only dS minima (and not critical points in general) are ruled out.
\end{conj}

This conjecture is very strong and not free of controversy. Currently, it has only been checked in the asymptotic regions of the moduli space (i.e. near infinite distance singularities) where it is more widely accepted to be true. We will discuss this asymptotic version of the dS conjecture in the next subsection. For reviews about the status of de Sitter vacua in string theory we refer the reader to \cite{Danielsson:2018ztv,Cicoli:2018kdo}.

Let us finally mention that there is another conjecture also constraining the scalar potential in de Sitter space; this is the Transplanckian Censorship conjecture (TCC) \cite{Bedroya:2019snp}. In these lectures, we will not discuss it in much detail, but we will at least give its definition and comment on the differences with respect to the original dS conjecture.\\ 

\begin{otherbox}[Transplanckian Censorship Conjecture]{box:transplanckian}
The expansion of the universe must slow down before all Planckian modes are stretched beyond Hubble size. It has two implications:
\begin{itemize}
\item No dS minima can exist at the asymptotic boundaries of the moduli space. In the asymptotic regimes, one recovers a similar bound to \eqref{dS} constraining the asymptotic behaviour of the potential, but with a fixed constant $c$ given by 
\begin{equation}
\label{TCC}
\frac{|\nabla V|}{V}\geq \frac{2}{\sqrt{(d-1)(d-2)}}~.
\end{equation}
\item A dS minimum can exist deep inside the bulk, but it must be short-lived. The lifetime $\tau$ for a metastable dS vacuum is bounded from above by
\begin{equation}
\tau\leq \frac{1}{H}\log \frac{M_{P}}{H} \,,
\end{equation}
where $ H $ is the Hubble scale.
\end{itemize}
\end{otherbox}
In a certain sense, the TCC is weaker than the dS conjecture as it does not completely forbid the existence of dS vacua, but only does so asymptotically. It also provides a lower bound on the exponential rate of the SDC tower, by assuming that the potential scales as $m^d$ with $d$ the space-time dimension \cite{Bedroya:2019snp} or as $m^2$ \cite{Andriot:2020lea}.

\subsection{Asymptotic de Sitter Conjecture and Relation to SDC}

\begin{figure}[ht]
	\centering
	\includegraphics[width=.65\textwidth]{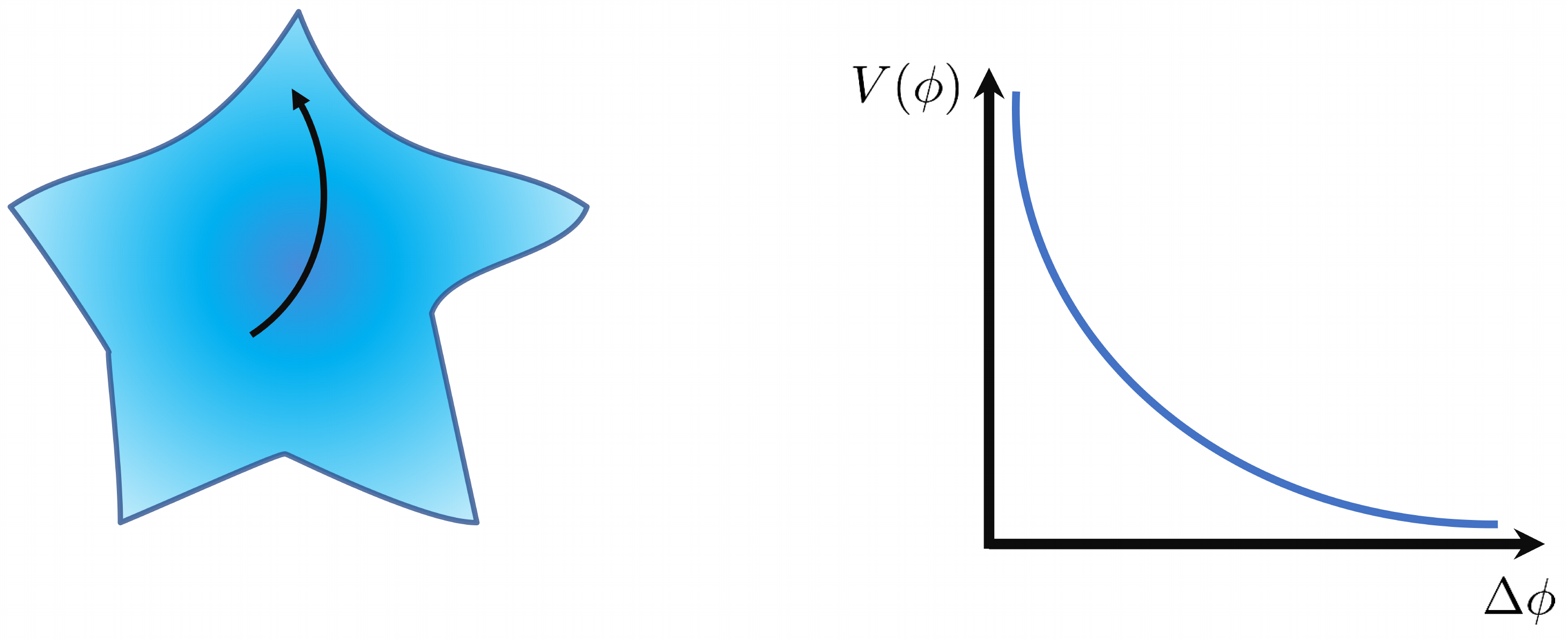}
	\caption{The runaway behavior of the potential as we approach an infinite distance point in the field space according to the asymptotic de Sitter conjecture.}
	\label{fig:potential}
\end{figure}

From the landscape of string theory compactifications, we can gather evidence for an asymptotic version of the dS conjecture. Although it has no official name, we will refer to it as the Asymptotic dS conjecture, see figure \ref{fig:potential}.\\
\begin{otherbox}[Asymptotic dS Conjecture]{box:asymp_dS}
A scalar potential of an EFT weakly coupled to Einstein gravity presents a runaway behavior when approaching an infinite field distance point,
\begin{equation}\label{runaway}
	\lvert \nabla V \rvert\geq c V~;~~~~c\sim\mathcal{O}(1)~.
\end{equation}	
\end{otherbox}
We can also motivate this claim by using the Distance conjecture, as we will do below. The parameter $c$ is undetermined although the relation to the SDC suggests that it is proportional to the exponential rate of the infinite tower. 

This can be thought of as a generalization of the ``Dine-Seiberg problem" for any direction in field space. Dine and Seiberg \cite{Dine:1985he} pointed out that the string coupling cannot be stabilized at weak coupling by classical effects. Classical limits are examples of infinite distance limits, and only away from these limits, when quantum corrections are important, can a minimum be generated. The asymptotic dS conjecture basically implies the same conclusion for any scalar direction when taking a large field limit, essentially forbidding the presence of dS vacua at parametric control. Notice that this asymptotic version of the conjecture does not constrain KKLT-like constructions \cite{Kachru:2003aw} since those vacua are not at parametric control.

This is a concrete claim that we can check in string theory, as these asymptotic regimes are precisely those regimes in which we have control of the corrections and we can trust the EFT. In fact, the majority of models in string phenomenology are constructed in these regimes as we are often dealing with large volume, weak couplings, etc. So most of the time we are at some asymptotic regime in the field space even if we do not declare it explicitly. To be clear, every limit represents some perturbative expansion of the EFT in $1/\phi$ where $\phi$ would be the scalar taken to be large. Whenever this perturbative expansion makes sense and the corrections are subleading with respect to the leading term, we say we are in the asymptotic regime. For the particular case of the string coupling, the corrections can be understood as quantum effects over a classical leading piece, although this interpretation in terms of classical vs quantum pieces does not extend to other limits. In section \ref{sec:dS-ST-evidence} we will discuss the string theory evidence in support of this asymptotic version of the dS conjecture. 

\medskip
According to the Distance conjecture of section \ref{sec:distconj}, the moment we are in some asymptotic regime of the moduli space, there is an infinite tower of states that becomes exponentially light. It was proposed in \cite{Ooguri:2018wrx} that the SDC tower is responsible for the runaway behaviour of the potential.
 The argument goes as follows. The entropy $ S_{\text{tower}} $ associated to the infinite tower of states depends on the number of states $ N $ as well as the de Sitter horizon radius $ R $. According to the SDC, the number of states will increases exponentially with the field distance
\begin{equation}
\label{Ntower}
N\sim e^{b\f}
\ .
\end{equation}
For a quasi-dS space in which we have a nearly flat potential, the contribution of the tower to the entropy cannot exceed the Gibbons-Hawking entropy, which is fixed by the value of $R$,
\begin{equation}
\label{boundS}
S_{\text{tower}}(N,R)\leq S_{\text{GH}}=R^{2}=\frac{1}{H^2}~,
\end{equation}
where $H$ is the Hubble scale. In the weak coupling limit,\footnote{Infinite distane limits are associated with weakly coupled descriptions of the theory.} one would expect that the light degrees of freedom
dominate the Hilbert space, so the above bound is saturated. 

Assuming that $ S_{\text{tower}} $ depends polynomially on $N$ and $R$, the previous two equations imply an exponential behaviour of $R$ in terms of the field distance. 
For a nearly flat potential, $H^2=R^{-2}$  scales the same way as the potential, implying
\begin{equation}
V(\f)\sim H^2\sim e^{-c\f}~.
\end{equation}
The constant $c$ will be proportional to $b$ in \eqref{Ntower} which is related to the exponential rate of the tower. More concretely, by parametrizing the entropy of the tower as 
\be
S_{\text{tower}}\simeq N^\gamma H^{-\delta}
\ee
the bound \eqref{boundS} implies
\be
\label{HS}
H\lesssim \frac{1}{N^{\frac{\gamma}{2-\delta}}}
\ee
in Planck units. The coefficient $c$ in the de Sitter conjecture is then given by
\be
\label{cS}
c=\frac{2\gamma}{2-\delta}b\ .
\ee

As a particular example, we can take a KK tower of a circle compactification, so the exponential rate $\lambda$ is given in \eqref{lambdaKK}. We can estimate the number of states as the number of species in \eqref{Nsepcies}, which can be written in Planck units as
\be
N=\Delta m^{\frac{2-d}{d-1}}
\ee
using \eqref{speciesm}. Hence, it implies
\be
\label{bKK}
b=\frac{d-2}{d-1}\lambda=\sqrt{\frac{d-2}{d-1}}\ .
\ee
Notice that \eqref{HS} actually resembles the species bound in \eqref{species}. In fact, if we impose that the Hubble scale is smaller than the species bound \eqref{species} we get
\begin{equation}
\label{HQG}
  H\lesssim \Lambda_{QG}\simeq \frac{1}{N^{\frac{1}{d-2}} }
\end{equation}
in Planck units, which is equivalent to \eqref{HS} upon identifying $\frac{\gamma}{2-\delta}=\frac{1}{d-2}$. An interesting coincidence occurs when using this species bound result and the growth of a KK tower. 
 By replacing  \eqref{Ntower} and \eqref{bKK} into \eqref{HQG}, we get that the potential decreases exponentially in the distance with the following rate
\be
c=\frac{2}{d-1}\lambda=\frac{2}{\sqrt{(d-1)(d-2)}}
\ee
which is precisely the value of the TCC factor in \eqref{TCC}. Hence, upon using the species bound, the exponential rate of a KK tower would give rise to the asymptotic behaviour of the scalar potential predicted by the TCC.

\subsection{String Theory Evidence} \label{sec:dS-ST-evidence}

Even before the formulation of this Swampland conjecture, there were several no-go theorems for de Sitter in the literature, valid in the classical regimes of certain string theory compactifications (see e.g. \cite{Maldacena:2000mw,Hertzberg:2007wc,Flauger:2008ad,Wrase:2010ew}). This has recently been extended to other setups, see e.g.  \cite{Danielsson:2018ztv,Andriot:2019wrs} for recent reviews, as well as more general infinite distance limits beyond the classical regime (i.e. beyond small $g_s$) \cite{Grimm:2019ixq}. 
Different no-go theorems apply to different compactification manifolds with fluxes and types of localized sources: D-branes, orientifolds, etc. Many of them are based on the asymptotic scaling of the potential with the moduli. As an example, we are going to discuss the original no-go theorem regarding type IIA on an orientifold CY$_3$ \cite{Hertzberg:2007wc}, as all the others can, in a sense, be considered as generalizations of this one.



Consider type IIA on a CY$_3$ with fluxes, D6-branes and O6-planes. 
In this example, all scalars from the closed string sector can be stabilized by an appropriate choice of RR and NS fluxes. For the no-go theorem it is enough to provide the dependence of the scalar  potential on the 4d dilaton $ \tau $ and the overall volume $ \r $ of the CY$_3$.  At weak coupling and large volume, the potential reads
\begin{equation}
V=\frac{A_{\text{NS}}(\f)}{\r^{3}\tau^{2}}+\sum_{p=0,2,4,6}\frac{A_{\text{RR},p}(\f)}{\r^{p-3}\tau^{4}}+\frac{A_{\text{D6}}(\f)}{\tau^{3}}-\frac{A_{\text{O6}}(\f)}{\tau^{3}}~,
\end{equation}
where each term corresponds to a different contribution from NS fluxes, RR $p$-form fluxes, D6-branes and O6-planes. The coefficients $A_i(\phi)$ are positive and depend on the axion partners of the dilaton and the volume, as well as other ingredients, such as fluxes or other scalars of the compactfication. 
One can check that the derivatives of the scalar potential satisfy
\begin{equation}\label{no extremum}
-\r\partial_{\r}V-3\tau\partial_{\tau}V=9V+\sum_{p=0,2,4,6} pV_{p}\geq 9V \,,
\end{equation}
where the last inequality holds because $ \sum p V_{p} $ is always positive. This implies that we cannot stabilize the moduli in dS, as this equation has no solution for $\partial_i V=0$ and $V>0$. In fact, no dS critical point is allowed.

Other no-go theorems will involve a similar inequality to \eqref{no extremum} but with different numerical coefficients coming from the moduli scaling of the different contributions to the potential. More generally, the no-go theorems based on the moduli scaling take the following form,
\begin{equation}\label{general no-go}
a V(\f)+\sum b_{i}\partial_{\f^{i}}V(\f)\leq 0~,~~a>0~,~~b_{i}\neq 0~,
\end{equation}
where the values of $a,b_i$ will depend on the specific moduli scaling of the potential and the field metric.
 If (\ref{general no-go}) is satisfied, then the dS conjecture is satisfied with an order one factor $c$ given by
\begin{equation}\label{dS conj.}
\lvert \nabla V\rvert\geq c V~;~~~~c^{2}=\frac{a^{2}}{\sum b^{2}_{i}}~.
\end{equation}
When leaving the asymptotic regime, the coefficients $A_i(\phi)$ receive corrections which can eventually change the sign of the coefficient, invalidating the no-go theorem.

Although many of these no-go theorems are formulated in the weak coupling limit of type II compactifications, 
it is also possible to generalize them beyond weak coupling by studying F-theory compactifications dual to M-theory with $G_4$ flux \cite{Grimm:2019ixq}. Employing the techniques of mixed Hodge structures, used to test the Distance conjecture in Calabi-Yau manifolds (see section \ref{sec:typeII-CY}), we can determine the asymptotic form of the flux potential near any type of infinite distance limit in the complex structure moduli space of the CY$ _{4} $, which also includes the type II dilaton. This way, one realizes that the coefficients $a,b_i$, and consequently $c$, are determined in terms of the type of infinite distance singularity under consideration, which is in turn fixed by the monodromy properties, as happened for the SDC. This supports that the dS conjecture is not associated to weak coupling but valid in more general infinite distance limits, and that indeed the SDC exponential rate and the dS conjecture factor $c$ are related, as they have the same geometrical origin.

There is still much more work to do and many questions to answer regarding this conjecture. For instance, we need to have a better understanding of the nature of the factor $ c $. The situation is analogous to the exponential rate of the SDC, where we can gather evidence from the geometry of different string theory compactifications, but a general derivation based on EFT data is missing. Interestingly, from a bottom-up perspective, the factor $ c $ may also be related to the extremality bound of the membranes that generate the scalar potential $ V(\f) $. Recall from \eqref{eq:CY-flux-potential} that the fluxes can be dualized to top-form gauge field strengths, so $ V(\f) $ becomes equivalent to the physical charge of a membrane whose quantized charges are equal to the fluxes.  By applying the WGC to these fluxes, we find that membranes \emph{saturating} the WGC generate a scalar potential that satisfy the de Sitter conjecture, as shown in \cite{Lanza:2020qmt}.





\section{Final Remarks}

To conclude these lectures, we can take a final look at the web of Swampland conjectures represented in figure \ref{fig:mapofconjectures}. Now, the meaning of the different connecting lines between the conjectures is hopefully more clear for the reader. Sometimes, they imply that a conjecture follows from the other in a concrete setup, or that a certain conjecture can inform another. In other cases, they imply generalizations or extensions of a given conjecture. Hence, proving one conjecture is \emph{not} enough to prove the ones related to it. However, the uncovered relations between a priori disconnected conjectures suggest that we are on the right path, and that some constraints along these lines should be require to guarantee consistency in quantum gravity.

In these notes, we have discussed in detail those conjectures for which the amount of evidence has significantly increased in the past few years, and with it, our confidence in the conjectures. These are the Absence of Global Symmetries, the Weak Gravity Conjecture and the Distance Conjecture.  We have also briefly explained the rest of the conjectures represented in figure \ref{fig:mapofconjectures}, although they can often be understood as extensions of the previous three. Let us remark that there are a few other swampland conjectures which have not been included in the figure, due to their recent formulation and lack of sufficient research by different groups. However, they might play an  important role in the future, in the same way that some of the present conjectures might get disproven and disappear. The swampland program is very rapidly evolving, so we cannot discard the possibility that  the diagram of the relevant conjectures might have a very different shape in the future.

Our job is to continue gathering evidence to prove or disprove the conjectures, as well as understanding their implications. In this sense, string theory is a perfect framework to sharpen and define the swampland constraints in a precise way. Even if our understanding of string theory is still in continuous evolution, we already have plenty of data from corners which are under computational control. Fortunately, it is precisely at these corners, typically boundaries of the field space, where weakly coupled Einstein gravity theories arise and, therefore, where the swampland conjectures should naturally hold and can be quantitatively tested. Any ambiguity or undetermined order one factor in the conjectures should then disappear upon collecting enough information about the precise statements that hold at these string theory corners. A more difficult obstacle will be to abandon supersymmetry, although some first steps to testing the conjectures in non-supersymmetric setups have already been taken.

As a final remark, let us emphasize that it is at the infinite distance boundaries of field space where approximate global symmetries, weakly coupled gauge theories, large field ranges and an Einstein gravity description arise. And it is precisely at these boundaries where quantum gravitational effects preventing too well-approximated global symmetries become important, and where the swampland conjectures seem to have implications for the IR physics. Our universe shares these features, as we observe Einstein gravity at low energies as well as a weakly coupled photon. So it might be precisely in universes like ours where we cannot ignore the UV imprint of quantum gravity at low energies. In fact, nature might be telling us with the naturalness issue of the cosmological constant and the EW hierarchy problem that it is time to break with the expectation of separation of scales originating from a Wilsonian quantum field theory approach. The UV/IR mixing induced by quantum gravity might be the missing piece to understanding these naturalness issues. In particular, our notion of naturalness should be modified if not the entire space of parameters is consistent with quantum gravity, as the Swampland program defends. Maybe, after all, Quantum Gravity really matters at low energies, even if the Planck scale happens to be so high in energies. And this is, certainly, a very exciting possibility!

\vspace{1cm}
\noindent{\bf{\large Acknowledgements}}\\

\noindent We would like to thank the organizers of the QFT and Geometry summer school 2020 for organizing such a wonderful school in such difficult times. Many thanks also to Miguel Montero for very helpful discussions and comments on the manuscript. The work of M.v.B. is in part supported by the ERC Consolidator Grant number 682608 “Higgs bundles: Supersymmetric Gauge Theories and Geometry (HIGGSBNDL)”. The work by J.C. is supported by the FPU grant no. FPU17/04181 from the
Spanish Ministry of Education and by the Spanish Research Agency (Agencia Espa\~nola de Investigaci\'on) through the grant IFT Centro de Excelencia Severo Ochoa  SEV-2016-0597. D.M. is supported by TUBITAK grant 117F376.  The research of I.V. was supported by a grant from the Simons Foundation (602883, CV).


\bibliography{References}

\providecommand{\href}[2]{#2}\begingroup\raggedright\begin{thebibliography}{100}

\bibitem{school}
``{QFT and Geometry Summer School 2020}.''
  \url{https://sites.google.com/view/qftandgeometrysummerschool}.
\newblock Videos of the lectures can be found there.

\bibitem{Palti:2019pca}
E.~Palti, \emph{{The Swampland: Introduction and Review}},
  \href{http://dx.doi.org/10.1002/prop.201900037}{\emph{Fortsch. Phys.} {\bf
  67} (2019) 1900037}, [\href{https://arxiv.org/abs/1903.06239}{{\tt
  1903.06239}}].

\bibitem{Brennan:2017rbf}
T.~D. Brennan, F.~Carta and C.~Vafa, \emph{{The String Landscape, the
  Swampland, and the Missing Corner}},
  \href{http://dx.doi.org/10.22323/1.305.0015}{\emph{PoS} {\bf TASI2017} (2017)
  015}, [\href{https://arxiv.org/abs/1711.00864}{{\tt 1711.00864}}].

\bibitem{Palti:2020mwc}
E.~Palti, \emph{{A Brief Introduction to the Weak Gravity Conjecture}},
  \href{http://dx.doi.org/10.31526/lhep.2020.176}{\emph{LHEP} {\bf 2020} (2020)
  176}.

\bibitem{Witten:1982fp}
E.~Witten, \emph{{An SU(2) Anomaly}},
  \href{http://dx.doi.org/10.1016/0370-2693(82)90728-6}{\emph{Phys. Lett. B}
  {\bf 117} (1982) 324--328}.

\bibitem{Vafa:2005ui}
C.~Vafa, \emph{{The String landscape and the swampland}},
  \href{https://arxiv.org/abs/hep-th/0509212}{{\tt hep-th/0509212}}.

\bibitem{Banks:2010zn}
T.~Banks and N.~Seiberg, \emph{{Symmetries and Strings in Field Theory and
  Gravity}}, \href{http://dx.doi.org/10.1103/PhysRevD.83.084019}{\emph{Phys.
  Rev. D} {\bf 83} (2011) 084019}, [\href{https://arxiv.org/abs/1011.5120}{{\tt
  1011.5120}}].

\bibitem{Gaiotto:2014kfa}
D.~Gaiotto, A.~Kapustin, N.~Seiberg and B.~Willett, \emph{{Generalized Global
  Symmetries}}, \href{http://dx.doi.org/10.1007/JHEP02(2015)172}{\emph{JHEP}
  {\bf 02} (2015) 172}, [\href{https://arxiv.org/abs/1412.5148}{{\tt
  1412.5148}}].

\bibitem{Susskind:1995da}
L.~Susskind, \emph{{Trouble for remnants}},
  \href{https://arxiv.org/abs/hep-th/9501106}{{\tt hep-th/9501106}}.

\bibitem{Israel:1967wq}
W.~Israel, \emph{{Event horizons in static vacuum space-times}},
  \href{http://dx.doi.org/10.1103/PhysRev.164.1776}{\emph{Phys. Rev.} {\bf 164}
  (1967) 1776--1779}.

\bibitem{Polchinski:1998rr}
J.~Polchinski, \emph{{String theory. Vol. 2: Superstring theory and beyond}}.
\newblock Cambridge Monographs on Mathematical Physics. Cambridge University
  Press, 12, 2007,
  \href{http://dx.doi.org/10.1017/CBO9780511618123}{10.1017/CBO9780511618123}.

\bibitem{Harlow:2018jwu}
D.~Harlow and H.~Ooguri, \emph{{Constraints on Symmetries from Holography}},
  \href{http://dx.doi.org/10.1103/PhysRevLett.122.191601}{\emph{Phys. Rev.
  Lett.} {\bf 122} (2019) 191601},
  [\href{https://arxiv.org/abs/1810.05337}{{\tt 1810.05337}}].

\bibitem{Harlow:2018tng}
D.~Harlow and H.~Ooguri, \emph{{Symmetries in quantum field theory and quantum
  gravity}},  \href{https://arxiv.org/abs/1810.05338}{{\tt 1810.05338}}.

\bibitem{Harlow:2018fse}
D.~Harlow, \emph{{TASI Lectures on the Emergence of Bulk Physics in AdS/CFT}},
  \href{http://dx.doi.org/10.22323/1.305.0002}{\emph{PoS} {\bf TASI2017} (2018)
  002}, [\href{https://arxiv.org/abs/1802.01040}{{\tt 1802.01040}}].

\bibitem{McNamara:2019rup}
J.~McNamara and C.~Vafa, \emph{{Cobordism Classes and the Swampland}},
  \href{https://arxiv.org/abs/1909.10355}{{\tt 1909.10355}}.

\bibitem{Harlow:2015lma}
D.~Harlow, \emph{{Wormholes, Emergent Gauge Fields, and the Weak Gravity
  Conjecture}}, \href{http://dx.doi.org/10.1007/JHEP01(2016)122}{\emph{JHEP}
  {\bf 01} (2016) 122}, [\href{https://arxiv.org/abs/1510.07911}{{\tt
  1510.07911}}].

\bibitem{Kim:2019vuc}
H.-C. Kim, G.~Shiu and C.~Vafa, \emph{{Branes and the Swampland}},
  \href{http://dx.doi.org/10.1103/PhysRevD.100.066006}{\emph{Phys. Rev. D} {\bf
  100} (2019) 066006}, [\href{https://arxiv.org/abs/1905.08261}{{\tt
  1905.08261}}].

\bibitem{Kim:2019ths}
H.-C. Kim, H.-C. Tarazi and C.~Vafa, \emph{{Four-dimensional
  $\mathbf{\mathcal{N}=4}$ SYM theory and the swampland}},
  \href{http://dx.doi.org/10.1103/PhysRevD.102.026003}{\emph{Phys. Rev. D} {\bf
  102} (2020) 026003}, [\href{https://arxiv.org/abs/1912.06144}{{\tt
  1912.06144}}].

\bibitem{Lee:2019skh}
S.-J. Lee and T.~Weigand, \emph{{Swampland Bounds on the Abelian Gauge
  Sector}}, \href{http://dx.doi.org/10.1103/PhysRevD.100.026015}{\emph{Phys.
  Rev. D} {\bf 100} (2019) 026015},
  [\href{https://arxiv.org/abs/1905.13213}{{\tt 1905.13213}}].

\bibitem{Fichet:2019ugl}
S.~Fichet and P.~Saraswat, \emph{{Approximate Symmetries and Gravity}},
  \href{http://dx.doi.org/10.1007/JHEP01(2020)088}{\emph{JHEP} {\bf 01} (2020)
  088}, [\href{https://arxiv.org/abs/1909.02002}{{\tt 1909.02002}}].

\bibitem{Daus:2020vtf}
T.~Daus, A.~Hebecker, S.~Leonhardt and J.~March-Russell, \emph{{Towards a
  Swampland Global Symmetry Conjecture using weak gravity}},
  \href{http://dx.doi.org/10.1016/j.nuclphysb.2020.115167}{\emph{Nucl. Phys. B}
  {\bf 960} (2020) 115167}, [\href{https://arxiv.org/abs/2002.02456}{{\tt
  2002.02456}}].

\bibitem{ArkaniHamed:2006dz}
N.~Arkani-Hamed, L.~Motl, A.~Nicolis and C.~Vafa, \emph{{The String landscape,
  black holes and gravity as the weakest force}},
  \href{http://dx.doi.org/10.1088/1126-6708/2007/06/060}{\emph{JHEP} {\bf 06}
  (2007) 060}, [\href{https://arxiv.org/abs/hep-th/0601001}{{\tt
  hep-th/0601001}}].

\bibitem{Kats:2006xp}
Y.~Kats, L.~Motl and M.~Padi, \emph{{Higher-order corrections to mass-charge
  relation of extremal black holes}},
  \href{http://dx.doi.org/10.1088/1126-6708/2007/12/068}{\emph{JHEP} {\bf 12}
  (2007) 068}, [\href{https://arxiv.org/abs/hep-th/0606100}{{\tt
  hep-th/0606100}}].

\bibitem{Cheung:2018cwt}
C.~Cheung, J.~Liu and G.~N. Remmen, \emph{{Proof of the Weak Gravity Conjecture
  from Black Hole Entropy}},
  \href{http://dx.doi.org/10.1007/JHEP10(2018)004}{\emph{JHEP} {\bf 10} (2018)
  004}, [\href{https://arxiv.org/abs/1801.08546}{{\tt 1801.08546}}].

\bibitem{Hamada:2018dde}
Y.~Hamada, T.~Noumi and G.~Shiu, \emph{{Weak Gravity Conjecture from Unitarity
  and Causality}},
  \href{http://dx.doi.org/10.1103/PhysRevLett.123.051601}{\emph{Phys. Rev.
  Lett.} {\bf 123} (2019) 051601},
  [\href{https://arxiv.org/abs/1810.03637}{{\tt 1810.03637}}].

\bibitem{Andriolo:2018lvp}
S.~Andriolo, D.~Junghans, T.~Noumi and G.~Shiu, \emph{{A Tower Weak Gravity
  Conjecture from Infrared Consistency}},
  \href{http://dx.doi.org/10.1002/prop.201800020}{\emph{Fortsch. Phys.} {\bf
  66} (2018) 1800020}, [\href{https://arxiv.org/abs/1802.04287}{{\tt
  1802.04287}}].

\bibitem{Crisford:2017gsb}
T.~Crisford, G.~T. Horowitz and J.~E. Santos, \emph{{Testing the Weak Gravity -
  Cosmic Censorship Connection}},
  \href{http://dx.doi.org/10.1103/PhysRevD.97.066005}{\emph{Phys. Rev. D} {\bf
  97} (2018) 066005}, [\href{https://arxiv.org/abs/1709.07880}{{\tt
  1709.07880}}].

\bibitem{Montero:2016tif}
M.~Montero, G.~Shiu and P.~Soler, \emph{{The Weak Gravity Conjecture in three
  dimensions}}, \href{http://dx.doi.org/10.1007/JHEP10(2016)159}{\emph{JHEP}
  {\bf 10} (2016) 159}, [\href{https://arxiv.org/abs/1606.08438}{{\tt
  1606.08438}}].

\bibitem{Heidenreich:2016aqi}
B.~Heidenreich, M.~Reece and T.~Rudelius, \emph{{Evidence for a sublattice weak
  gravity conjecture}},
  \href{http://dx.doi.org/10.1007/JHEP08(2017)025}{\emph{JHEP} {\bf 08} (2017)
  025}, [\href{https://arxiv.org/abs/1606.08437}{{\tt 1606.08437}}].

\bibitem{Hod:2017uqc}
S.~Hod, \emph{{A proof of the weak gravity conjecture}},
  \href{http://dx.doi.org/10.1142/S0218271817420044}{\emph{Int. J. Mod. Phys.
  D} {\bf 26} (2017) 1742004}, [\href{https://arxiv.org/abs/1705.06287}{{\tt
  1705.06287}}].

\bibitem{Urbano:2018kax}
A.~Urbano, \emph{{Towards a proof of the Weak Gravity Conjecture}},
  \href{https://arxiv.org/abs/1810.05621}{{\tt 1810.05621}}.

\bibitem{Montero:2018fns}
M.~Montero, \emph{{A Holographic Derivation of the Weak Gravity Conjecture}},
  \href{http://dx.doi.org/10.1007/JHEP03(2019)157}{\emph{JHEP} {\bf 03} (2019)
  157}, [\href{https://arxiv.org/abs/1812.03978}{{\tt 1812.03978}}].

\bibitem{Lee:2018urn}
S.-J. Lee, W.~Lerche and T.~Weigand, \emph{{Tensionless Strings and the Weak
  Gravity Conjecture}},
  \href{http://dx.doi.org/10.1007/JHEP10(2018)164}{\emph{JHEP} {\bf 10} (2018)
  164}, [\href{https://arxiv.org/abs/1808.05958}{{\tt 1808.05958}}].

\bibitem{Cheung:2014vva}
C.~Cheung and G.~N. Remmen, \emph{{Naturalness and the Weak Gravity
  Conjecture}},
  \href{http://dx.doi.org/10.1103/PhysRevLett.113.051601}{\emph{Phys. Rev.
  Lett.} {\bf 113} (2014) 051601}, [\href{https://arxiv.org/abs/1402.2287}{{\tt
  1402.2287}}].

\bibitem{Heidenreich:2015nta}
B.~Heidenreich, M.~Reece and T.~Rudelius, \emph{{Sharpening the Weak Gravity
  Conjecture with Dimensional Reduction}},
  \href{http://dx.doi.org/10.1007/JHEP02(2016)140}{\emph{JHEP} {\bf 02} (2016)
  140}, [\href{https://arxiv.org/abs/1509.06374}{{\tt 1509.06374}}].

\bibitem{Lee:2019tst}
S.-J. Lee, W.~Lerche and T.~Weigand, \emph{{Modular Fluxes, Elliptic Genera,
  and Weak Gravity Conjectures in Four Dimensions}},
  \href{http://dx.doi.org/10.1007/JHEP08(2019)104}{\emph{JHEP} {\bf 08} (2019)
  104}, [\href{https://arxiv.org/abs/1901.08065}{{\tt 1901.08065}}].

\bibitem{ArkaniHamed:2005yv}
N.~Arkani-Hamed, S.~Dimopoulos and S.~Kachru, \emph{{Predictive landscapes and
  new physics at a TeV}},  \href{https://arxiv.org/abs/hep-th/0501082}{{\tt
  hep-th/0501082}}.

\bibitem{Dvali:2007wp}
G.~Dvali and M.~Redi, \emph{{Black Hole Bound on the Number of Species and
  Quantum Gravity at LHC}},
  \href{http://dx.doi.org/10.1103/PhysRevD.77.045027}{\emph{Phys. Rev. D} {\bf
  77} (2008) 045027}, [\href{https://arxiv.org/abs/0710.4344}{{\tt
  0710.4344}}].

\bibitem{Dvali:2007hz}
G.~Dvali, \emph{{Black Holes and Large N Species Solution to the Hierarchy
  Problem}}, \href{http://dx.doi.org/10.1002/prop.201000009}{\emph{Fortsch.
  Phys.} {\bf 58} (2010) 528--536},
  [\href{https://arxiv.org/abs/0706.2050}{{\tt 0706.2050}}].

\bibitem{Dvali:2010vm}
G.~Dvali and C.~Gomez, \emph{{Species and Strings}},
  \href{https://arxiv.org/abs/1004.3744}{{\tt 1004.3744}}.

\bibitem{Rudelius:2015xta}
T.~Rudelius, \emph{{Constraints on Axion Inflation from the Weak Gravity
  Conjecture}}, \href{http://dx.doi.org/10.1088/1475-7516/2015/09/020,
  10.1088/1475-7516/2015/9/020}{\emph{JCAP} {\bf 1509} (2015) 020},
  [\href{https://arxiv.org/abs/1503.00795}{{\tt 1503.00795}}].

\bibitem{Montero:2015ofa}
M.~Montero, A.~M. Uranga and I.~Valenzuela, \emph{{Transplanckian axions!?}},
  \href{http://dx.doi.org/10.1007/JHEP08(2015)032}{\emph{JHEP} {\bf 08} (2015)
  032}, [\href{https://arxiv.org/abs/1503.03886}{{\tt 1503.03886}}].

\bibitem{Brown:2015iha}
J.~Brown, W.~Cottrell, G.~Shiu and P.~Soler, \emph{{Fencing in the Swampland:
  Quantum Gravity Constraints on Large Field Inflation}},
  \href{http://dx.doi.org/10.1007/JHEP10(2015)023}{\emph{JHEP} {\bf 10} (2015)
  023}, [\href{https://arxiv.org/abs/1503.04783}{{\tt 1503.04783}}].

\bibitem{Hebecker:2015rya}
A.~Hebecker, P.~Mangat, F.~Rompineve and L.~T. Witkowski, \emph{{Winding out of
  the Swamp: Evading the Weak Gravity Conjecture with F-term Winding
  Inflation?}},
  \href{http://dx.doi.org/10.1016/j.physletb.2015.07.026}{\emph{Phys. Lett. B}
  {\bf 748} (2015) 455--462}, [\href{https://arxiv.org/abs/1503.07912}{{\tt
  1503.07912}}].

\bibitem{Palti:2017elp}
E.~Palti, \emph{{The Weak Gravity Conjecture and Scalar Fields}},
  \href{http://dx.doi.org/10.1007/JHEP08(2017)034}{\emph{JHEP} {\bf 08} (2017)
  034}, [\href{https://arxiv.org/abs/1705.04328}{{\tt 1705.04328}}].

\bibitem{Heidenreich:2019zkl}
B.~Heidenreich, M.~Reece and T.~Rudelius, \emph{{Repulsive Forces and the Weak
  Gravity Conjecture}},
  \href{http://dx.doi.org/10.1007/JHEP10(2019)055}{\emph{JHEP} {\bf 10} (2019)
  055}, [\href{https://arxiv.org/abs/1906.02206}{{\tt 1906.02206}}].

\bibitem{Lee:2018spm}
S.-J. Lee, W.~Lerche and T.~Weigand, \emph{{A Stringy Test of the Scalar Weak
  Gravity Conjecture}},
  \href{http://dx.doi.org/10.1016/j.nuclphysb.2018.11.001}{\emph{Nucl. Phys. B}
  {\bf 938} (2019) 321--350}, [\href{https://arxiv.org/abs/1810.05169}{{\tt
  1810.05169}}].

\bibitem{Gendler:2020dfp}
N.~Gendler and I.~Valenzuela, \emph{{Merging the Weak Gravity and Distance
  Conjectures Using BPS Extremal Black Holes}},
  \href{https://arxiv.org/abs/2004.10768}{{\tt 2004.10768}}.

\bibitem{Ooguri:2016pdq}
H.~Ooguri and C.~Vafa, \emph{{Non-supersymmetric AdS and the Swampland}},
  \href{http://dx.doi.org/10.4310/ATMP.2017.v21.n7.a8}{\emph{Adv. Theor. Math.
  Phys.} {\bf 21} (2017) 1787--1801},
  [\href{https://arxiv.org/abs/1610.01533}{{\tt 1610.01533}}].

\bibitem{Craig:2018yvw}
N.~Craig, I.~Garcia~Garcia and S.~Koren, \emph{{Discrete Gauge Symmetries and
  the Weak Gravity Conjecture}},
  \href{http://dx.doi.org/10.1007/JHEP05(2019)140}{\emph{JHEP} {\bf 05} (2019)
  140}, [\href{https://arxiv.org/abs/1812.08181}{{\tt 1812.08181}}].

\bibitem{Buratti:2020kda}
G.~Buratti, J.~Calderon, A.~Mininno and A.~M. Uranga, \emph{{Discrete
  Symmetries, Weak Coupling Conjecture and Scale Separation in AdS Vacua}},
  \href{http://dx.doi.org/10.1007/JHEP06(2020)083}{\emph{JHEP} {\bf 06} (2020)
  083}, [\href{https://arxiv.org/abs/2003.09740}{{\tt 2003.09740}}].

\bibitem{Saraswat:2016eaz}
P.~Saraswat, \emph{{Weak gravity conjecture and effective field theory}},
  \href{http://dx.doi.org/10.1103/PhysRevD.95.025013}{\emph{Phys. Rev. D} {\bf
  95} (2017) 025013}, [\href{https://arxiv.org/abs/1608.06951}{{\tt
  1608.06951}}].

\bibitem{Montero:2019ekk}
M.~Montero, T.~Van~Riet and G.~Venken, \emph{{Festina Lente: EFT Constraints
  from Charged Black Hole Evaporation in de Sitter}},
  \href{http://dx.doi.org/10.1007/JHEP01(2020)039}{\emph{JHEP} {\bf 01} (2020)
  039}, [\href{https://arxiv.org/abs/1910.01648}{{\tt 1910.01648}}].

\bibitem{Ooguri:2006in}
H.~Ooguri and C.~Vafa, \emph{{On the Geometry of the String Landscape and the
  Swampland}},
  \href{http://dx.doi.org/10.1016/j.nuclphysb.2006.10.033}{\emph{Nucl. Phys. B}
  {\bf 766} (2007) 21--33}, [\href{https://arxiv.org/abs/hep-th/0605264}{{\tt
  hep-th/0605264}}].

\bibitem{Bedroya:2019snp}
A.~Bedroya and C.~Vafa, \emph{{Trans-Planckian Censorship and the Swampland}},
  \href{http://dx.doi.org/10.1007/JHEP09(2020)123}{\emph{JHEP} {\bf 09} (2020)
  123}, [\href{https://arxiv.org/abs/1909.11063}{{\tt 1909.11063}}].

\bibitem{Andriot:2020lea}
D.~Andriot, N.~Cribiori and D.~Erkinger, \emph{{The web of swampland
  conjectures and the TCC bound}},
  \href{https://arxiv.org/abs/2004.00030}{{\tt 2004.00030}}.

\bibitem{Lanza:2020qmt}
S.~Lanza, F.~Marchesano, L.~Martucci and I.~Valenzuela, \emph{{Swampland
  Conjectures for Strings and Membranes}},
  \href{https://arxiv.org/abs/2006.15154}{{\tt 2006.15154}}.

\bibitem{Lee:2019wij}
S.-J. Lee, W.~Lerche and T.~Weigand, \emph{{Emergent Strings from Infinite
  Distance Limits}},  \href{https://arxiv.org/abs/1910.01135}{{\tt
  1910.01135}}.

\bibitem{Gonzalo:2019gjp}
E.~Gonzalo and L.~E. Ib\'a\~nez, \emph{{A Strong Scalar Weak Gravity Conjecture
  and Some Implications}},
  \href{http://dx.doi.org/10.1007/JHEP08(2019)118}{\emph{JHEP} {\bf 08} (2019)
  118}, [\href{https://arxiv.org/abs/1903.08878}{{\tt 1903.08878}}].

\bibitem{Freivogel:2019mtr}
B.~Freivogel, T.~Gasenzer, A.~Hebecker and S.~Leonhardt, \emph{{A Conjecture on
  the Minimal Size of Bound States}},
  \href{http://dx.doi.org/10.21468/SciPostPhys.8.4.058}{\emph{SciPost Phys.}
  {\bf 8} (2020) 058}, [\href{https://arxiv.org/abs/1912.09485}{{\tt
  1912.09485}}].

\bibitem{DallAgata:2020ino}
G.~Dall'Agata and M.~Morittu, \emph{{Covariant formulation of BPS black holes
  and the scalar weak gravity conjecture}},
  \href{http://dx.doi.org/10.1007/JHEP03(2020)192}{\emph{JHEP} {\bf 03} (2020)
  192}, [\href{https://arxiv.org/abs/2001.10542}{{\tt 2001.10542}}].

\bibitem{Benakli:2020pkm}
K.~Benakli, C.~Branchina and G.~Lafforgue-Marmet, \emph{{Revisiting the Scalar
  Weak Gravity Conjecture}},
  \href{http://dx.doi.org/10.1140/epjc/s10052-020-8268-0}{\emph{Eur. Phys. J.
  C} {\bf 80} (2020) 742}, [\href{https://arxiv.org/abs/2004.12476}{{\tt
  2004.12476}}].

\bibitem{Gonzalo:2020kke}
E.~Gonzalo and L.~E. Ib{\'a}{\~n}ez, \emph{{Pair Production and Gravity as the
  Weakest Force}},  \href{https://arxiv.org/abs/2005.07720}{{\tt 2005.07720}}.

\bibitem{Grimm:2018ohb}
T.~W. Grimm, E.~Palti and I.~Valenzuela, \emph{{Infinite Distances in Field
  Space and Massless Towers of States}},
  \href{http://dx.doi.org/10.1007/JHEP08(2018)143}{\emph{JHEP} {\bf 08} (2018)
  143}, [\href{https://arxiv.org/abs/1802.08264}{{\tt 1802.08264}}].

\bibitem{Grimm:2018cpv}
T.~W. Grimm, C.~Li and E.~Palti, \emph{{Infinite Distance Networks in Field
  Space and Charge Orbits}},
  \href{http://dx.doi.org/10.1007/JHEP03(2019)016}{\emph{JHEP} {\bf 03} (2019)
  016}, [\href{https://arxiv.org/abs/1811.02571}{{\tt 1811.02571}}].

\bibitem{Corvilain:2018lgw}
P.~Corvilain, T.~W. Grimm and I.~Valenzuela, \emph{{The Swampland Distance
  Conjecture for K\"ahler moduli}},
  \href{http://dx.doi.org/10.1007/JHEP08(2019)075}{\emph{JHEP} {\bf 08} (2019)
  075}, [\href{https://arxiv.org/abs/1812.07548}{{\tt 1812.07548}}].

\bibitem{schmid}
W.~Schmid, \emph{Variation of hodge structure: The singularities of the period
  mapping}, \href{http://dx.doi.org/10.1007/BF01389674}{\emph{Inventiones
  mathematicae} {\bf 22} (1973) 211--319}.

\bibitem{Marchesano:2019ifh}
F.~Marchesano and M.~Wiesner, \emph{{Instantons and infinite distances}},
  \href{http://dx.doi.org/10.1007/JHEP08(2019)088}{\emph{JHEP} {\bf 08} (2019)
  088}, [\href{https://arxiv.org/abs/1904.04848}{{\tt 1904.04848}}].

\bibitem{Baume:2019sry}
F.~Baume, F.~Marchesano and M.~Wiesner, \emph{{Instanton Corrections and
  Emergent Strings}},
  \href{http://dx.doi.org/10.1007/JHEP04(2020)174}{\emph{JHEP} {\bf 04} (2020)
  174}, [\href{https://arxiv.org/abs/1912.02218}{{\tt 1912.02218}}].

\bibitem{Lee:2019xtm}
S.-J. Lee, W.~Lerche and T.~Weigand, \emph{{Emergent Strings, Duality and Weak
  Coupling Limits for Two-Form Fields}},
  \href{https://arxiv.org/abs/1904.06344}{{\tt 1904.06344}}.

\bibitem{Lee:2020gvu}
S.-J. Lee, W.~Lerche, G.~Lockhart and T.~Weigand, \emph{{Quasi-Jacobi Forms,
  Elliptic Genera and Strings in Four Dimensions}},
  \href{https://arxiv.org/abs/2005.10837}{{\tt 2005.10837}}.

\bibitem{Font:2019cxq}
A.~Font, A.~Herr\'aez and L.~E. Ib\'a\~nez, \emph{{The Swampland Distance
  Conjecture and Towers of Tensionless Branes}},
  \href{http://dx.doi.org/10.1007/JHEP08(2019)044}{\emph{JHEP} {\bf 08} (2019)
  044}, [\href{https://arxiv.org/abs/1904.05379}{{\tt 1904.05379}}].

\bibitem{Herraez:2020tih}
A.~Herraez, \emph{{A Note on Membrane Interactions and the Scalar potential}},
  \href{http://dx.doi.org/10.1007/JHEP10(2020)009}{\emph{JHEP} {\bf 10} (2020)
  009}, [\href{https://arxiv.org/abs/2006.01160}{{\tt 2006.01160}}].

\bibitem{Bielleman:2015ina}
S.~Bielleman, L.~E. Ibanez and I.~Valenzuela, \emph{{Minkowski 3-forms, Flux
  String Vacua, Axion Stability and Naturalness}},
  \href{http://dx.doi.org/10.1007/JHEP12(2015)119}{\emph{JHEP} {\bf 12} (2015)
  119}, [\href{https://arxiv.org/abs/1507.06793}{{\tt 1507.06793}}].

\bibitem{Herraez:2018vae}
A.~Herraez, L.~E. Ibanez, F.~Marchesano and G.~Zoccarato, \emph{{The Type IIA
  Flux Potential, 4-forms and Freed-Witten anomalies}},
  \href{http://dx.doi.org/10.1007/JHEP09(2018)018}{\emph{JHEP} {\bf 09} (2018)
  018}, [\href{https://arxiv.org/abs/1802.05771}{{\tt 1802.05771}}].

\bibitem{Klaewer:2016kiy}
D.~Klaewer and E.~Palti, \emph{{Super-Planckian Spatial Field Variations and
  Quantum Gravity}},
  \href{http://dx.doi.org/10.1007/JHEP01(2017)088}{\emph{JHEP} {\bf 01} (2017)
  088}, [\href{https://arxiv.org/abs/1610.00010}{{\tt 1610.00010}}].

\bibitem{Aghanim:2018eyx}
{\scshape Planck} collaboration, N.~Aghanim et~al., \emph{{Planck 2018 results.
  VI. Cosmological parameters}},
  \href{http://dx.doi.org/10.1051/0004-6361/201833910}{\emph{Astron.
  Astrophys.} {\bf 641} (2020) A6},
  [\href{https://arxiv.org/abs/1807.06209}{{\tt 1807.06209}}].

\bibitem{Scalisi:2018eaz}
M.~Scalisi and I.~Valenzuela, \emph{{Swampland distance conjecture, inflation
  and $\alpha$-attractors}},
  \href{http://dx.doi.org/10.1007/JHEP08(2019)160}{\emph{JHEP} {\bf 08} (2019)
  160}, [\href{https://arxiv.org/abs/1812.07558}{{\tt 1812.07558}}].

\bibitem{linde_chaotic_1983}
A.~D. Linde, \emph{{Chaotic inflation}},
  \href{http://dx.doi.org/{https://doi.org/10.1016/0370-2693(83)90837-7}}{\emph{Physics
  Letters B} {\bf 129} (1983) 177 -- 181}.

\bibitem{Akrami:2018odb}
{\scshape Planck} collaboration, Y.~Akrami et~al., \emph{{Planck 2018 results.
  X. Constraints on inflation}},
  \href{http://dx.doi.org/10.1051/0004-6361/201833887}{\emph{Astron.
  Astrophys.} {\bf 641} (2020) A10},
  [\href{https://arxiv.org/abs/1807.06211}{{\tt 1807.06211}}].

\bibitem{Calderon-Infante:2020dhm}
J.~Calder\'on-Infante, A.~M. Uranga and I.~Valenzuela, \emph{{The Convex Hull
  Swampland Distance Conjecture and Bounds on Non-geodesics}},
  \href{https://arxiv.org/abs/2012.00034}{{\tt 2012.00034}}.

\bibitem{Baume:2016psm}
F.~Baume and E.~Palti, \emph{{Backreacted Axion Field Ranges in String
  Theory}}, \href{http://dx.doi.org/10.1007/JHEP08(2016)043}{\emph{JHEP} {\bf
  08} (2016) 043}, [\href{https://arxiv.org/abs/1602.06517}{{\tt 1602.06517}}].

\bibitem{Valenzuela:2016yny}
I.~Valenzuela, \emph{{Backreaction Issues in Axion Monodromy and Minkowski
  4-forms}}, \href{http://dx.doi.org/10.1007/JHEP06(2017)098}{\emph{JHEP} {\bf
  06} (2017) 098}, [\href{https://arxiv.org/abs/1611.00394}{{\tt 1611.00394}}].

\bibitem{Blumenhagen:2017cxt}
R.~Blumenhagen, I.~Valenzuela and F.~Wolf, \emph{{The Swampland Conjecture and
  F-term Axion Monodromy Inflation}},
  \href{http://dx.doi.org/10.1007/JHEP07(2017)145}{\emph{JHEP} {\bf 07} (2017)
  145}, [\href{https://arxiv.org/abs/1703.05776}{{\tt 1703.05776}}].

\bibitem{Grimm:2019ixq}
T.~W. Grimm, C.~Li and I.~Valenzuela, \emph{{Asymptotic Flux Compactifications
  and the Swampland}},
  \href{http://dx.doi.org/10.1007/JHEP06(2020)009}{\emph{JHEP} {\bf 06} (2020)
  009}, [\href{https://arxiv.org/abs/1910.09549}{{\tt 1910.09549}}].

\bibitem{Lust:2019zwm}
D.~L\"ust, E.~Palti and C.~Vafa, \emph{{AdS and the Swampland}},
  \href{http://dx.doi.org/10.1016/j.physletb.2019.134867}{\emph{Phys. Lett. B}
  {\bf 797} (2019) 134867}, [\href{https://arxiv.org/abs/1906.05225}{{\tt
  1906.05225}}].

\bibitem{DeWolfe:2005uu}
O.~DeWolfe, A.~Giryavets, S.~Kachru and W.~Taylor, \emph{{Type IIA moduli
  stabilization}},
  \href{http://dx.doi.org/10.1088/1126-6708/2005/07/066}{\emph{JHEP} {\bf 07}
  (2005) 066}, [\href{https://arxiv.org/abs/hep-th/0505160}{{\tt
  hep-th/0505160}}].

\bibitem{McOrist:2012yc}
J.~McOrist and S.~Sethi, \emph{{M-theory and Type IIA Flux Compactifications}},
  \href{http://dx.doi.org/10.1007/JHEP12(2012)122}{\emph{JHEP} {\bf 12} (2012)
  122}, [\href{https://arxiv.org/abs/1208.0261}{{\tt 1208.0261}}].

\bibitem{Junghans:2020acz}
D.~Junghans, \emph{{O-Plane Backreaction and Scale Separation in Type IIA Flux
  Vacua}}, \href{http://dx.doi.org/10.1002/prop.202000040}{\emph{Fortsch.
  Phys.} {\bf 68} (2020) 2000040},
  [\href{https://arxiv.org/abs/2003.06274}{{\tt 2003.06274}}].

\bibitem{Marchesano:2020qvg}
F.~Marchesano, E.~Palti, J.~Quirant and A.~Tomasiello, \emph{{On supersymmetric
  AdS$_{4}$ orientifold vacua}},
  \href{http://dx.doi.org/10.1007/JHEP08(2020)087}{\emph{JHEP} {\bf 08} (2020)
  087}, [\href{https://arxiv.org/abs/2003.13578}{{\tt 2003.13578}}].

\bibitem{Baume:2020dqd}
F.~Baume and J.~Calder\'on~Infante, \emph{{Tackling the SDC in AdS with CFTs}},
   \href{https://arxiv.org/abs/2011.03583}{{\tt 2011.03583}}.

\bibitem{Perlmutter:2020buo}
E.~Perlmutter, L.~Rastelli, C.~Vafa and I.~Valenzuela, \emph{{A CFT Distance
  Conjecture}},  \href{https://arxiv.org/abs/2011.10040}{{\tt 2011.10040}}.

\bibitem{Draper:2019utz}
P.~Draper and S.~Farkas, \emph{{Transplanckian Censorship and the Local
  Swampland Distance Conjecture}},
  \href{http://dx.doi.org/10.1007/JHEP01(2020)133}{\emph{JHEP} {\bf 01} (2020)
  133}, [\href{https://arxiv.org/abs/1910.04804}{{\tt 1910.04804}}].

\bibitem{Heidenreich:2017sim}
B.~Heidenreich, M.~Reece and T.~Rudelius, \emph{{The Weak Gravity Conjecture
  and Emergence from an Ultraviolet Cutoff}},
  \href{http://dx.doi.org/10.1140/epjc/s10052-018-5811-3}{\emph{Eur. Phys. J.
  C} {\bf 78} (2018) 337}, [\href{https://arxiv.org/abs/1712.01868}{{\tt
  1712.01868}}].

\bibitem{Heidenreich:2018kpg}
B.~Heidenreich, M.~Reece and T.~Rudelius, \emph{{Emergence of Weak Coupling at
  Large Distance in Quantum Gravity}},
  \href{http://dx.doi.org/10.1103/PhysRevLett.121.051601}{\emph{Phys. Rev.
  Lett.} {\bf 121} (2018) 051601},
  [\href{https://arxiv.org/abs/1802.08698}{{\tt 1802.08698}}].

\bibitem{Strominger:1995cz}
A.~Strominger, \emph{{Massless black holes and conifolds in string theory}},
  \href{http://dx.doi.org/10.1016/0550-3213(95)00287-3}{\emph{Nucl. Phys. B}
  {\bf 451} (1995) 96--108}, [\href{https://arxiv.org/abs/hep-th/9504090}{{\tt
  hep-th/9504090}}].

\bibitem{Freivogel:2016qwc}
B.~Freivogel and M.~Kleban, \emph{{Vacua Morghulis}},
  \href{https://arxiv.org/abs/1610.04564}{{\tt 1610.04564}}.

\bibitem{Maldacena:1998uz}
J.~M. Maldacena, J.~Michelson and A.~Strominger, \emph{{Anti-de Sitter
  fragmentation}},
  \href{http://dx.doi.org/10.1088/1126-6708/1999/02/011}{\emph{JHEP} {\bf 02}
  (1999) 011}, [\href{https://arxiv.org/abs/hep-th/9812073}{{\tt
  hep-th/9812073}}].

\bibitem{witten_instability_1982}
E.~Witten, \emph{{Instability of the Kaluza-Klein vacuum}},
  \href{http://dx.doi.org/https://doi.org/10.1016/0550-3213(82)90007-4}{\emph{Nuclear
  Physics B} {\bf 195} (1982) 481 -- 492}.

\bibitem{GarciaEtxebarria:2020xsr}
I.~n. Garc\'\i{}a~Etxebarria, M.~Montero, K.~Sousa and I.~Valenzuela,
  \emph{{Nothing is certain in string compactifications}},
  \href{https://arxiv.org/abs/2005.06494}{{\tt 2005.06494}}.

\bibitem{Ibanez:2017kvh}
L.~E. Ibanez, V.~Martin-Lozano and I.~Valenzuela, \emph{{Constraining Neutrino
  Masses, the Cosmological Constant and BSM Physics from the Weak Gravity
  Conjecture}}, \href{http://dx.doi.org/10.1007/JHEP11(2017)066}{\emph{JHEP}
  {\bf 11} (2017) 066}, [\href{https://arxiv.org/abs/1706.05392}{{\tt
  1706.05392}}].

\bibitem{Obied:2018sgi}
G.~Obied, H.~Ooguri, L.~Spodyneiko and C.~Vafa, \emph{{De Sitter Space and the
  Swampland}},  \href{https://arxiv.org/abs/1806.08362}{{\tt 1806.08362}}.

\bibitem{Ooguri:2018wrx}
H.~Ooguri, E.~Palti, G.~Shiu and C.~Vafa, \emph{{Distance and de Sitter
  Conjectures on the Swampland}},
  \href{http://dx.doi.org/10.1016/j.physletb.2018.11.018}{\emph{Phys. Lett. B}
  {\bf 788} (2019) 180--184}, [\href{https://arxiv.org/abs/1810.05506}{{\tt
  1810.05506}}].

\bibitem{Garg:2018reu}
S.~K. Garg and C.~Krishnan, \emph{{Bounds on Slow Roll and the de Sitter
  Swampland}}, \href{http://dx.doi.org/10.1007/JHEP11(2019)075}{\emph{JHEP}
  {\bf 11} (2019) 075}, [\href{https://arxiv.org/abs/1807.05193}{{\tt
  1807.05193}}].

\bibitem{Danielsson:2018ztv}
U.~H. Danielsson and T.~Van~Riet, \emph{{What if string theory has no de Sitter
  vacua?}}, \href{http://dx.doi.org/10.1142/S0218271818300070}{\emph{Int. J.
  Mod. Phys. D} {\bf 27} (2018) 1830007},
  [\href{https://arxiv.org/abs/1804.01120}{{\tt 1804.01120}}].

\bibitem{Cicoli:2018kdo}
M.~Cicoli, S.~De~Alwis, A.~Maharana, F.~Muia and F.~Quevedo, \emph{{De Sitter
  vs Quintessence in String Theory}},
  \href{http://dx.doi.org/10.1002/prop.201800079}{\emph{Fortsch. Phys.} {\bf
  67} (2019) 1800079}, [\href{https://arxiv.org/abs/1808.08967}{{\tt
  1808.08967}}].

\bibitem{Dine:1985he}
M.~Dine and N.~Seiberg, \emph{{Is the Superstring Weakly Coupled?}},
  \href{http://dx.doi.org/10.1016/0370-2693(85)90927-X}{\emph{Phys. Lett. B}
  {\bf 162} (1985) 299--302}.

\bibitem{Kachru:2003aw}
S.~Kachru, R.~Kallosh, A.~D. Linde and S.~P. Trivedi, \emph{{De Sitter vacua in
  string theory}},
  \href{http://dx.doi.org/10.1103/PhysRevD.68.046005}{\emph{Phys. Rev. D} {\bf
  68} (2003) 046005}, [\href{https://arxiv.org/abs/hep-th/0301240}{{\tt
  hep-th/0301240}}].

\bibitem{Maldacena:2000mw}
J.~M. Maldacena and C.~Nunez, \emph{{Supergravity description of field theories
  on curved manifolds and a no go theorem}},
  \href{http://dx.doi.org/10.1142/S0217751X01003937}{\emph{Int. J. Mod. Phys.
  A} {\bf 16} (2001) 822--855},
  [\href{https://arxiv.org/abs/hep-th/0007018}{{\tt hep-th/0007018}}].

\bibitem{Hertzberg:2007wc}
M.~P. Hertzberg, S.~Kachru, W.~Taylor and M.~Tegmark, \emph{{Inflationary
  Constraints on Type IIA String Theory}},
  \href{http://dx.doi.org/10.1088/1126-6708/2007/12/095}{\emph{JHEP} {\bf 12}
  (2007) 095}, [\href{https://arxiv.org/abs/0711.2512}{{\tt 0711.2512}}].

\bibitem{Flauger:2008ad}
R.~Flauger, S.~Paban, D.~Robbins and T.~Wrase, \emph{{Searching for slow-roll
  moduli inflation in massive type IIA supergravity with metric fluxes}},
  \href{http://dx.doi.org/10.1103/PhysRevD.79.086011}{\emph{Phys. Rev. D} {\bf
  79} (2009) 086011}, [\href{https://arxiv.org/abs/0812.3886}{{\tt
  0812.3886}}].

\bibitem{Wrase:2010ew}
T.~Wrase and M.~Zagermann, \emph{{On Classical de Sitter Vacua in String
  Theory}}, \href{http://dx.doi.org/10.1002/prop.201000053}{\emph{Fortsch.
  Phys.} {\bf 58} (2010) 906--910},
  [\href{https://arxiv.org/abs/1003.0029}{{\tt 1003.0029}}].

\bibitem{Andriot:2019wrs}
D.~Andriot, \emph{{Open problems on classical de Sitter solutions}},
  \href{http://dx.doi.org/10.1002/prop.201900026}{\emph{Fortsch. Phys.} {\bf
  67} (2019) 1900026}, [\href{https://arxiv.org/abs/1902.10093}{{\tt
  1902.10093}}].

\end{thebibliography}\endgroup
\bibliographystyle{JHEP}

\end{document}